\documentclass{elsart}
\usepackage{microtype}
\usepackage{ragged2e}
\usepackage{graphics}
\usepackage{graphicx}
\usepackage{epsfig}
\usepackage{amssymb}
\usepackage{lineno}
\usepackage{blindtext}
\usepackage{mathtools}
\usepackage[breaklinks=true]{hyperref}
\usepackage{breakcites}
\usepackage{cite}
\begin{document}
\begin{frontmatter}
\title{Nonlinear energy loss in the oscillations of coated and uncoated bubbles: Role of thermal, radiation damping and encapsulating shell at various excitation pressures}
\author{A.J. Sojahrood \thanksref{a},\thanksref{b}} \footnote{Email: amin.jafarisojahrood@ryerson.ca},
\author{H. Haghi \thanksref{a},\thanksref{b}}
\author{Q. Li \thanksref{c}}
\author{T.M. Porter \thanksref{c}}
\author{R. Karshfian \thanksref{a},\thanksref{b}}
\author{and  M. C. Kolios \thanksref{a},\thanksref{b}}
\address[a]{Department of Physics, Ryerson University, Toronto, Canada}
\address[b]{Institute for Biomedical Engineering, Science and Technology (IBEST) a partnership between Ryerson University and St. Mike's Hospital, Toronto, Ontario, Canada}
\address[c]{Department of Biomedical Engineering, Boston University, Boston, MA, USA.}
\vspace{-0.5cm}
\begin{abstract}
A simple generalized model (GM) for coated bubbles accounting for the effect of compressibility of the liquid is presented. The GM was then coupled with nonlinear ODEs that account for the thermal effects. Starting with mass and momentum conservation equations for a bubbly liquid and using the GM, nonlinear pressure dependent terms were derived for energy dissipation due to thermal damping (Td), radiation damping (Rd) and dissipation due to the viscosity of liquid (Ld) and coating (Cd). The dissipated energies were solved for uncoated and coated 2- 20  $\mu m$ bubbles over a frequency range of $0.25f_r-2.5f_r$ ($f_r$ is the bubble resonance) and for various acoustic pressures (1kPa-300kPa). Thermal effects were examined for air and C3F8 gas cores in each case. For uncoated bubbles with an air gas core and a diameter larger than 4 $\mu m$, thermal damping is the strongest damping factor. When pressure increases, the contributions of Rd grow faster and become the dominant damping mechanism for pressure dependent resonance frequencies (e.g. fundamental and super harmonic resonances). For coated bubbles, Cd is the strongest damping mechanism. As pressure increases Rd contributes more to damping compared to Ld and Td. In case of air bubbles, as pressure increases, the linear thermal model largely deviates from the nonlinear model and accurate modeling requires inclusion of the full thermal model. However, for coated C3F8 bubbles of diameter 1-8 $\mu m$, typically used in medical ultrasound, thermal effects maybe neglected even at higher pressures. We show that the scattering to damping ratio (STDR), a measure of the effectiveness of the bubble as contrast agent, is pressure dependent and can be maximized for specific frequency ranges and pressures.
\end{abstract}
\end{frontmatter}
\section{Introduction}
\justify
Acoustically excited bubbles are oscillating gas or vapor cores that are found in free (uncoated) or coated form (e.g. encapsulated by  a  lipid  or  polymer shell). They form the core of several applications in liquids \cite{1}, and in soft and palpable matter (e.g. tissue \cite{1,2,3} or sediments[1,4,5]). When excited by a sound field they can oscillate with amplitudes large enough to destroy most materials \cite{1,6,7}; enhance chemical reactions \cite{8,9,10}, and act as healing or diagnostic agents in medicine \cite{11,12,13}. During these high amplitude oscillations, temperatures reached in the gas core are high enough to turn the bubbles into tiny light bulbs \cite{9,10,14}.\\
Dynamics of the acoustically excited bubbles are nonlinear and chaotic. These dynamics have been the subject of numerous experimental \cite{1,15,17,18,19,20} and numerical \cite{21,22,23,24,25,26,27,28,29,30} studies . Achieving the full potential of bubbles in applications and understanding their role in the associated phenomena not only requires a detailed knowledge over their complex behavior but also on the effect of bubble oscillations on the propagation of acoustic waves.\\Propagation properties of acoustic waves in bubbly media are considerably different from those in single-phase media (e.g. pure water) \cite{31,32,33,34,35,36}. Derivation of the correct wave properties are essential for the understanding and predicting of bubble oscillations in different locations of the media and consequently optimization of bubble related applications.\\
One of the characteristic properties of bubble oscillations in a medium is the increased attenuation. Bubbles attenuate ultrasound through viscous damping due to liquid (Ld) and coating (Cd) viscous effects, radiation damping (Rd) and thermal damping (Td) \cite{33,34,35,36,37,38,39}. The changes in attenuation of bubbly media has significance in studies related to oceanography \cite{4,5} and acoustic characterization of coated bubbles used as ultrasound contrast agents (UCAs) \cite{38,39,40,41,42,43}. Furthermore, knowledge of the nonlinear attenuation in the medium will help in optimization of applications related to sonochemistry \cite{35,36,37} and medical ultrasound \cite{11,12,13,44,45} by enhancing and controlling the acoustic energy at the target.\\ The majority of the previous studies employed linear approximations \cite{33,38,39,40,41,42} (limited to very small bubble oscillation amplitudes) to calculate the damping parameters during bubble oscillations. Thus, linear models neglect the dependence of the dissipated energy on the local pressure \cite{33,34,35,36} and are not applicable to conditions under which bubbles are excited in most applications. To properly account for the pressure dependence of attenuation new approaches are developed that only consider the pressure dependence of the radiation damping while using linear terms for the rest of the damping factors (thermal and viscous damping) \cite{39}. However other damping factors are also pressure dependent and linear approximations reduce the accuracy of model predictions.\\ To account for the pressure dependence of all of the damping parameters, mass and momentum conservation equations were used by Louisnard \cite{34} and nonlinear energy terms for Td and Ld were derived for a bubbly liquid using the Rayleigh-Plesset equation \cite{46}. Jamshidi and Brenner \cite{35} used Louisnards approach \cite{34} and the Keller-Miksis (KM) equation \cite{47} to capture the nonlinear damping term for re-radiated energy by bubbles (Rd). Louisnard \cite{34} and Jamshidi’s \cite{35} study showed that damping from nonlinear oscillations of bubbles can be several orders of magnitude higher than the damping estimated by linear models.\\ We have shown in \cite{36}, that the derived terms in \cite{35} are incorrect as they can lead to non-physical values for Rd; moreover, predictions of their model were not consistent with the results of radiation damping due to re-radiated pressure (Sd) by bubbles \cite{36}. We have corrected the nonlinear damping terms in \cite{36} and predictions of Rd were in excellent agreement with Sd. We have shown that as pressure increases, Rd grows faster than other damping factors and there exist frequency and pressure domains in which Rd is stronger than other damping parameters.\\ Despite the importance of nonlinear thermal losses when it comes to modeling the bubble oscillations, most studies simplify the role of thermal losses on bubble behavior by using models that are derived based on linear approximations\cite{4,5,33,34,35,36}. In addition, for coated bubbles, the role of radiation damping is not completely captured as effects of the compressibility of the liquid are often neglected or simplified \cite{38,48}. It is shown in \cite{35,36,37} that radiation damping is pressure dependent and can be a major contributor to the total damping and should not be neglected. A more complete estimation of the wave attenuation in bubbly media must incorporate the effects of thermal and radiation damping on bubble oscillations and the total dissipated power \cite{34,35,36,37}.\\ In this work we present a generalized model (GM) for the oscillations of coated bubbles. Compressibility effects (similar to the KM model \cite{47}) are added to the Church-Hoff model \cite{38}. We can call this model the KMCH model. KMCH is then coupled with the ordinary differential equations (ODEs) \cite{49} that take into account the thermal effects. In case of uncoated bubbles, to capture the thermal and radiation effects on the oscillations, the Keller-Miksis (KM) model is also coupled with thermal ODEs \cite{49}. Using the equations for conservation of mass and momentum in bubbly liquids and applying the KMCH model, our proposed approach in \cite{36} was applied to derive all the damping terms Rd, Cd, Td and Ld for the oscillations of a coated bubble.\\
The total dissipated power by bubble oscillations was then studied as a function of frequency for various excitation pressures. In each case (free uncoated and coated bubbles) three models were considered; these models were solved for an a) air and b) C3F8 gas core that are generally used in UCAs. In the 3 models, the first model neglects the thermal effects, the second model includes the thermal effects using linear approximations by introducing an artificial thermal viscosity \cite{50} and the third model includes full thermal effects \cite{49,51,52,53} (full thermal model). The second model has widely been used in studies related to oceanography \cite{4,5} or ultrasound contrast agent characterization \cite{38,39,40,41,42}). In this paper we call this the linear thermal model. The third model includes full thermal effects \cite{49,51,52,53} (full thermal model) by solving the ODEs related to thermal loss during the bubble oscillations.\\
Here we will show that for uncoated bubbles, thermal effects are very important; at very low acoustic pressures ($P_A \approx 1 kPa$) predictions of the linear thermal model are in good agreement with the full thermal model. However, as pressure increases predictions of the full thermal and linear thermal model deviate for both gas cores; thus the full thermal model must be used for accurate bubble modeling. In case of C3F8 coated bubbles, however, thermal effects are masked by large dissipation due to the bubble coating and are negligible. Even at high pressures (e.g. 100 kPa), predictions of the linear thermal, full thermal and non-thermal model are in good agreement. However, for coated bubbles with air, thermal effects are important and the full thermal model must be used for accurate prediction of bubble behavior at higher pressures and at frequencies and pressures in which subharmonic (SH) oscillations are generated.\\
For both cases (uncoated and coated bubble) and for both gas types, mechanisms of energy dissipation were studied as a function of frequency and pressure. We show that increasing the excitation pressure leads to a faster growth in Rd compared to other damping mechanisms; thus, optimum frequency and pressure ranges exist in which scattering (STDR) to total damping ratio is maximized. 
\section{Methods}
\subsection{Coated bubble model}
The dynamics of a coated bubble oscillator can be modeled using the Church-Hoff model \cite{38}:
\begin{equation}
\begin{gathered}
\rho \left[R\ddot{R}+\frac{3}{2}\dot{R}^2\right]=\\
\left(P_g-\frac{4\mu_L\dot{R}}{R}-\frac{12\mu_{sh}\epsilon R_0^2\dot{R}}{R^4}-12G_s\epsilon R_0^2 \left(\frac{1}{R^3}-\frac{R_0}{R^4}\right)-P_0-P\right)
\end{gathered}
\end{equation}
Where $\rho$ is the density of the medium, R is the radius at time t, $\dot{R}$ is the bubble wall velocity, $\ddot{R}$ is the bubble wall acceleration, $R_0$ is the initial radius of the bubble, $\mu$ and $\mu_{sh}$ are the viscosity of the liquid and shell (coating) respectively, $\epsilon$ is the thickness of the coating, $G_s$ is the shell shear modulus, $P_g$ is the gas pressure inside the bubble, $P_0$ is the atmospheric pressure (101.325 kPa) and P is the acoustic pressure given by $P=P_asin(2\pi ft)$ with $P_a$ and $f$ are respectively the excitation pressure and frequency. Church-Hoff model (Eq. 1) does not incorporate the effects of the compressibility of the medium. Similar to the KM model \cite{47} we added the effects of the compressibility of the  medium to the first order of Mach number. The generalized model (GM) can be called Keller-Miksis-Church-Hoff (KMCH) model and is written as:
\begin{equation}
\begin{gathered}
\rho \left[\left(1-\frac{\dot{R}}{c}\right)R\ddot{R}+\frac{3}{2}\dot{R}^2\left(1-\frac{\dot{R}}{3c}\right)\right]=\\
\left(1+\frac{\dot{R}}{c}+\frac{R}{c}\frac{d}{dt}\right)\left(P_g-\frac{4\mu_L\dot{R}}{R}-\frac{12\mu_{sh}\epsilon R_0^2\dot{R}}{R^4}-12G_s\epsilon R_0^2 \left(\frac{1}{R^3}-\frac{R_0}{R^4}\right)-P_0-P\right)
\end{gathered}
\end{equation}
Here c is the sound speed in the liquid. In this paper for all of the simulations related to coated bubbles $G_s$=45 MPa and $\mu_{sh}=\frac{1.49(R_0(\mu m)-0.86)}{\theta (nm)}$ \cite{54} ($sh$ stands for shell (coating)) with $\theta=4 nm$.
\subsection{Uncoated Bubble model}
The dynamics of the bubble model including the compressibility effects to the first order of Mach number can be modeled using Keller-Miksis (KM) equation \cite{47}:
\justifying
\begin{equation}
\rho\left[\left(1-\frac{\dot{R}}{c}\right)R\ddot{R}+\frac{3}{2}\dot{R}\left(1-\frac{R}{3c}\right)\right]=\left(1+\frac{\dot{R}}{c}\right)\left(G\right)+\frac{R}{c}\frac{d}{dt}\left(G\right)
\end{equation}
where $G=P_g-\frac{4\mu_L\dot{R}}{R}-\frac{2\sigma}{R}-P_0-P_A sin(2 \pi f t)$.\\
In this equation, R is radius at time t, $R_0$ is the initial bubble radius, $\dot{R}$ is the wall velocity of the bubble, $\ddot{R}$ is the wall acceleration,	$\rho{}$ is the liquid density (998 $\frac{kg}{m^3}$), c is the sound speed (1481 m/s), $P_g$ is the gas pressure, $\sigma{}$ is the surface tension (0.0725 $\frac{N}{m}$), $\mu{}$ is the liquid viscosity (0.001 Pa.s), and $P_A$ and \textit{f} are the amplitude and frequency of the applied acoustic pressure. The values in the parentheses are for pure water at 293$^0$K. In this paper the gas inside the bubble is either air or C3F8 and water is the host media. Depending on the which model is used for the simulation of the thermal effects $P_g$ will be a function that will be defined in the next 3 subsections.\\
\subsection{Non-thermal model}
If the terms related to thermal damping are neglected, $P_g$ in Eq. 2 and 3 can be written in the form of:
\begin{equation}
P_g=P_{g0}\times(\frac{R_0}{R})^{3\gamma}
\end{equation}
Where $\gamma$ is the polytropic exponent and is given by $\frac{C_p}{C_v}$. According to Church-Hoff Model \cite{38} for a coated bubble $P_{g0}=P_0$ where $P_0$ is the atmospheric pressure. For uncoated bubble as given by Keller-Miksis equation \cite{47}. $P_{g0}=P_0-\frac{2\sigma}{R_0}$. In this work we have neglected the small effect of vapor pressure.
\subsection{Full thermal effects}
If thermal effects are considered, $P_g$ is given by Eq. 5 \cite{49,50,51,52,53}:
\begin{equation}
P_g=\frac{N_gKT}{\frac{4}{3}\pi R(t)^3-N_g B}
\end{equation}
Where $N_g$ is the total number of the gas molecules, $K$ is the Boltzman constant and B is the molecular co-volume. The average temperature inside the gas can be calculated using Eq. 6 \cite{49}:
\begin{equation}
\dot{T}=\frac{4\pi R(t)^2}{C_v} \left(\frac{L\left(T_0-T\right)}{L_{th}}-\dot{R}P_g\right)
\end{equation}
Where $C_v$ is the specific heat at constant volume, $T_0$=$293^0$K is the initial gas temperature, $L_{th}$ is the thickness of the thermal boundary layer. $L_{th}$ is given by $L_{th}=min(\sqrt{\frac{aR(t)}{|\dot{R(t)}|}},\frac{R(t)}{\pi})$ where $a$ is the thermal diffusivity of the gas which can be calculated using $a=\frac{L}{C_p \rho_g}$ where L is the gas thermal conductivity and $C_p$ is specific heat at constant pressure and $\rho_g$ is the gas density.\\
\begin{table}
\begin{tabular}{ |p{2cm}||p{3.5cm}|p{2cm}|p{2cm}|p{2cm}|  }
	\hline
	\multicolumn{5}{|c|}{Thermal parameters of the gases at 1 atm} \\
	\hline
	Gas type  & L ($\frac{W}{m^0K}$) &$C_p$$\frac{kJ}{kg^0C}$ &$C_v$ $\frac{kJ}{kg^0C}$&$\rho_g$ $\frac{kg}{m^3}$\\
	\hline
	Air \cite{55}   & 0.01165+C*T \footnotemark[2] &1.0049&   0.7187&1.025\\
	C3F8 \cite{56} &   0.012728  & 0.79   &0.7407&8.17\\
	\hline
\end{tabular}
\caption{Thermal properties of the gases used in simulations.{$^2$ C=$5.528*10^{25}$ $\frac{W}{m^0K^2}$}.}
\label{table:1}
\end{table}
Predictions of the full thermal model have been shown to be in good agreement with predictions of the models that incorporate the thermal effects using the PDEs \cite{51}. To calculate the radial oscillations of the coated bubble and uncoated bubble while including the full thermal effects Eqs. 2 (coated bubble) or Eq. 3 (uncoated bubble) are 
to calculate the radial oscillations of the coated bubble and uncoated bubble while including the full thermal effects Eqs. 2 and Eq. 3 are respectively coupled with Eq. 5 and then solved using the ode45 solver of Matlab.
\subsection{Linear thermal model}
The linear thermal model \cite{33,50} is a popular model that has been widely used in studies related to oceanography \cite{4,5} and the modeling and charecterization of coated bubble oscillations\cite{38,39,40,41,42}. In this model through linearization thermal damping is approximated by adding an artificial viscosity term to the liquid viscosity. Furthermore, a variable isoentropic index is used instead of the polytropic exponent of the gas.\\ In this model $P_g$ is given by:
\begin{equation}
P_g=P_{g0}\left(\frac{R_0}{R}\right)^{3k}
\end{equation} 
Where the polytropic exponent $\gamma$ is replaced by isoentropic indice ($k$):
\begin{equation}
k=\frac{1}{3}\Re(\phi)
\end{equation}
liquid viscosity is artificially increased by adding a thermal viscosity ($\mu_{th}$) to the liquid viscosity. This thermal viscosity ($\mu_{th}$) is given by:
\begin{equation}
\mu_{th}=\frac{P_{g0} \Im(\phi)}{\omega}
\end{equation}

In the above equations the complex term $\phi$ is calculated from

\begin{equation}
\phi=\frac{3\gamma}{1-3\left(\gamma-1\right)i\chi\left[\left(\frac{i}{\chi}\right)^{\frac{1}{2}}coth\left(\frac{i}{\chi}\right)^{\frac{1}{2}}-1\right]}
\end{equation}
where $\gamma$ is the polytropic exponent and $\chi=\frac{D}{\omega R_0^2}$ represents the thermal diffusion length where $D$ is the thermal diffusivity of the gas. $D= \frac{L}{\gamma C_p \rho_g}$ where $C_p$, $\rho_g$, and $L$ are specific heat in constant pressure, density and thermal conductivity of the gas inside the bubble.\\
To calculate the radial oscillations of the coated bubble and uncoated bubble while including the linear thermal effects Eqs. 2 and Eq. 3 are respectively coupled with Eq. 7 and liquid viscosity is increased by $\mu_{th}$. 
\subsection{Derivation of the nonlinear terms of dissipation for GM}
van Wijngaardan \cite{32} and Caflish et al. \cite{31} presented the mass and momentum conservation equations for a
bubbly liquid as:
\begin{equation}
\frac{1}{\rho c^2}\frac{\partial P}{\partial t}+\nabla.v=\frac{\partial \beta}{\partial t}
\end{equation}
and
\begin{equation}
\rho\frac{\partial v}{\partial t}=-\nabla P
\end{equation}
where $c$ is the sound speed, $\rho$ is the density of the medium, $v(r,t)$ is the velocity field, $P(r,t)$ is acoustic pressure, $\beta=\frac{4}{3}N\pi R(t)^3$ is the void fraction where N is number of bubbles per unit volume, and $R(t)$ is the radius of the bubble at time $t$. These two equations can be re-written into an equation of energy conservation, by multiplying (1) by $P$ and (2) by $v$:
\begin{equation}
\frac{\partial}{\partial t}\left(\frac{1}{2}\frac{P^2}{\rho c^2}+\frac{1}{2}\rho v^2\right)=NP\frac{\partial V}{\partial t}
\end{equation}
Multiplying Eq.2 by $N\frac{\partial V}{\partial t}$ results in:
\begin{equation}
\begin{gathered}
\rho N\left(R\ddot{R}+\frac{3}{2}\dot{R}^2\right)\frac{\partial V}{\partial t}-\rho N\frac{\dot{R}}{c}\left(R\ddot{R}+\frac{1}{2}\dot{R}^2\right)\frac{\partial V}{\partial t}\\
=N\left(P_g+\frac{\dot{R}}{c}P_g+\frac{R}{c}\frac{dP_g}{dt}\right)\frac{\partial V}{\partial t}-N\left(\frac{4\mu_L\dot{R}}{R}+\frac{\dot{R}}{c}\frac{4\mu_L \dot{R}}{R}+\frac{R}{c}\frac{d}{dt}\left(\frac{4\mu_L \dot{R}}{R}\right)\right)\frac{\partial V}{\partial t}\\
-N\left(12\mu_{sh}\epsilon R_0^2\frac{\dot{R}}{R^4}+\frac{\dot{R}}{c}12\mu_{sh}\epsilon R_0^2\frac{\dot{R}}{R^4}+\frac{R}{c}\frac{d}{dt}\left(12\mu_{sh}\epsilon R_0^2\frac{\dot{R}}{R^4}\right)\right)\frac{\partial V}{\partial t}\\
-N\left(12G_s\epsilon R_0^2\left(\frac{1}{R^3}-\frac{R_0}{R^4}\right)+\frac{\dot{R}}{c}12G_s\epsilon R_0^2\left(\frac{1}{R^3}-\frac{R_0}{R^4}\right)+\frac{R}{c}\frac{d}{dt}\left(12G_s\epsilon R_0^2\left(\frac{1}{R^3}-\frac{R_0}{R^4}\right)\right)\right)\frac{\partial V}{\partial t}\\
-N\left(P+\frac{\dot{R}}{c}P+\frac{R}{c}\frac{dP}{dt}\right)
\end{gathered}
\end{equation}

if we add Eq.13 and Eq.14:
\begin{equation}
\begin{gathered}
\rho N\left(R\ddot{R}+\frac{3}{2}\dot{R}^2\right)\frac{\partial V}{\partial t}-\rho N\frac{\dot{R}}{c}\left(R\ddot{R}+\frac{1}{2}\dot{R}^2\right)\frac{\partial V}{\partial t}+\frac{\partial}{\partial t}\left(\frac{1}{2}\frac{P^2}{\rho c^2}+\frac{1}{2}\rho v^2\right)\\
=N\left(P_g+\frac{\dot{R}}{c}P_g+\frac{R}{c}\frac{dP_g}{dt}\right)\frac{\partial V}{\partial t}-N\left(\frac{4\mu_L\dot{R}}{R}+\frac{\dot{R}}{c}\frac{4\mu_L \dot{R}}{R}+\frac{R}{c}\frac{d}{dt}\left(\frac{4\mu_L \dot{R}}{R}\right)\right)\frac{\partial V}{\partial t}\\
-N\left(12\mu_{sh}\epsilon R_0^2\frac{\dot{R}}{R^4}+\frac{\dot{R}}{c}12\mu_{sh}\epsilon R_0^2\frac{\dot{R}}{R^4}+\frac{R}{c}\frac{d}{dt}\left(12\mu_{sh}\epsilon R_0^2\frac{\dot{R}}{R^4}\right)\right)\frac{\partial V}{\partial t}\\
-N\left(12G_s\epsilon R_0^2\left(\frac{1}{R^3}-\frac{R_0}{R^4}\right)+\frac{\dot{R}}{c}12G_s\epsilon R_0^2\left(\frac{1}{R^3}-\frac{R_0}{R^4}\right)+\frac{R}{c}\frac{d}{dt}\left(12G_s\epsilon R_0^2\left(\frac{1}{R^3}-\frac{R_0}{R^4}\right)\right)\right)\frac{\partial V}{\partial t}\\
-N\left(\frac{\dot{R}}{c}P+\frac{R}{c}\frac{dP}{dt}\right)
\end{gathered}
\end{equation}

The mass of the liquid around the bubble can be calculated as
\begin{equation}
M_l=\frac{1}{2}\int_{R}^{\infty} \rho 4\pi r^2 dr
\end{equation}
Keller-Miksis type models are derived by assuming that the flow of the liquid \cite{47} around the bubble is non-rotational $\vec{\nabla}\times v=0$. This assumption leads to definition of the velocity potential $\varphi$, which is given by:
\begin{equation}
v=\frac{\partial \varphi}{\partial r}
\end{equation}
Furthermore the kinetic energy of the liquid around the bubble can be written as \cite{35}:
\begin{equation}
K_l=\frac{1}{2}\int_{R}^{\infty} \rho \left(\frac{\partial \varphi}{\partial r}\right)^2 4\pi r^2 dr = 2\pi \rho R^3 \dot{R}^2 
\end{equation}

inserting $K_l$ in to Eq. 15:

\begin{equation}
\begin{gathered}
\frac{\partial}{\partial t}\left(\frac{1}{2}\frac{P^2}{\rho c^2}+\frac{1}{2}\rho v^2\right)+N\left(\left(1-\frac{\dot{R}}{c}\right)\frac{\partial K_l}{\partial t}+\frac{\rho\dot{R}^3}{c}\frac{\partial V}{\partial t}\right)+\nabla.(Pv)\\
=N\left(P_g+\frac{\dot{R}}{c}P_g+\frac{R}{c}\frac{dP_g}{dt}\right)\frac{\partial V}{\partial t}-N\left(\frac{4\mu_L\dot{R}}{R}+\frac{\dot{R}}{c}\frac{4\mu_L \dot{R}}{R}+\frac{R}{c}\frac{d}{dt}\left(\frac{4\mu_L \dot{R}}{R}\right)\right)\frac{\partial V}{\partial t}\\
-N\left(12\mu_{sh}\epsilon R_0^2\frac{\dot{R}}{R^4}+\frac{\dot{R}}{c}12\mu_{sh}\epsilon R_0^2\frac{\dot{R}}{R^4}+\frac{R}{c}\frac{d}{dt}\left(12\mu_{sh}\epsilon R_0^2\frac{\dot{R}}{R^4}\right)\right)\frac{\partial V}{\partial t}\\
-N\left(12G_s\epsilon R_0^2\left(\frac{1}{R^3}-\frac{R_0}{R^4}\right)+\frac{\dot{R}}{c}12G_s\epsilon R_0^2\left(\frac{1}{R^3}-\frac{R_0}{R^4}\right)+\frac{R}{c}\frac{d}{dt}\left(12G_s\epsilon R_0^2\left(\frac{1}{R^3}-\frac{R_0}{R^4}\right)\right)\right)\frac{\partial V}{\partial t}\\
-N\left(\frac{\dot{R}}{c}P+\frac{R}{c}\frac{dP}{dt}\right)
\end{gathered}
\end{equation}

Eq.19 can be re-arranged as:
\begin{equation}
\begin{gathered}
\frac{\partial}{\partial t}\left(\frac{1}{2}\frac{P^2}{\rho c^2}+\frac{1}{2}\rho v^2+NK_l\right)+\nabla.(Pv)=\\
N\left(P_g\right)\frac{\partial V}{\partial t}-N\left(\frac{4\mu_L\dot{R}}{R}\right)\frac{\partial V}{\partial t}-N\left(12\mu_{sh}\varepsilon R_0^2\frac{\dot{R}}{R^4}\right)\frac{\partial V}{\partial t}-N\left(12G_s\varepsilon R_0^2\left(\frac{1}{R^3}-\frac{R_0}{R^4}\right)\right)\frac{\partial V}{\partial t}\\
-N\left(\left[\frac{\dot{R}}{c}P+\frac{R}{c}\dot{P}-\frac{\rho \dot{R}^3}{c}-\frac{\dot{R}}{c}P_g-\frac{R}{c}\dot{P}_g+\frac{\dot{R}}{c}\frac{4\mu_L\dot{R}}{R}+\frac{R}{c}\left(\frac{4\mu_L\ddot{R}}{R}-\frac{4\mu_L\dot{R}^2}{R^2}\right)\right.\right.\\
+\frac{\dot{R}}{c}12\mu_{sh}\varepsilon R_0^2\frac{\dot{R}}{R^4}+\frac{R}{c}12\mu_{sh}\varepsilon R_0^2\left(\frac{\ddot{R}}{R^4}-\frac{4\dot{R}^2}{R^5}\right)\\
\left. \left. \frac{\dot{R}}{c}12G_s\varepsilon R_0^2\left(\frac{1}{R^3}-\frac{R_0}{R^4}\right)+\frac{R}{c}12G_s\varepsilon R_0^2\left(\frac{-3\dot{R}}{R^4}+\frac{4R_0\dot{R}}{R^5}\right)\right]\frac{\partial V}{\partial t}-\frac{\dot{R}}{c}\frac{\partial K_l}{\partial t}\right)
\end{gathered}
\end{equation}
We then can simplify Eq. 20 as
\begin{equation}
\begin{gathered}
\frac{\partial}{\partial t}\left(\frac{1}{2}\frac{P^2}{\rho c^2}+\frac{1}{2}\rho v^2\right)+\nabla.(Pv)=-N\left(\pi_{Thermal}+\pi_{Liquid}+\pi_{Shell}+\pi_{Gs}+\pi_{Radiation}\right)
\end{gathered}
\end{equation}	
where
\begin{equation}
\begin{dcases}
\pi_{Thermal}=-P_g\frac{\partial V}{\partial t}\\ \\
\pi_{Liquid}=\left(\frac{4\mu_L\dot{R}}{R}\right)\frac{\partial V}{\partial t}\\ \\
\pi_{coating}=\left(12\mu_{sh}\varepsilon R_0^2\frac{\dot{R}}{R^4}\right)\frac{\partial V}{\partial t}\\ \\
\pi_{Gs}=\left(12G_s\varepsilon R_0^2\left(\frac{1}{R^3}-\frac{R_0}{R^4}\right)\right)\frac{\partial V}{\partial t}\\ \\
\begin{gathered}
\pi_{Radiation}=\left(\left[\frac{\dot{R}}{c}P+\frac{R}{c}\dot{P}-\frac{\rho \dot{R}^3}{c}-\frac{\dot{R}}{c}P_g-\frac{R}{c}\dot{P}_g+\frac{\dot{R}}{c}\frac{4\mu_L\dot{R}}{R}+\frac{R}{c}\left(\frac{4\mu_L\ddot{R}}{R}-\frac{4\mu_L\dot{R}^2}{R^2}\right)\right.\right.\\
+\frac{\dot{R}}{c}12\mu_{sh}\varepsilon R_0^2\frac{\dot{R}}{R^4}+\frac{R}{c}12\mu_{sh}\varepsilon R_0^2\left(\frac{\ddot{R}}{R^4}-\frac{4\dot{R}^2}{R^5}\right)\\
\left. \left. \frac{\dot{R}}{c}12G_s\varepsilon R_0^2\left(\frac{1}{R^3}-\frac{R_0}{R^4}\right)+\frac{R}{c}12G_s\varepsilon R_0^2\left(\frac{-3\dot{R}}{R^4}+\frac{4R_0\dot{R}}{R^5}\right)\right]\frac{\partial V}{\partial t}-\frac{\dot{R}}{c}\frac{\partial K_l}{\partial t}\right)
\end{gathered}
\end{dcases}
\end{equation}
where time dependent $\pi_{Thermal}$, $\pi_{Liquid}$, $\pi_{coating}$ and $\pi_{Radiation}$ describe damping due to thermal, liquid viscosity, coating viscosity and re-radiation. The term  $\pi_{Gs}$ can be referred to as the damping due to the stiffness of the coating. As you will see further, when averaged over full driving acoustic cycles this damping term always return zero.\\ 
Averaging Eq. 21 over a time period $T$ yields:
\begin{equation}
\begin{gathered}
\frac{1}{T}\int_{0}^{T}\frac{\partial}{\partial t}\left(\frac{1}{2}\frac{P^2}{\rho c^2}+\frac{1}{2}\rho v^2\right)dt+\nabla.<Pv>
=-N\left(Td+Ld+Cd+Gd+Rd\right)
\end{gathered}
\end{equation}	
Where Td,Ld,Cd,Rd and Gd are the dissipated power due to thermal, Liquid viscosity , coating viscosity, re-radiation  and stiffness of the coating. 
\begin{equation}
\begin{dcases}
Td=\frac{-4\pi}{T}\int_{0}^{T}R^2\dot{R}P_g dt\\ \\
Ld=\frac{16\pi\mu_L}{T}\int_{0}^{T}R\dot{R}^2dt\\ \\
Cd=\frac{48\pi\mu_{sh}\varepsilon R_0^2}{T}\int_{0}^{T}\frac{\dot{R}^2}{R^2}dt\\ \\
Gd=\frac{48\pi G_s\varepsilon R_0^2}{T}\int_{0}^{T}\left(\frac{\dot{R}}{R}-\frac{R_0\dot{R}}{R^2}\right)dt\\ \\
\begin{gathered}
Rd=\frac{1}{T}\int_{0}^{T}\left(4\pi\left[\frac{R^2\dot{R}^2}{c}\left(P-P_g\right)+\frac{R^3\dot{R}}{c}\left(\dot{P}-\dot{P}_g\right)+\frac{4\mu_LR^2\dot{R}\ddot{R}}{c}\right.\right.\\
\left.+12\mu_{sh}\varepsilon R0^2 \left(\frac{\dot{R}\ddot{R}}{cR}-\frac{3\dot{R}^3}{c R^2}\right)+12G_s\varepsilon R0^2\left(\frac{-2\dot{R}^2}{cR}+\frac{3R_0\dot{R}^2}{cR^2}\right) \right]\\
\left. -\frac{\rho R^2\dot{R}^4}{2c}-\frac{\rho R^3 \dot{R}^2\ddot{R}}{c}\right)dt
\end{gathered}
\end{dcases}
\end{equation}
In case of using the model with no thermal damping effects Td=0 and in case of incorporating the linear thermal effects $Td=\frac{16\pi\mu_{th}}{T}\int_{0}^{T}\left(R\dot{R}^2\right)dt$
\subsection{Nonlinear terms of dissipation for the KM model (the uncoated bubble)}
Using the same approach as above we have derived the dissipation terms for an uncoated bubble as follows \cite{36}:
\begin{equation}
\begin{dcases}
Td=\frac{-1}{T}\int_{0}^{T}\left(P_g\right)\frac{\partial V}{\partial t}dt\\ \\
Ld=\frac{16\pi\mu_L}{T}\int_{0}^{T}\left(R\dot{R}^2\right)dt\\ \\
\begin{gathered}
Rd=\frac{1}{T}\int_{0}^{T} \left[\frac{4\pi}{c}\left(R^2\dot{R}\left(\dot{R}P+R\dot{P}-\frac{1}{2}\rho \dot{R}^3-\rho R\dot{R}\ddot{R}\right)\right)\right.\\
\left.-\left(\frac{\dot{R}}{c}P_g+\frac{R}{c}\dot{P}_g\right)\frac{\partial V}{\partial t}+\frac{16\pi\mu_LR^2\dot{R}\ddot{R}}{c}\right]dt
\end{gathered}
\end{dcases}
\end{equation}
In case of using the model with no thermal damping effects Td=0 and in case of incorporating the linear thermal effects $Td=\frac{16\pi\mu_{th}}{T}\int_{0}^{T}\left(R\dot{R}^2\right)dt$
All the dissipated powers were calculated for the last 20 cycles of pulses with 200 cycles length. Simulations were carried out in Matlab using ODE45. The minimum time size for integration in each simulation was $\frac{10^{-5}}{f}$ where $f$ is the excitation frequency.  
\subsection{Acoustic power due to scattered pressure by bubbles}
Radiation damping is due to the re-radiated (scattered) energy by the bubble.
The acoustic energy scattered by an oscillating bubble can be calculated using \cite{57,58}:
\begin{equation}
W_{sc}=\frac{4\pi r^2}{\rho c}P_{sc}^2
\end{equation}
where $P_{sc}$ is the pressure scattered (re-radiated) by the oscillating bubble \cite{57,58}:
\begin{equation}
P_{sc}=\rho\frac{R}{r}\left(R\ddot{R}+2\dot{R}^2\right)
\end{equation}
here $r$ is the distance from the bubble center. Using Eq.15 and Eq.16 we can write:
\begin{equation}
W_{sc}=\frac{4\pi\rho}{c}R^2\left(R\ddot{R}+2\dot{R}^2\right)^2
\end{equation}
The dissipated power due to radiation should have the same value of the acoustic scattered power by the bubble. Equation 28 was used to validate the predictions of the Rd in Eq. 25 in \cite{36}. In this work, We used Eq. 28 to validate the predictions of the Rd in Eq. 24.
\section{Results}
\subsection{Total dissipated power by uncoated bubbles}
\begin{figure*}
	\begin{center}
		\scalebox{0.43}{\includegraphics{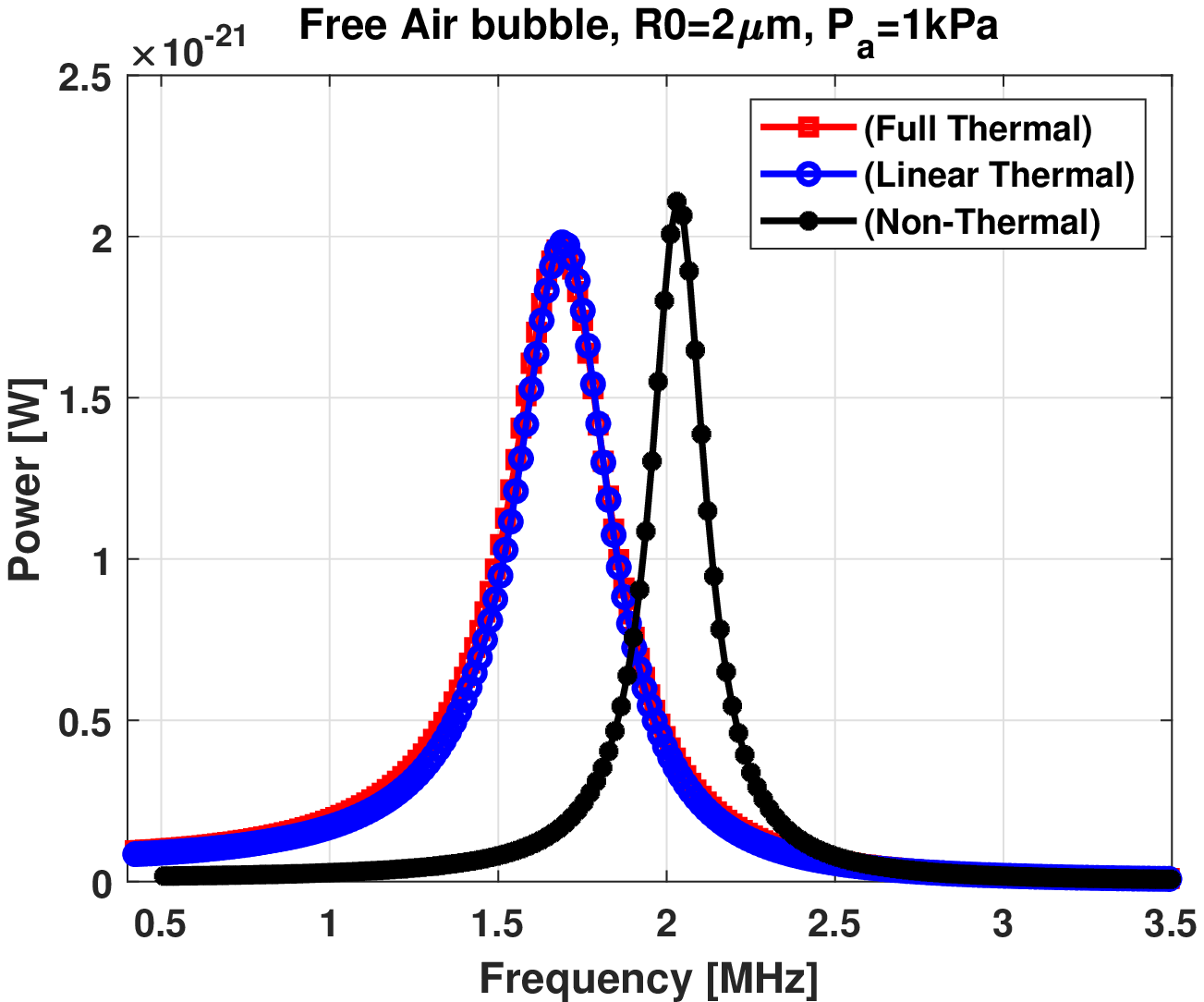}} \scalebox{0.43}{\includegraphics{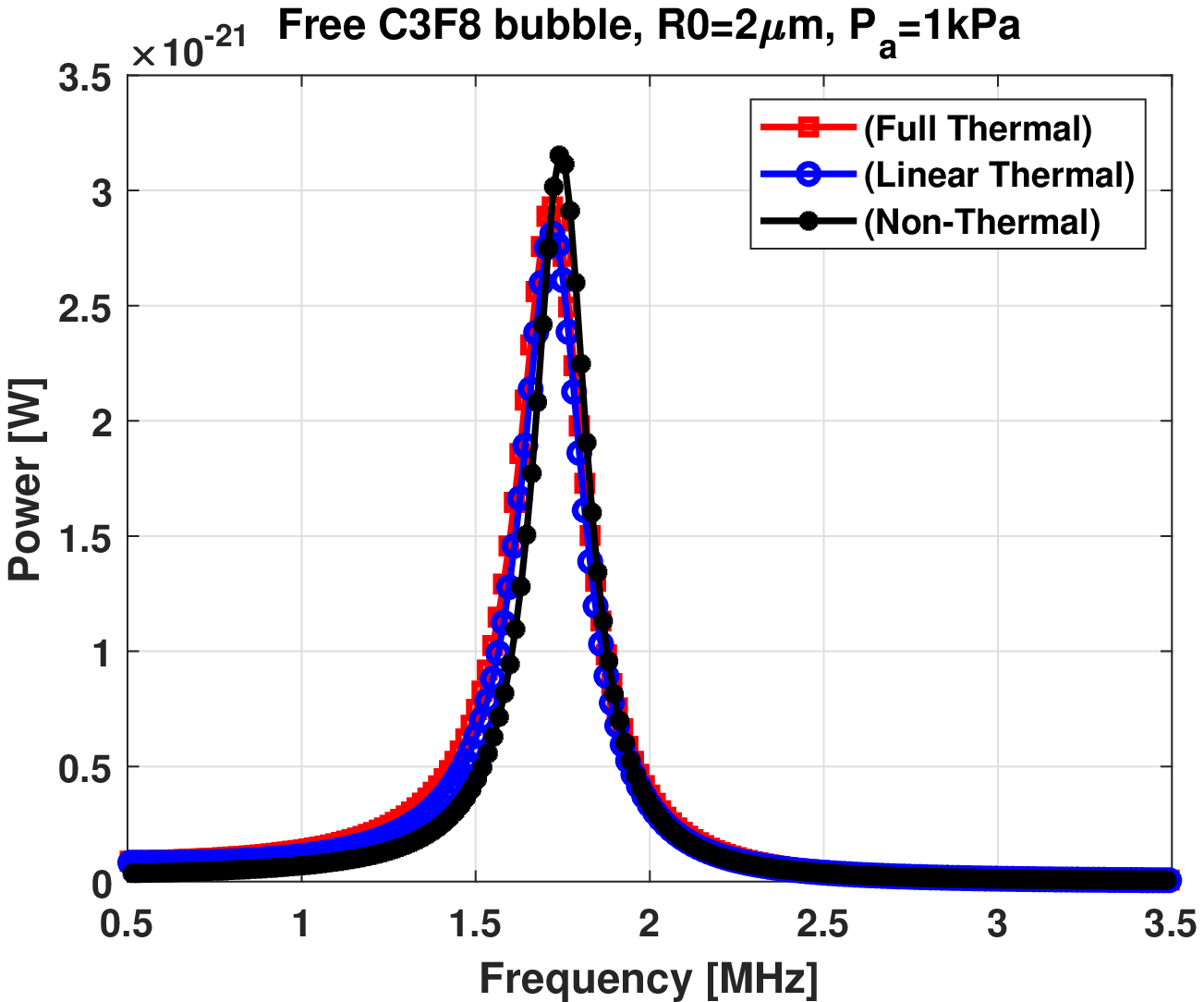}}\\
		\hspace{0.5cm} (a) \hspace{6cm} (b)\\
		\scalebox{0.43}{\includegraphics{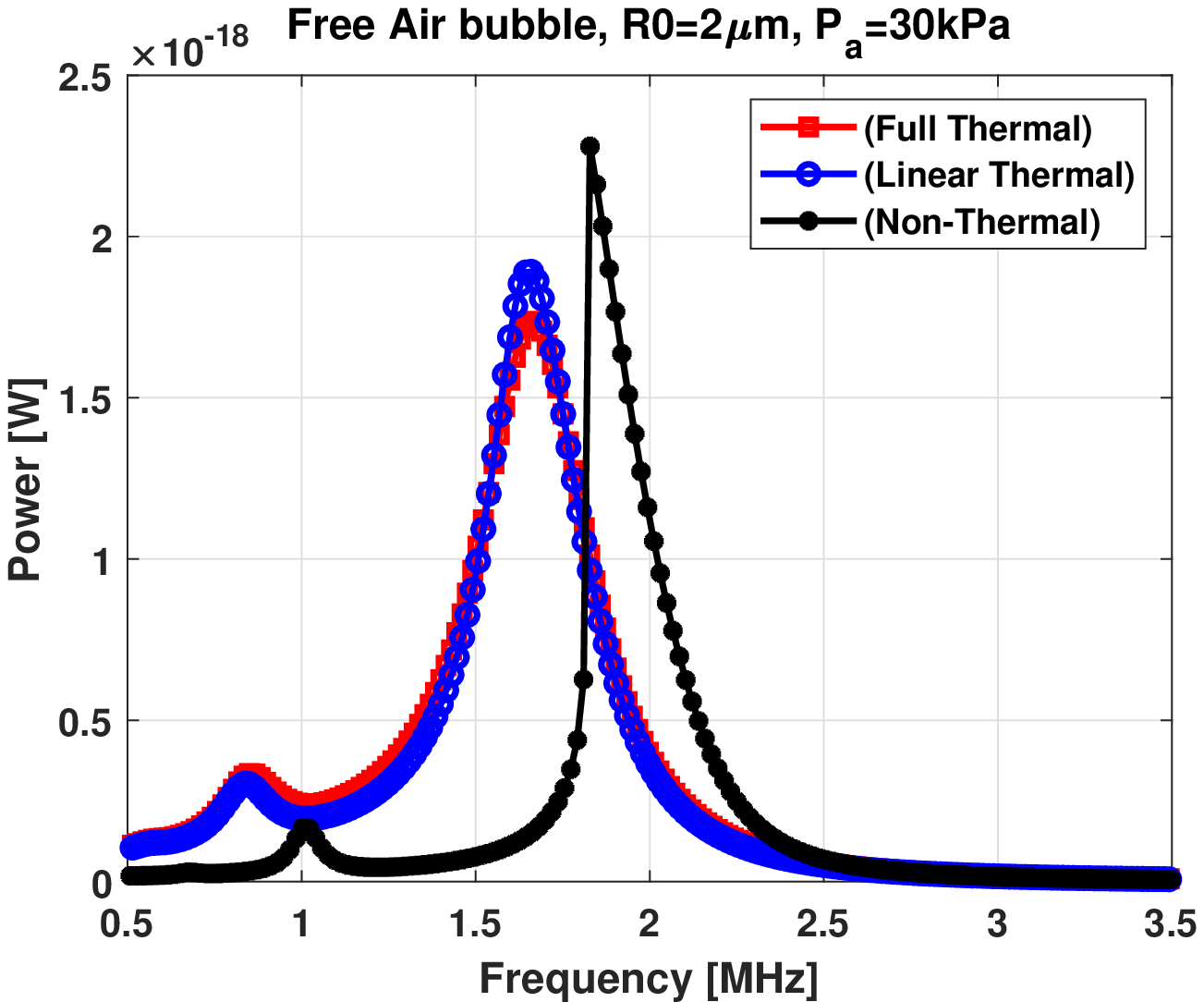}} \scalebox{0.43}{\includegraphics{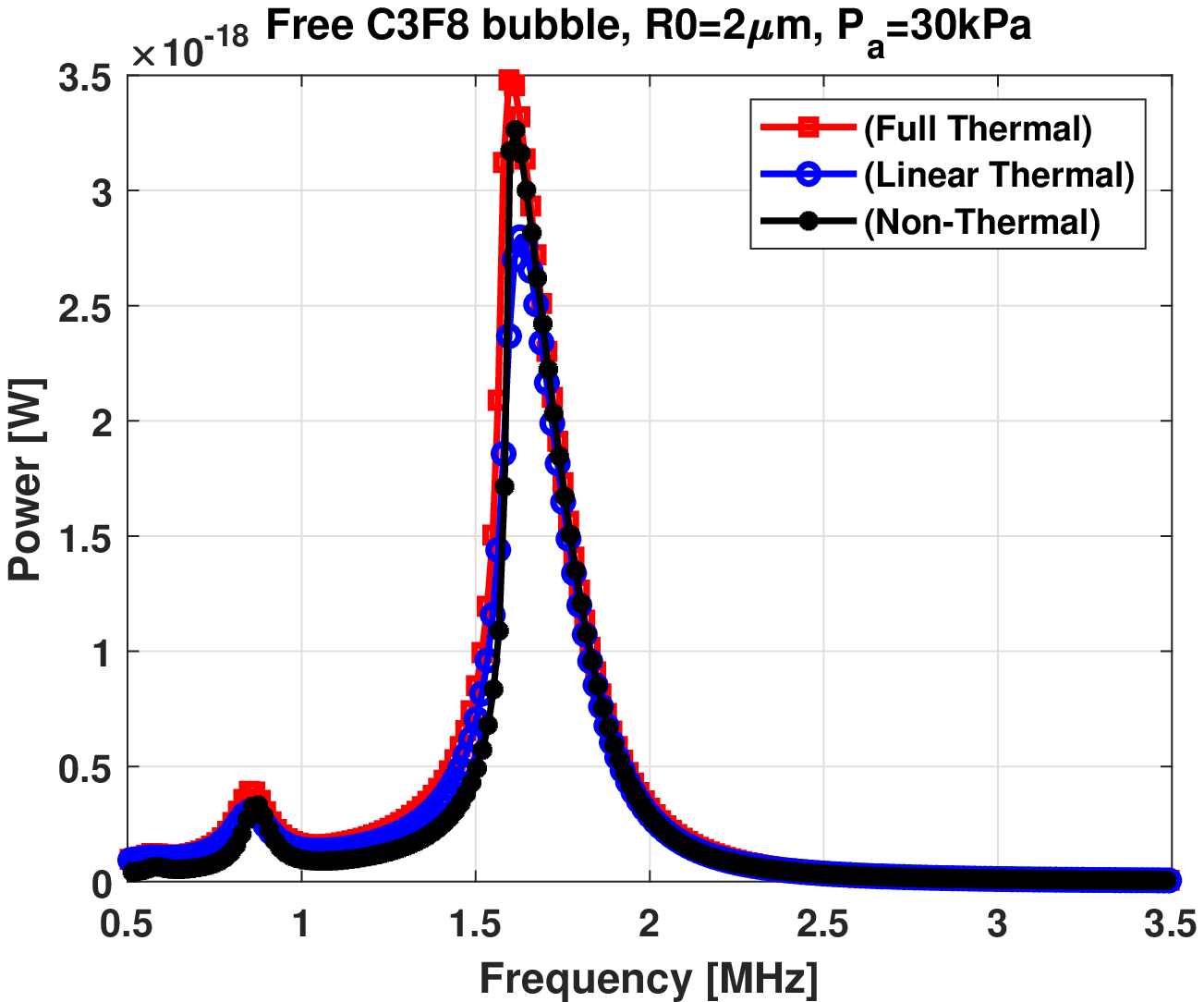}}\\
		\hspace{0.5cm} (c) \hspace{6cm} (d)\\
		\scalebox{0.43}{\includegraphics{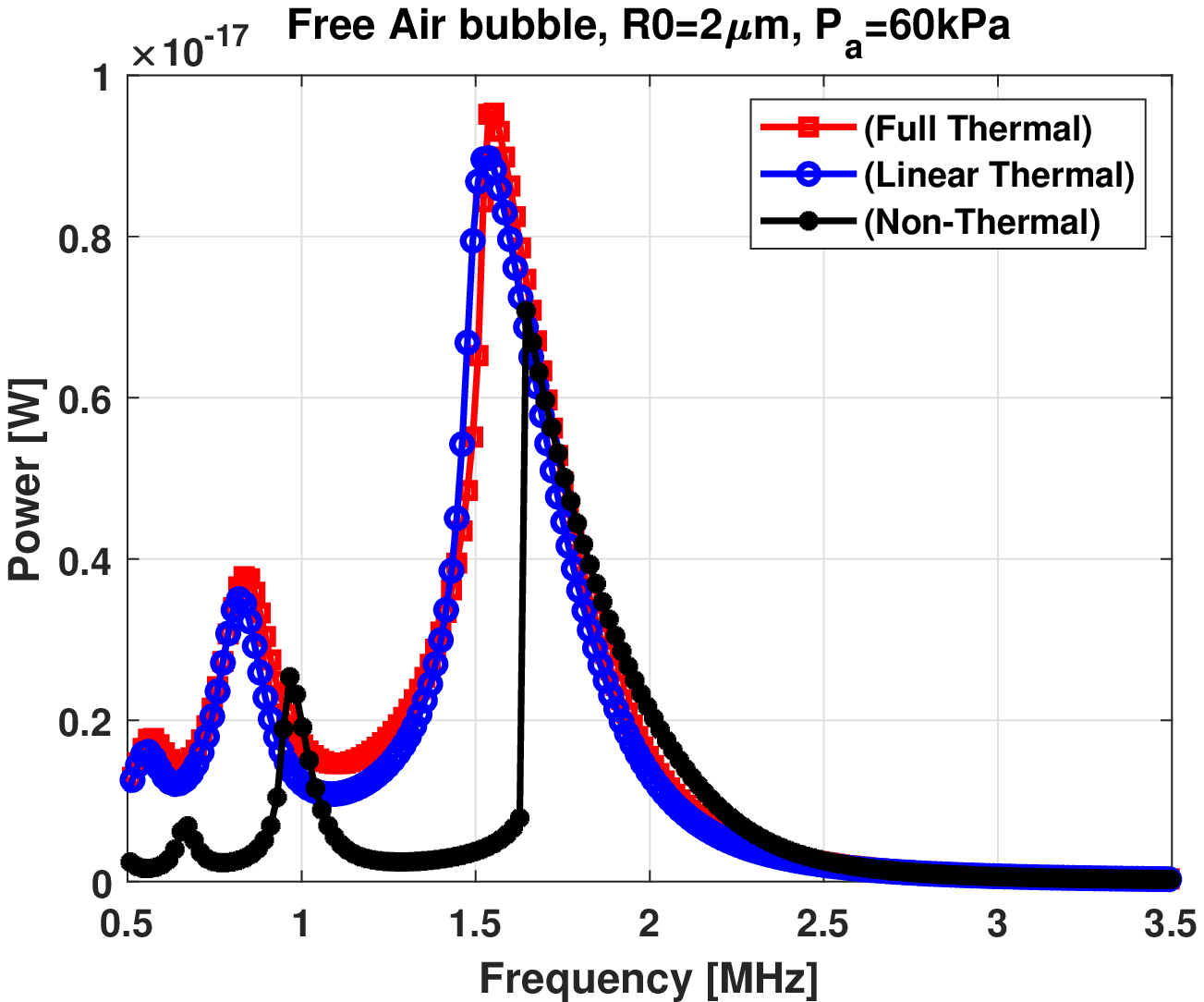}} \scalebox{0.43}{\includegraphics{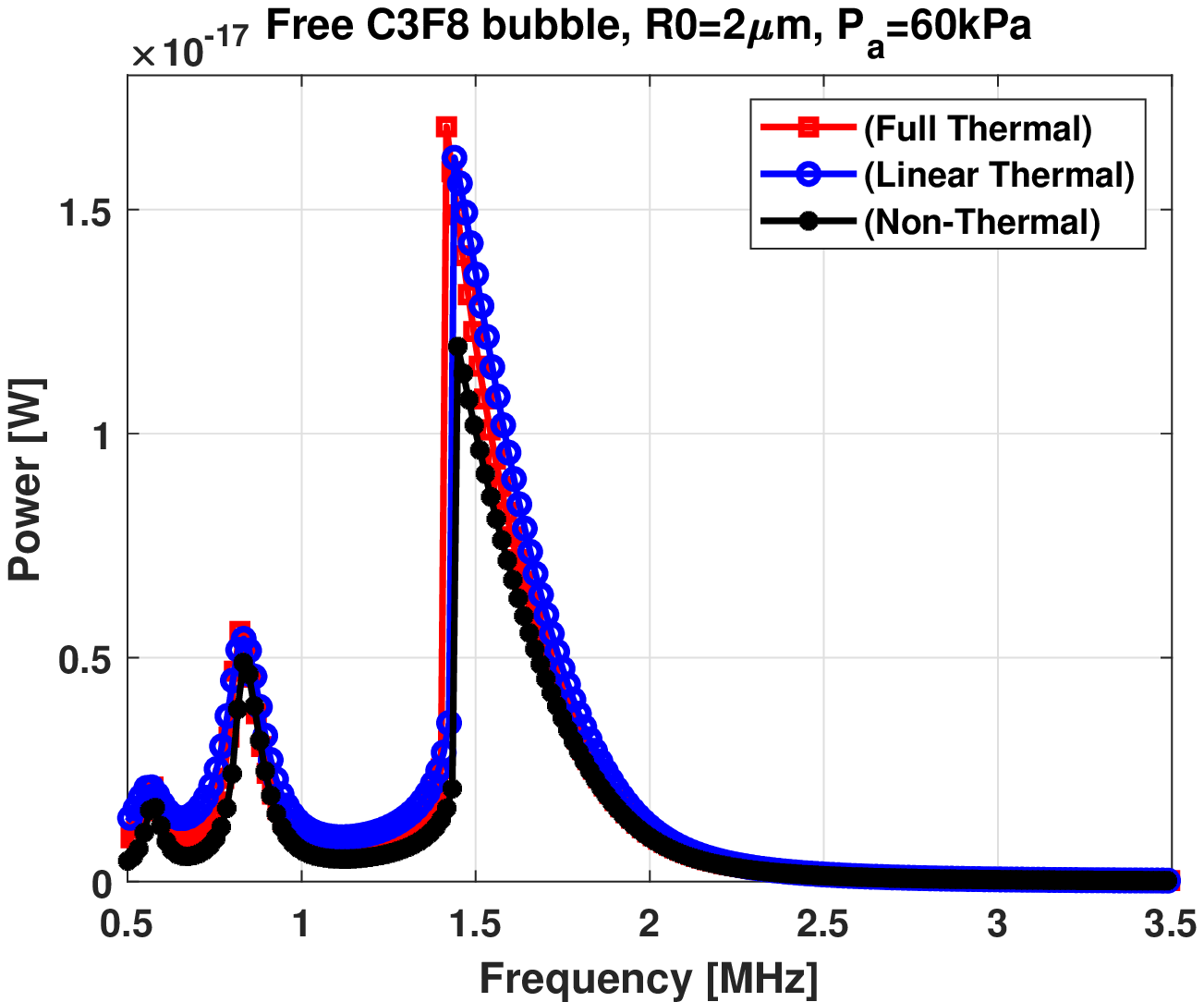}}\\
		\hspace{0.5cm} (e) \hspace{6cm} (f)\\
		\scalebox{0.43}{\includegraphics{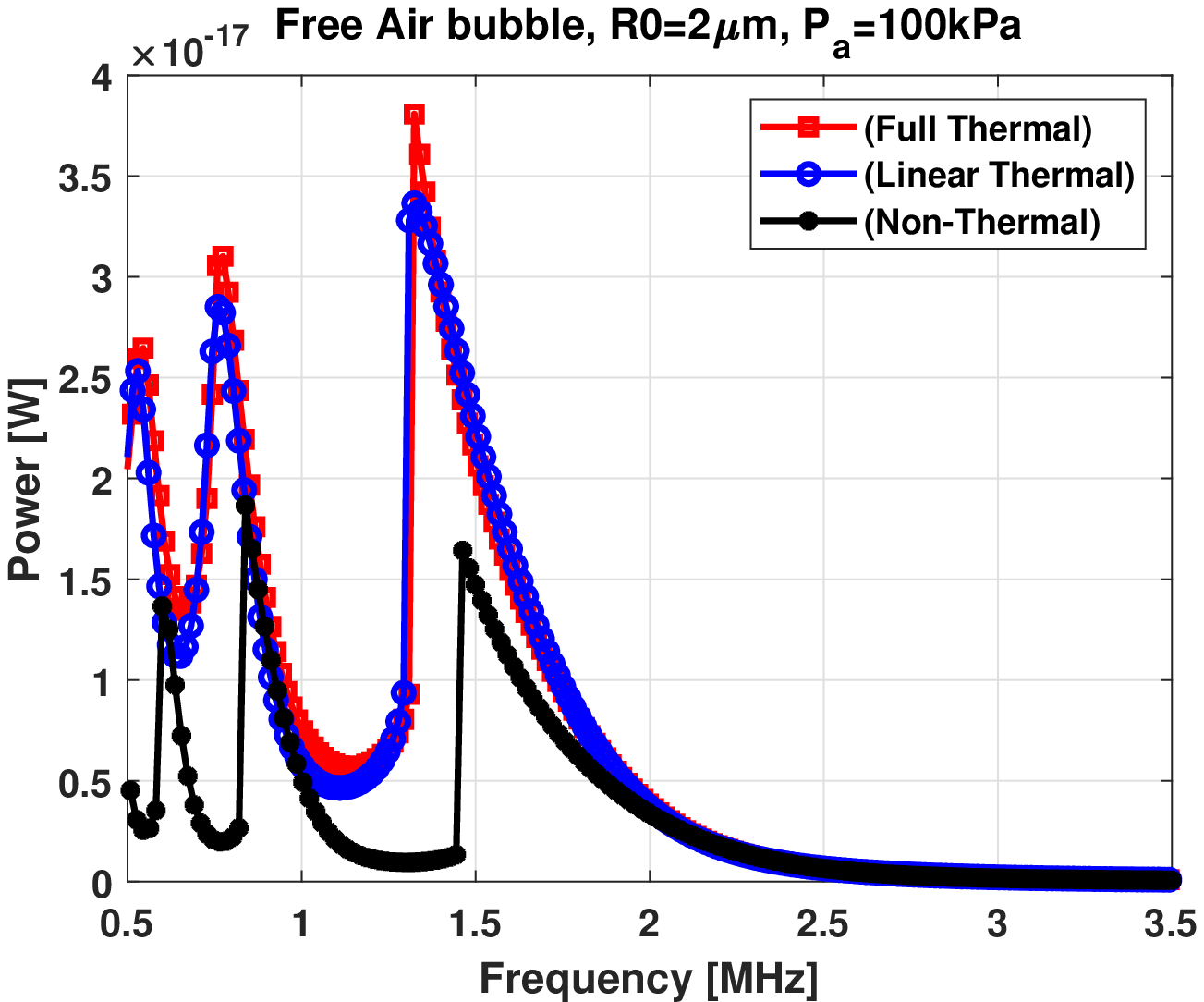}} \scalebox{0.43}{\includegraphics{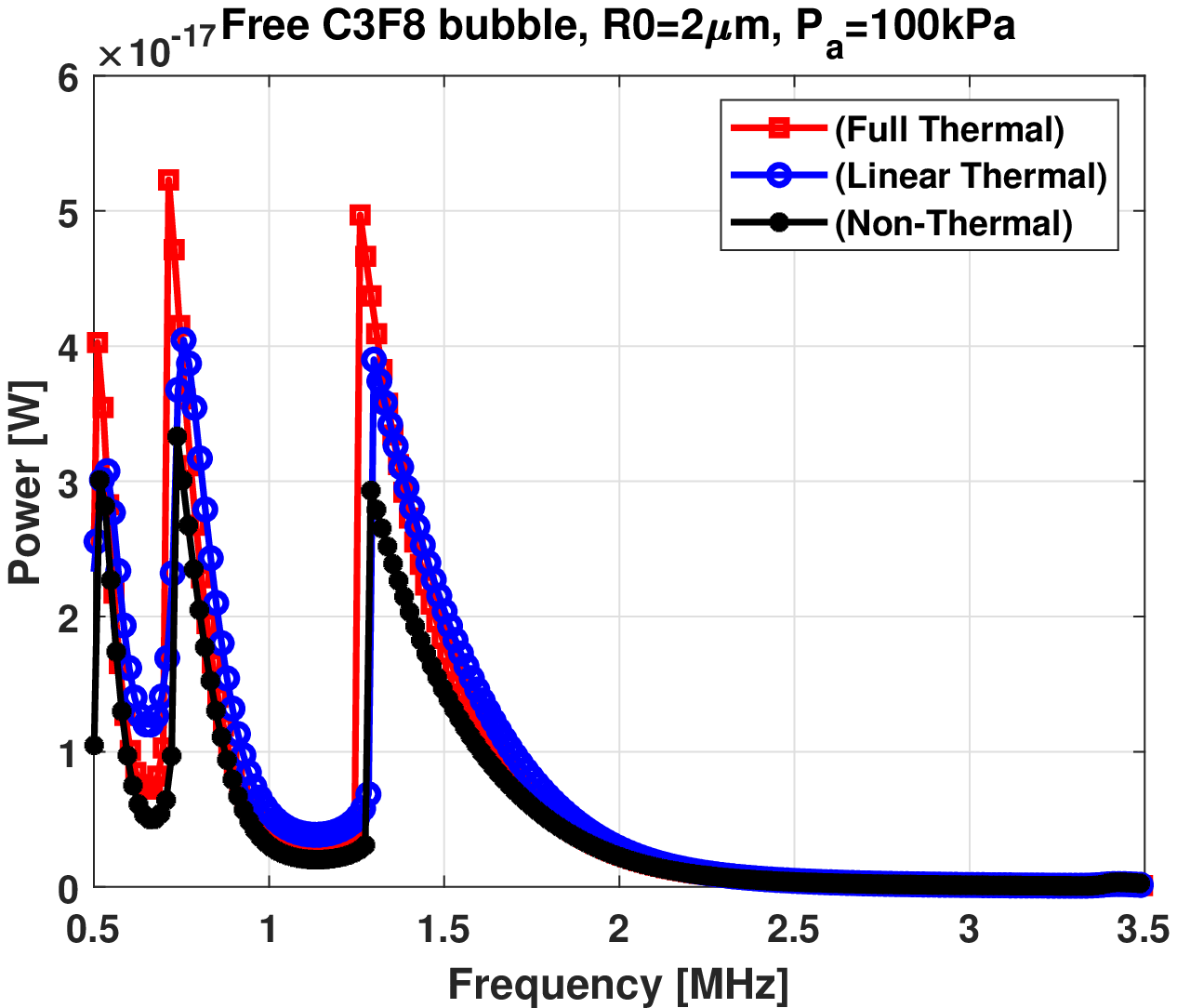}}\\
		\hspace{0.5cm} (g) \hspace{6cm} (h)\\
		\caption{Total dissipated power as predicted by the non-thermal, linear thermal, and full thermal model as a function of frequency for a free bubble with $R_0$= 2 $\mu$ m at various pressures (left column is for Air gas core and right column is for C3F8 gas core).}
	\end{center}
\end{figure*}

\begin{figure*}
	\begin{center}
		\scalebox{0.43}{\includegraphics{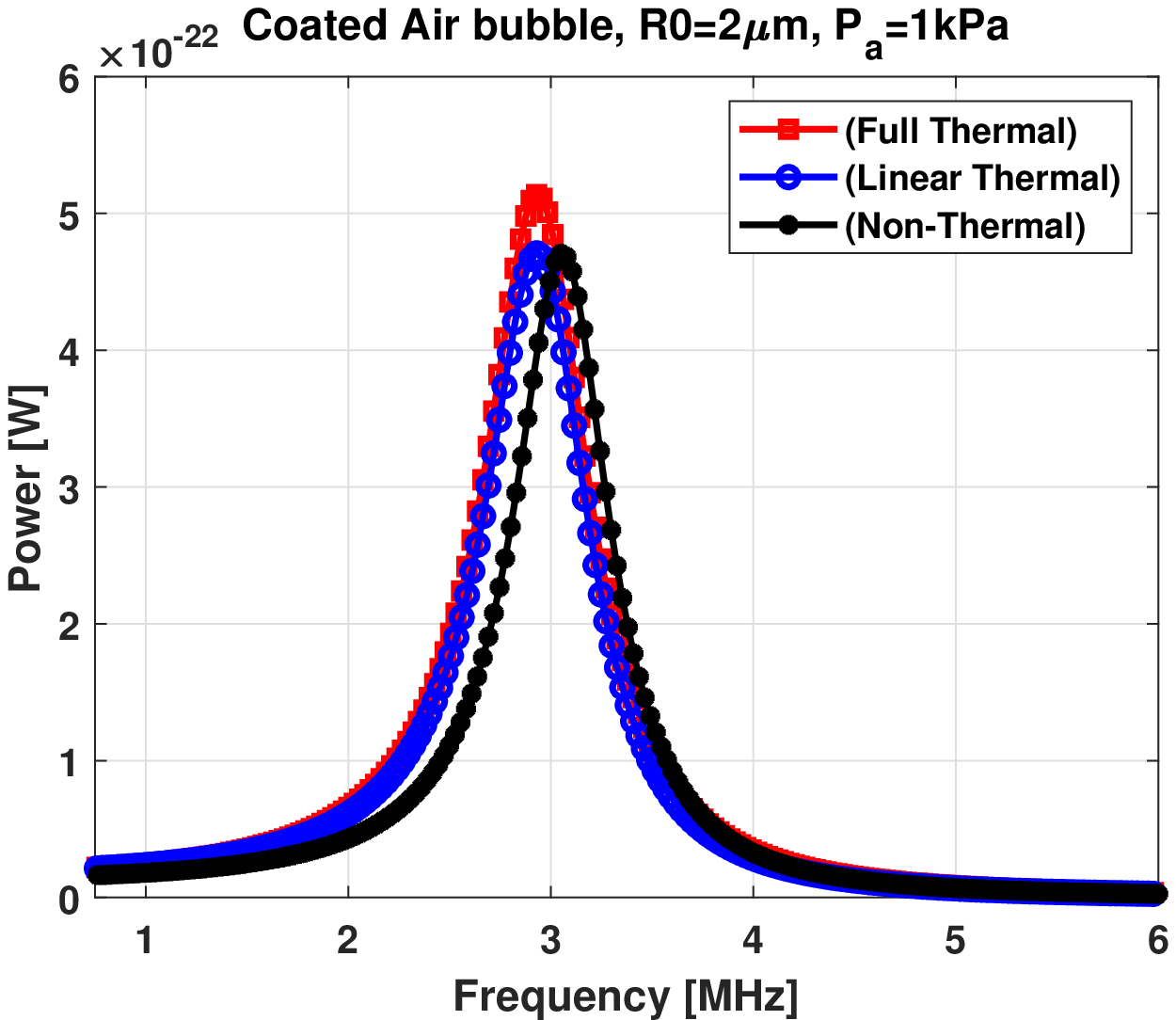}} \scalebox{0.43}{\includegraphics{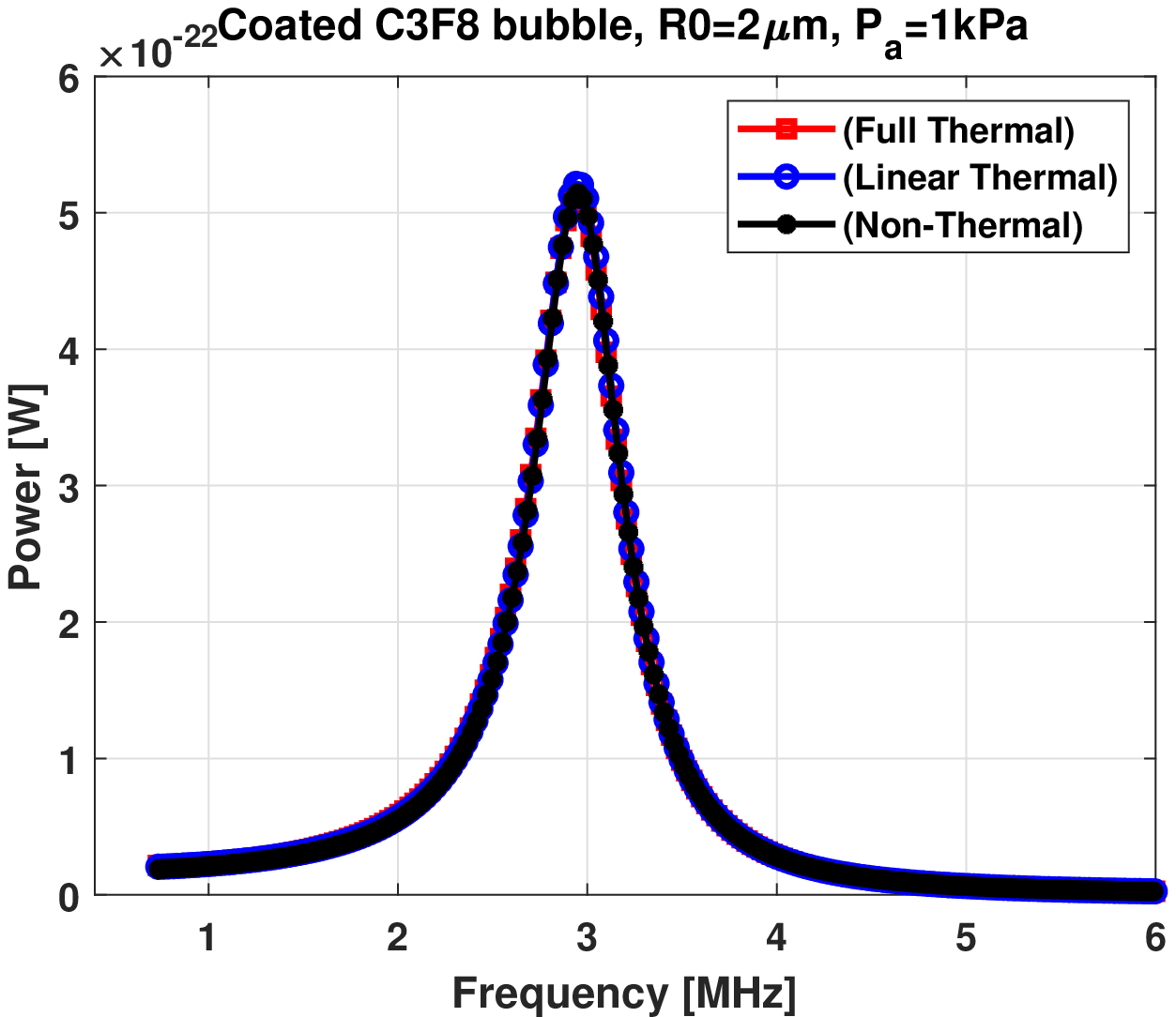}}\\
		\hspace{0.5cm} (a) \hspace{6cm} (b)\\
		\scalebox{0.43}{\includegraphics{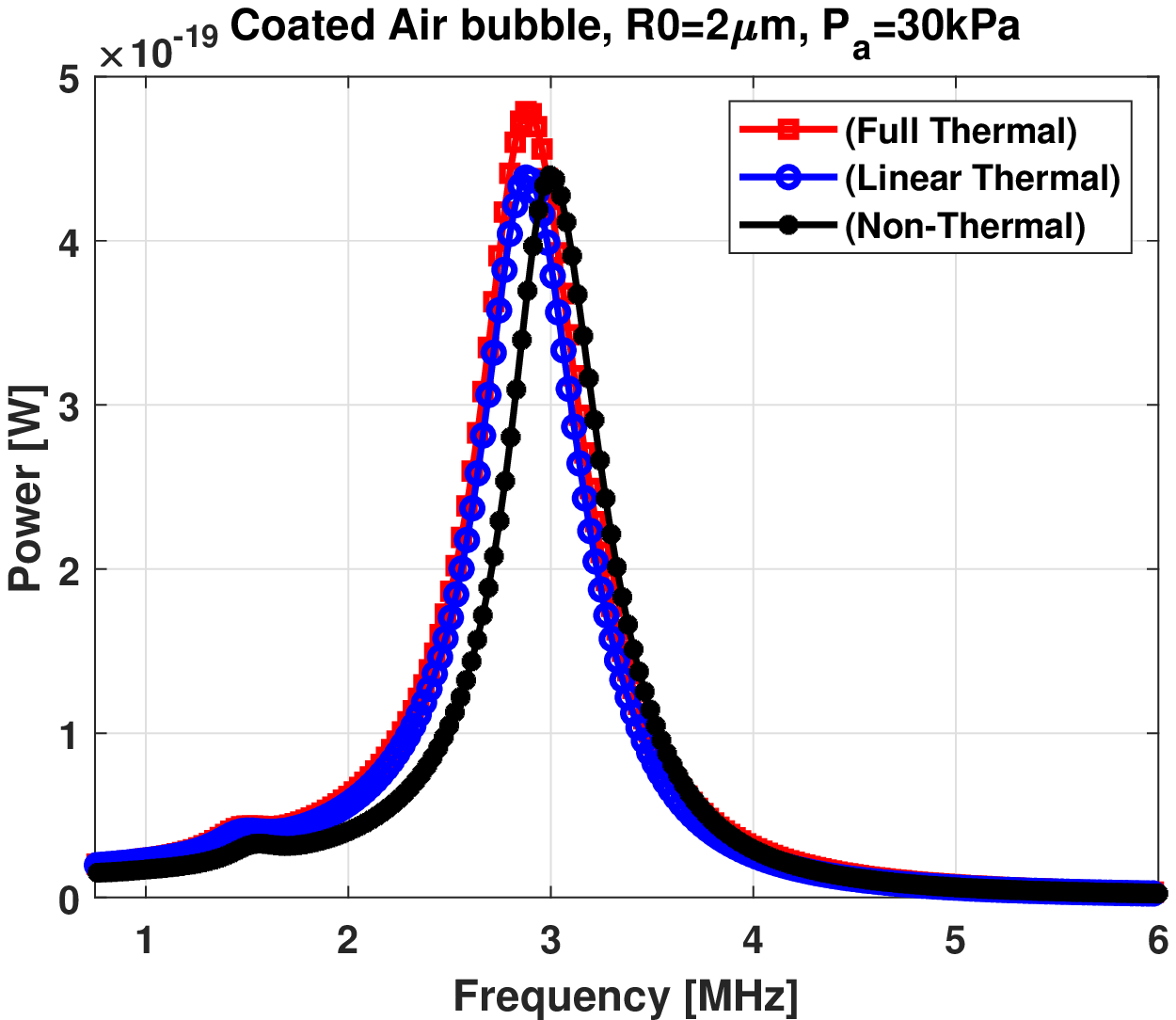}} \scalebox{0.43}{\includegraphics{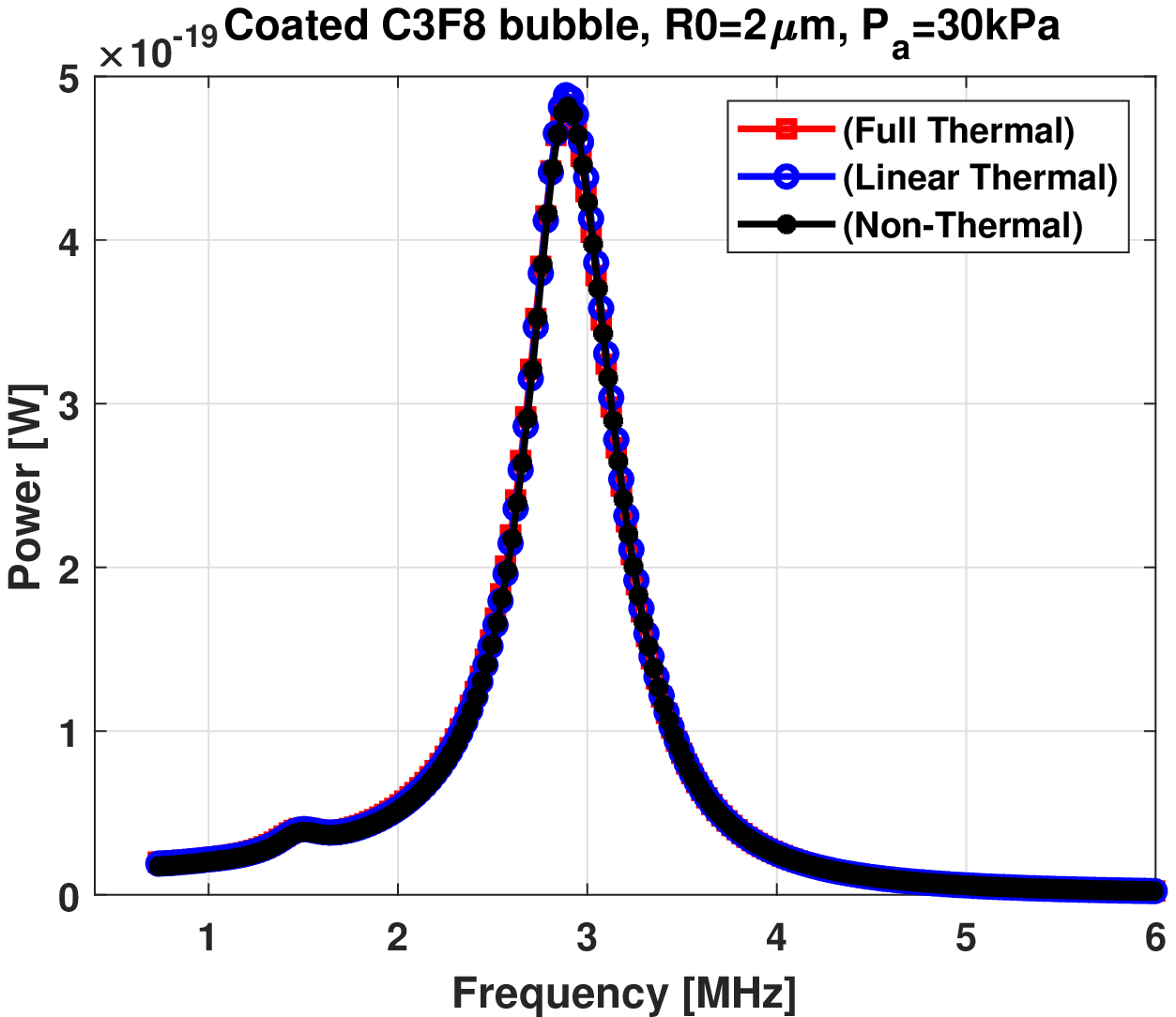}}\\
		\hspace{0.5cm} (c) \hspace{6cm} (b)\\
		\scalebox{0.43}{\includegraphics{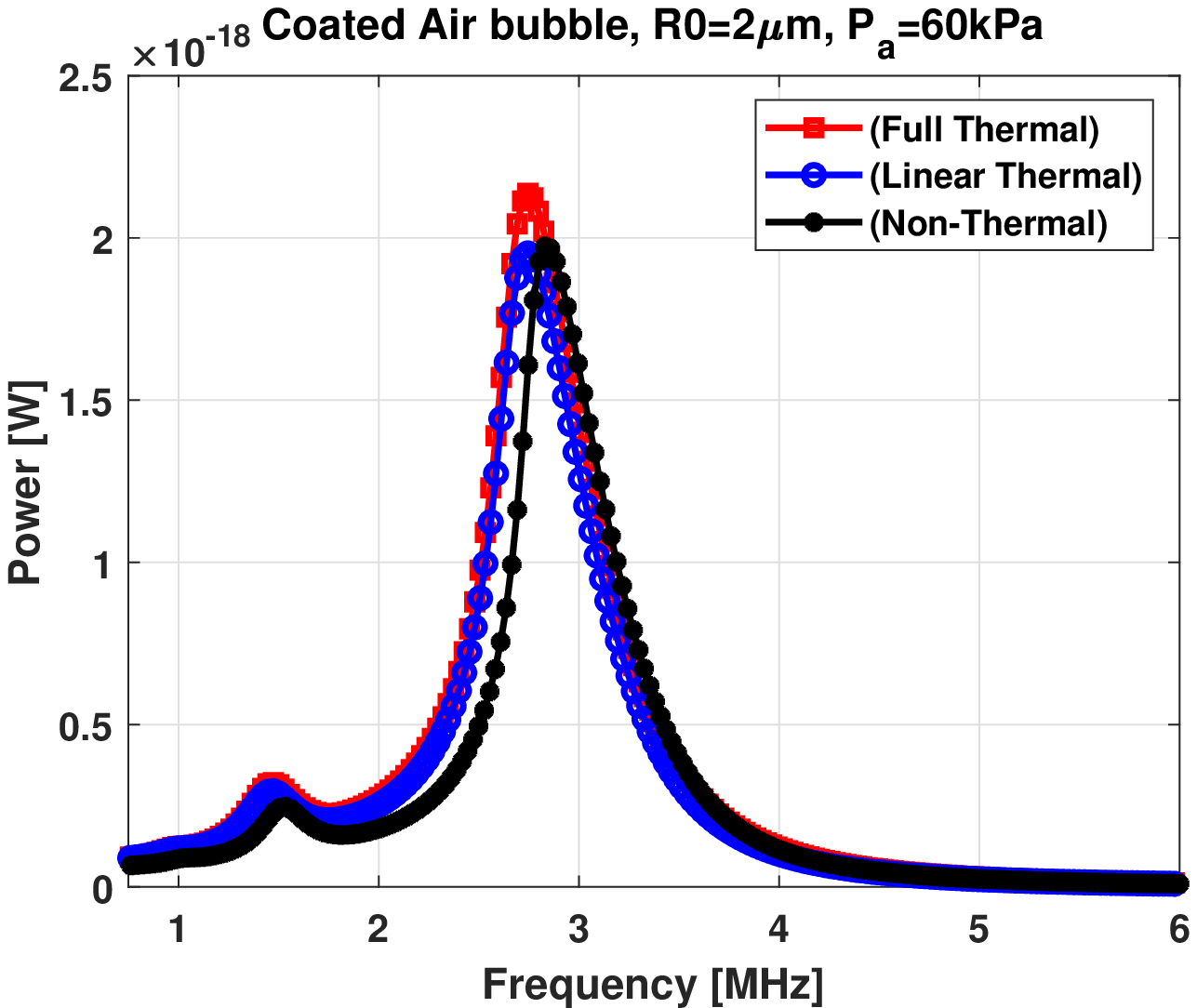}} \scalebox{0.43}{\includegraphics{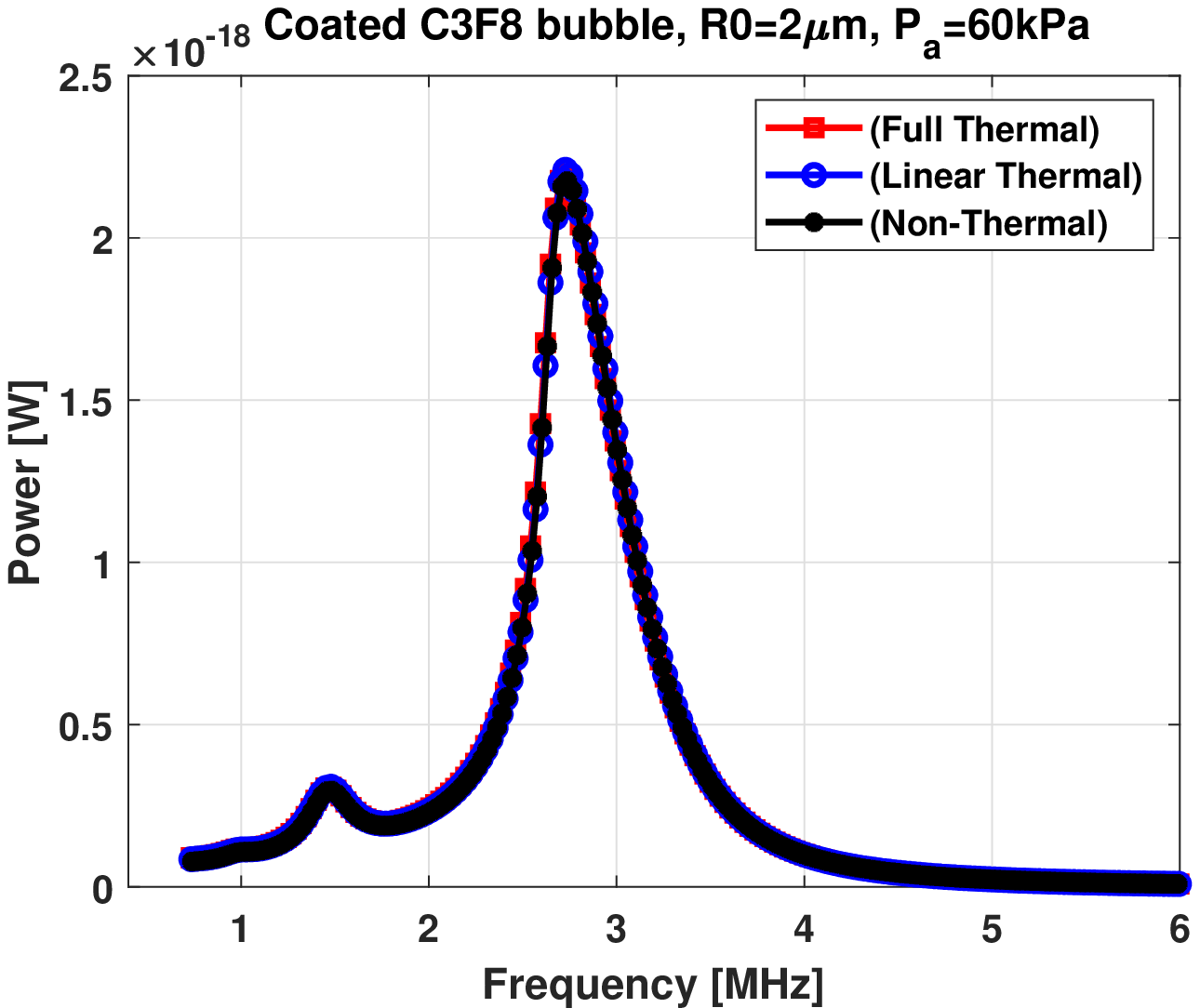}}\\
		\hspace{0.5cm} (e) \hspace{6cm} (f)\\
		\scalebox{0.43}{\includegraphics{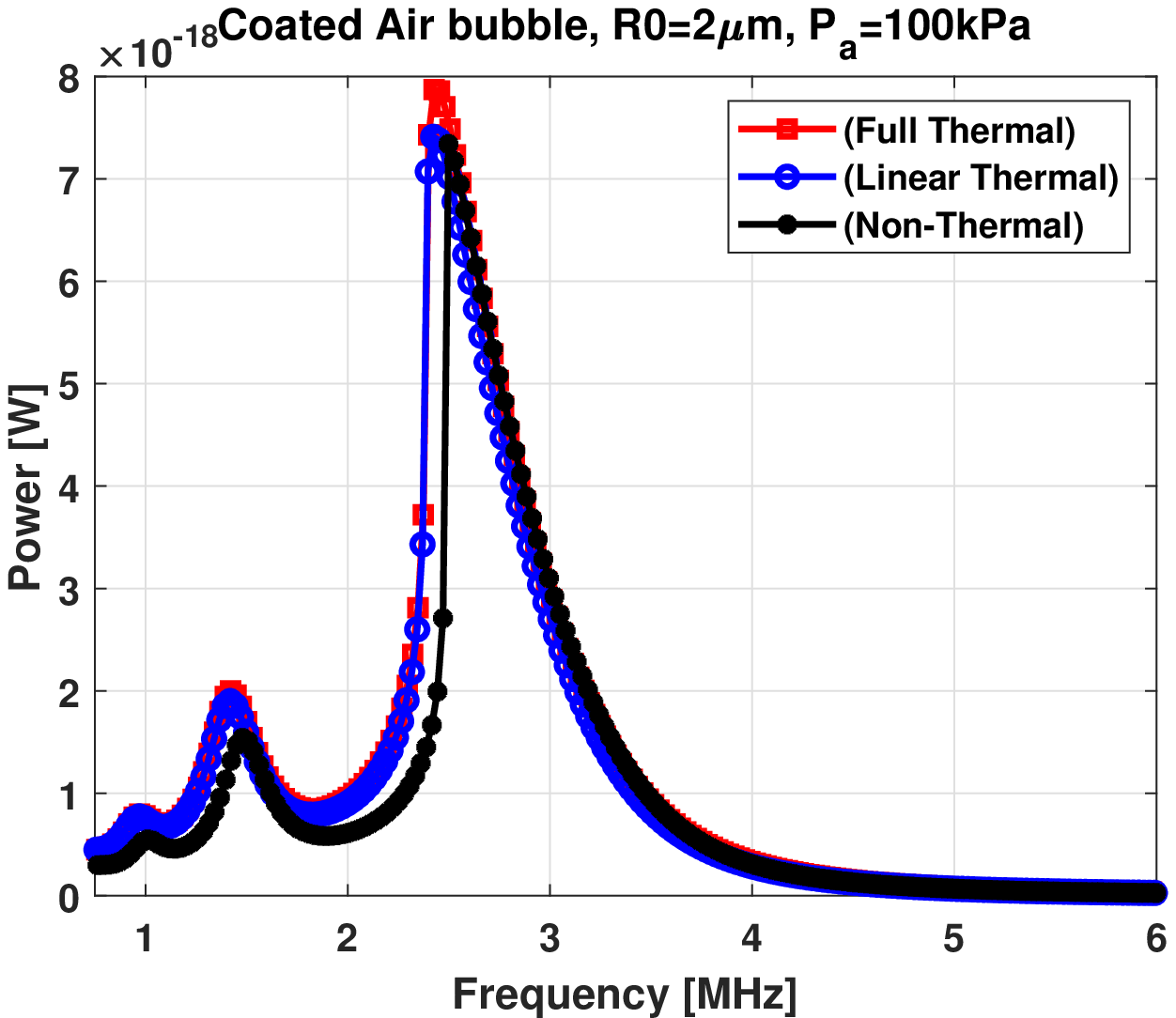}} \scalebox{0.43}{\includegraphics{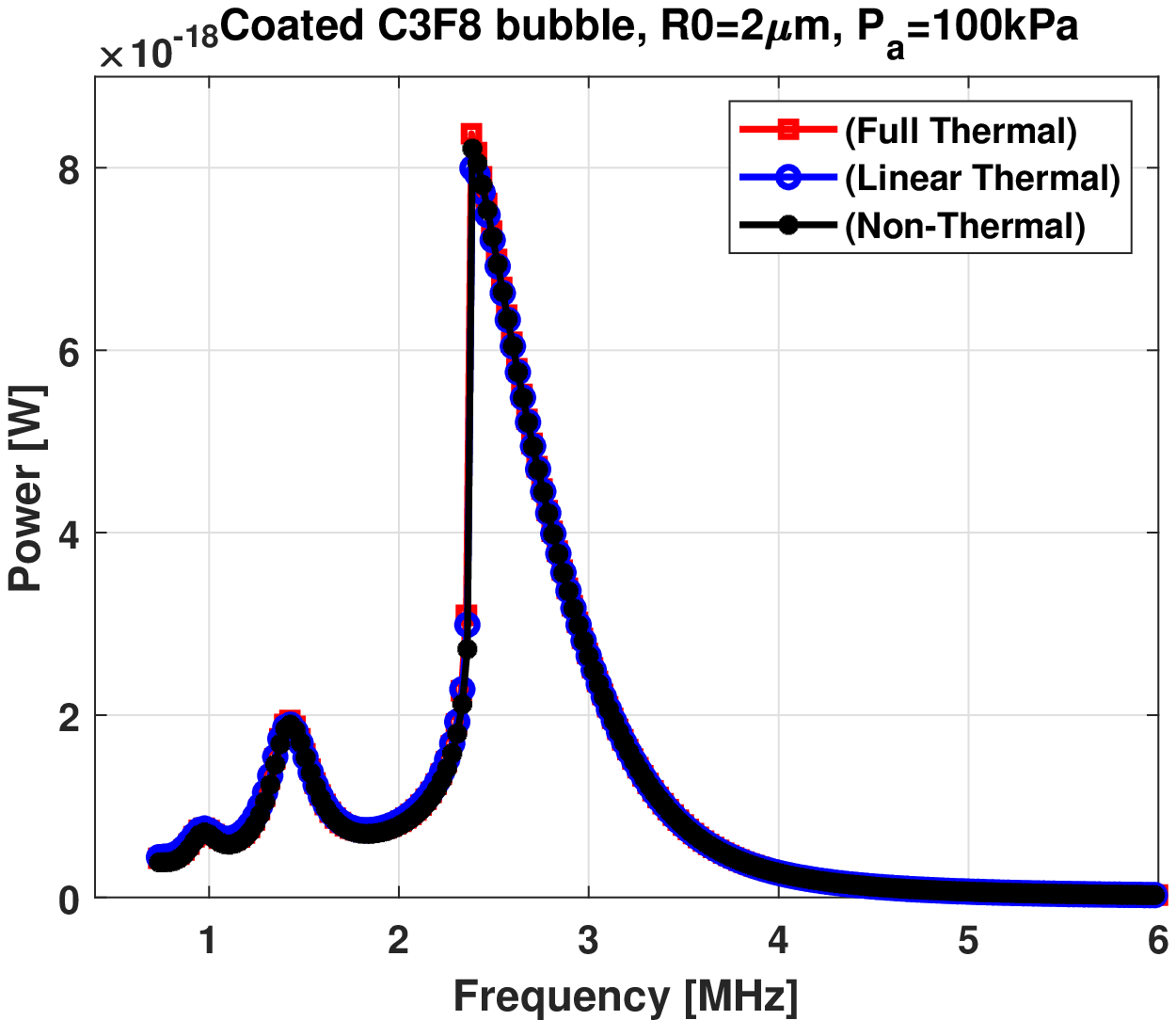}}\\
		\hspace{0.5cm} (g) \hspace{6cm} (h)\\
		\caption{Total dissipated power as predicted by the non-thermal, linear thermal, and full thermal model as a function of frequency for a coated bubble with $R_0$= 2 $\mu$ m at various pressures (left column is a free Air bubble and right column is a C3F8 coated (left column is for air gas core and right column is for C3F8 gas core)).}
	\end{center}
\end{figure*}
In order to get a better understanding on the effect of thermal dissipation on the uncoated bubble oscillations, Fig. 1 presents the total dissipated power as a function of frequency for different excitation pressures. The bubble have $R_0=2 \mu m$ and the dissipated power is calculated by coupling the KM model with the Full Thermal (FTM) (Eqs. 5-6), linear Thermal (LTM) (Eqs. 7-10) and Non-Thermal model (NTM) (Eq. 4). The left column is for an air gas core and right column represents the C3F8 gas core. At $P_a=1 KPa$ (Figs 1a-b), predictions of FTM and LTM are in good agreement ; however, the NTM over estimates the dissipated power and also over estimates the resonance frequency. The resonance energy curves are wider when thermal effects are present which is due to the increased damping. Furthermore, due to lower thermal damping in the case of a C3F8 gas core, the resonance energy curves in Fig. 1b are narrower compared to Fig. 1a.\\
Nonlinear effects increase with increasing pressure and at  $P_a=30 kPa$ (Figs 2c-d)  predictions of the FTM and LTM start deviating. We also observe a 2nd superharmonic (SuH) resonance peak below resonance frequency. At $P_a=60 kPa$ (Figs. 2e-f) nonlinear effects result in further disagreement between the FTM and the LTM. Morever, the damping predicted by LTM and FTM are $\approx 30\%$ higher than the predictions of the NTM models. At higher pressure the total dissipated power grows at resonance and the 2nd SuH resonance while both resonance frequencies decrease in magnitude; this is a phenomenon known as pressure dependent (PD) resonance \cite{24}. We also observe the generation of a third peak below 2nd SuH resonance which indicates the 3rd SuH resonance frequency. NTM underestimates the total dissipated power by about 250 $\%$ at the main resonance. NTM predicts the correct value for the resonance frequencies of the bubble with C3F8 gas core; however, it over-predicts the resonance frequencies of the air bubble.\\
At $P_a=100 kPa$ predictions of the FTM for the total dissipated power deviate by up to 20 $\%$ from those made by the LTM model. However, for all pressures, the FTM and the LTM predict the same  resonance frequencies. The NTM underestimates the total dissipated power and over-predicts the resonance frequencies.\\We see here that for pressures above 1 kPa and for air and C3F8 uncoated bubbles FTM must be applied for more accurate predictions of the dissipated powers and bubble oscillations. In case of gases with higher thermal damping like Ar usage of the FTM becomes more important as thermal effects become even more significant.\\
\subsection{Total dissipated power by coated bubbles}
Figure 2 shows the total dissipated power as a function of frequency and gas core composition for a coated bubble with $R_0=2\mu m$, $G_s$=45 MPa and $\mu_{sh}=\frac{1.49(R_0(\mu m)-0.86)}{\theta (nm)}$ with $\theta=4 nm$. The right column represents an air gas core and the left column illustrates the dissipated power for a C3F8 gas core. When $P_a=1kPa$ (Fig 2a-b) and with air as the gas core, the NTM over estimates the value of the resonance frequency while the FTM and LTM predict the same resonance frequency. However, the LTM slightly under-predicts ($\approx 6 \%$ at resonance peak) the dissipated power. In case of a C3F8 gas core, the NTM, FTM and LTM predict the same value for resonance frequency and total dissipated power.  The total dissipated power increases with an increase in the excitation pressure (Fig. 2c-h), and the main resonance frequency shifting to lower frequencies. In case of an air gas core, the LTM slightly underestimates the total dissipated power at resonance; however, the LTM and FTM predict the same value for the dissipated power at other frequencies (Fig. 2). For the studied pressures in Fig.2 (1-100 kPa) even when 2nd order SuH occurs, predictions of the LTM and FTM are in good agreement. This is because unlike the uncoated bubble where the thermal damping is the major contributor to the total dissipated power, in case of coated bubble thermal effects are largely masked by the strong dissipation due to Cd. In case of a C3F8 gas core, predictions of the NTM, FTM and LTM are in excellent agreement at all excitation pressures examined. This is because C3F8 has smaller Td compared to air, therefore Td is masked by the strong Cd. For the characterization of UCAs  that enclose gases like C3F8 (coated bubbles with $R_0 < 4 \mu m$), thermal effects can fully be neglected. For larger bubbles, Td increases as the surface area for Td increases; however, UCA sizes are limited to bubbles smaller than 8 $\mu m$ \cite{12}. Fig. 1 however shows that FTM  is the appropriate model for more accurate prediction of the behavior of uncoated bubbles.\\

\begin{figure*}
	\begin{center}
		\scalebox{0.43}{\includegraphics{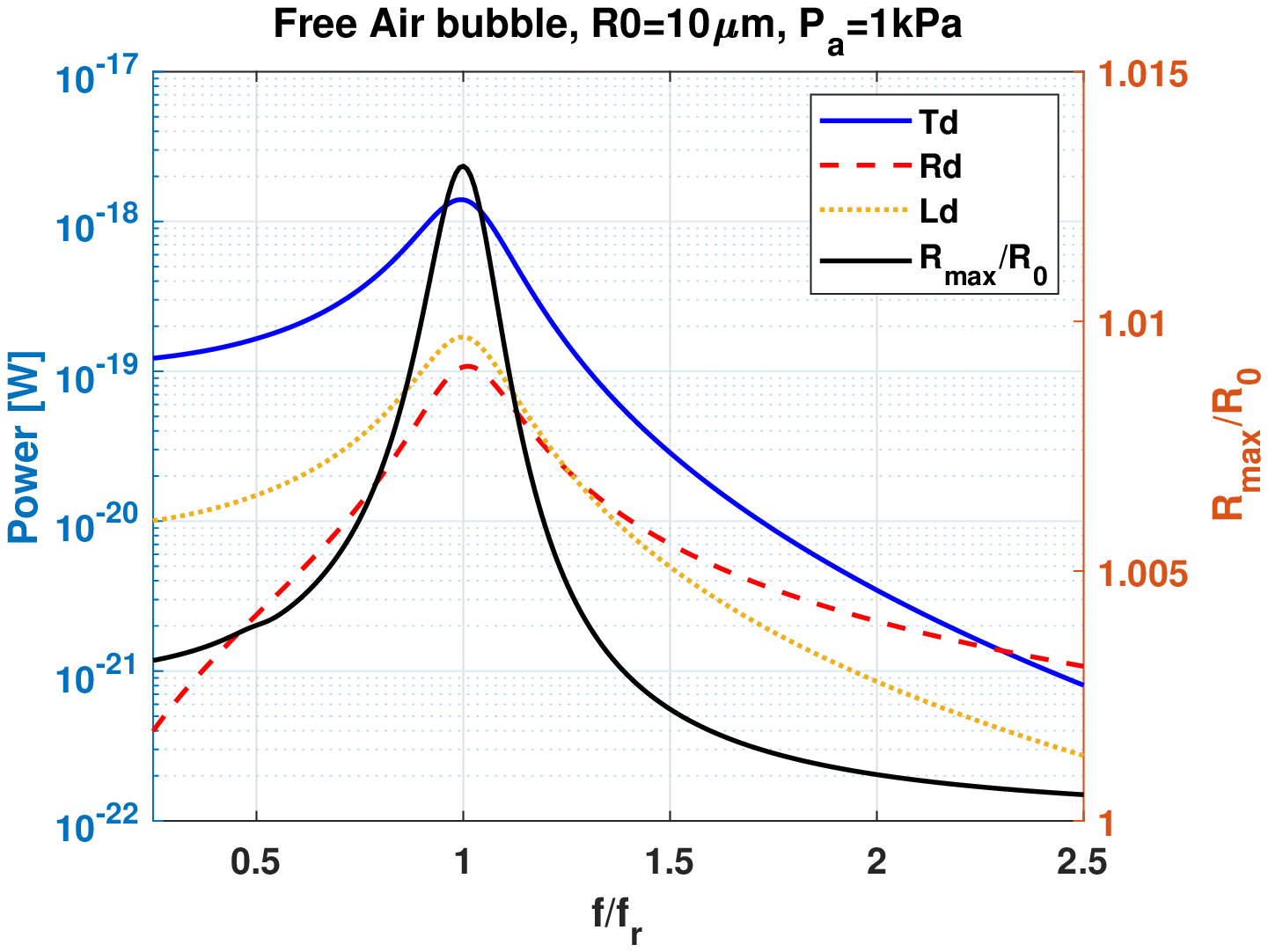}} \scalebox{0.43}{\includegraphics{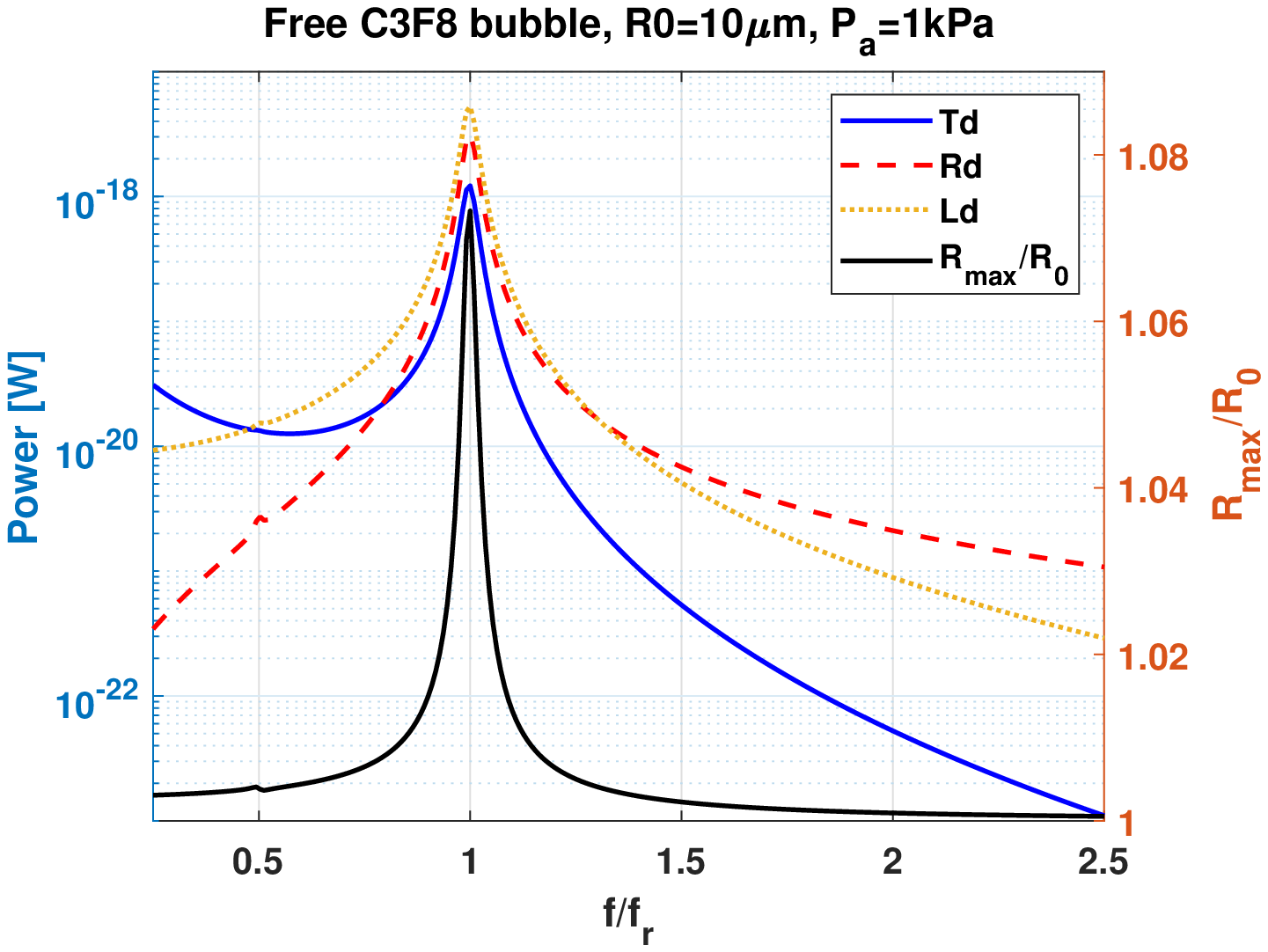}}\\
		\hspace{0.5cm} (a) \hspace{6cm} (b)\\
		\scalebox{0.43}{\includegraphics{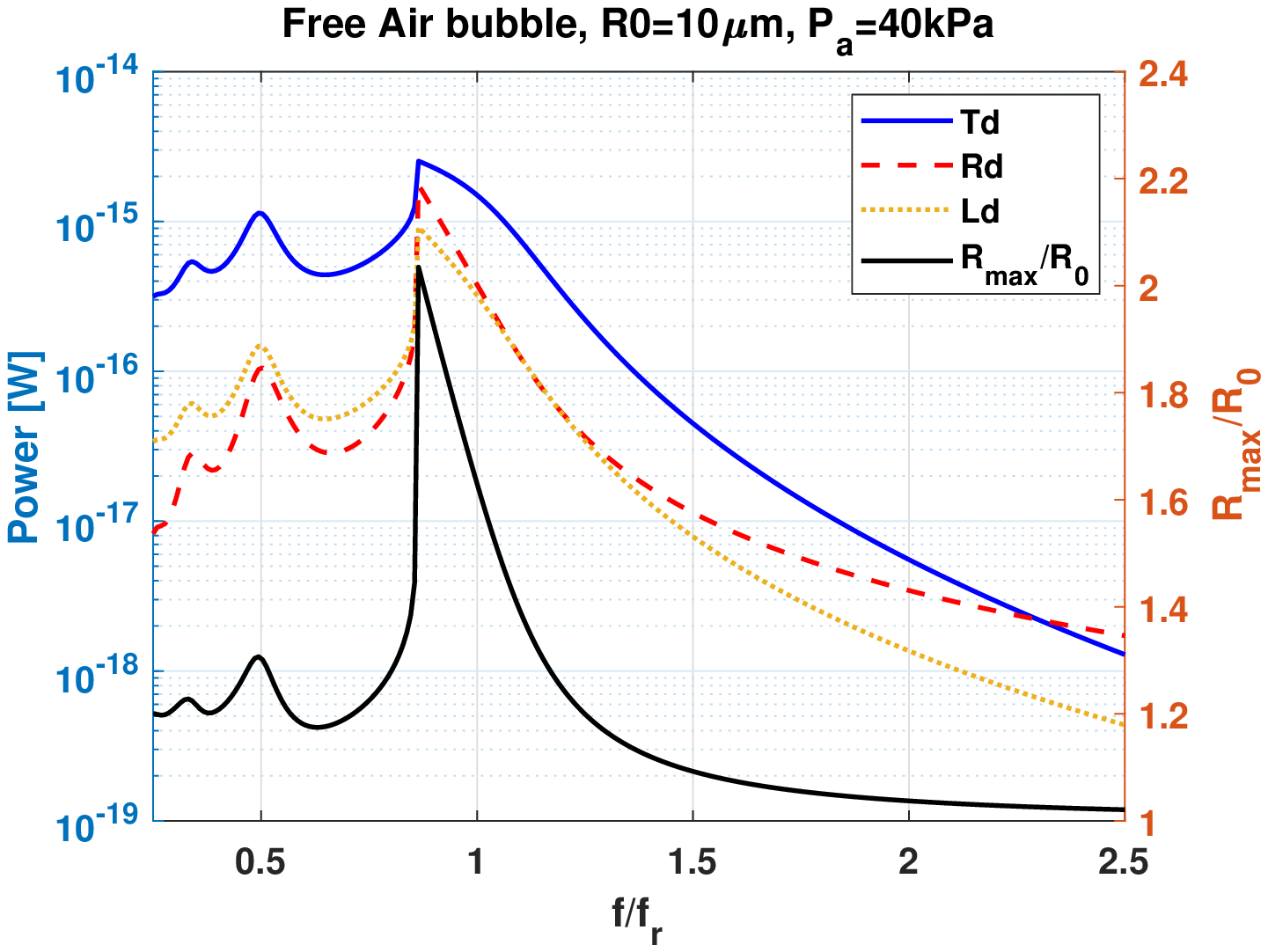}} \scalebox{0.43}{\includegraphics{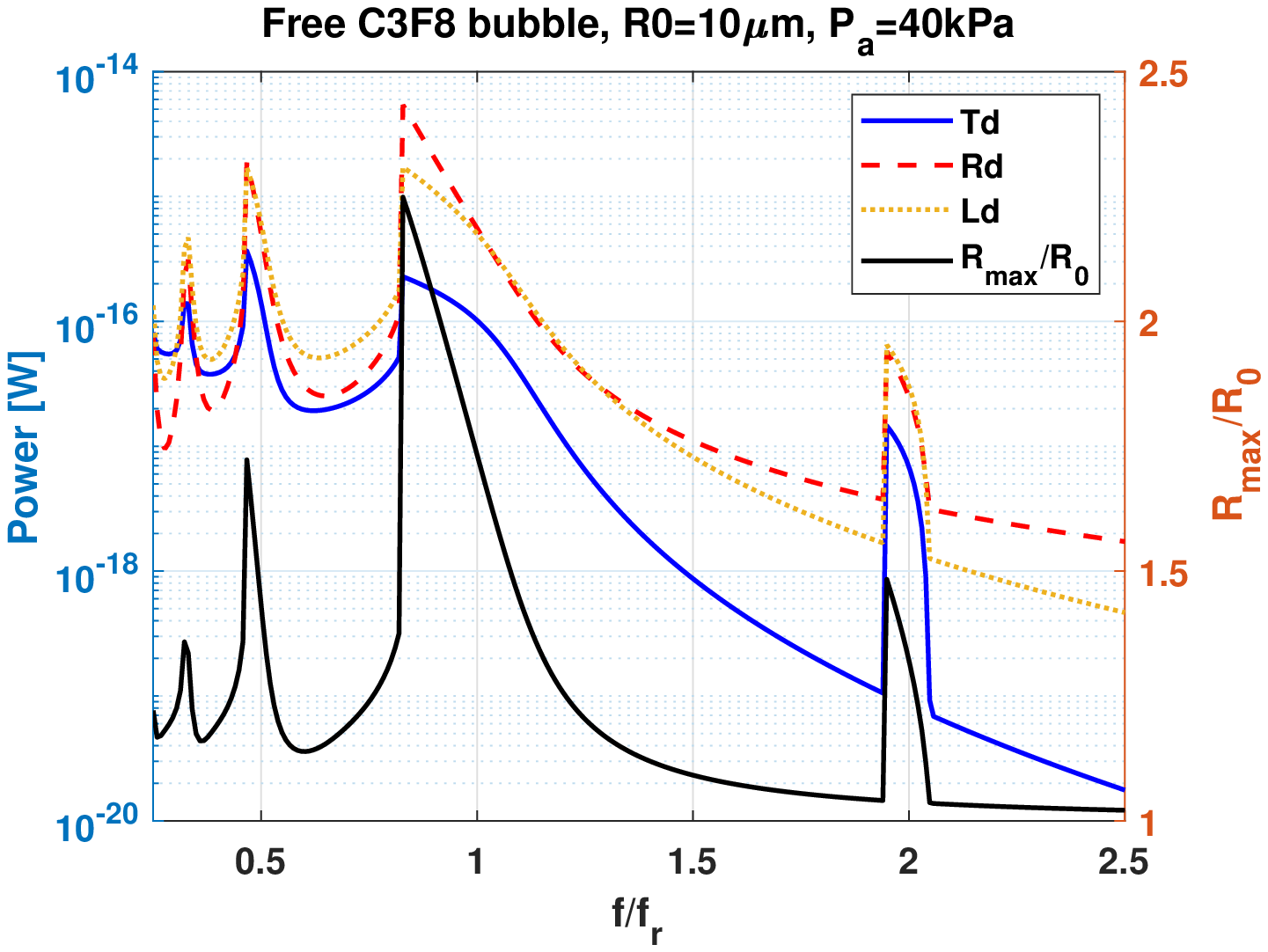}}\\
		\hspace{0.5cm} (c) \hspace{6cm} (d)\\
		\scalebox{0.43}{\includegraphics{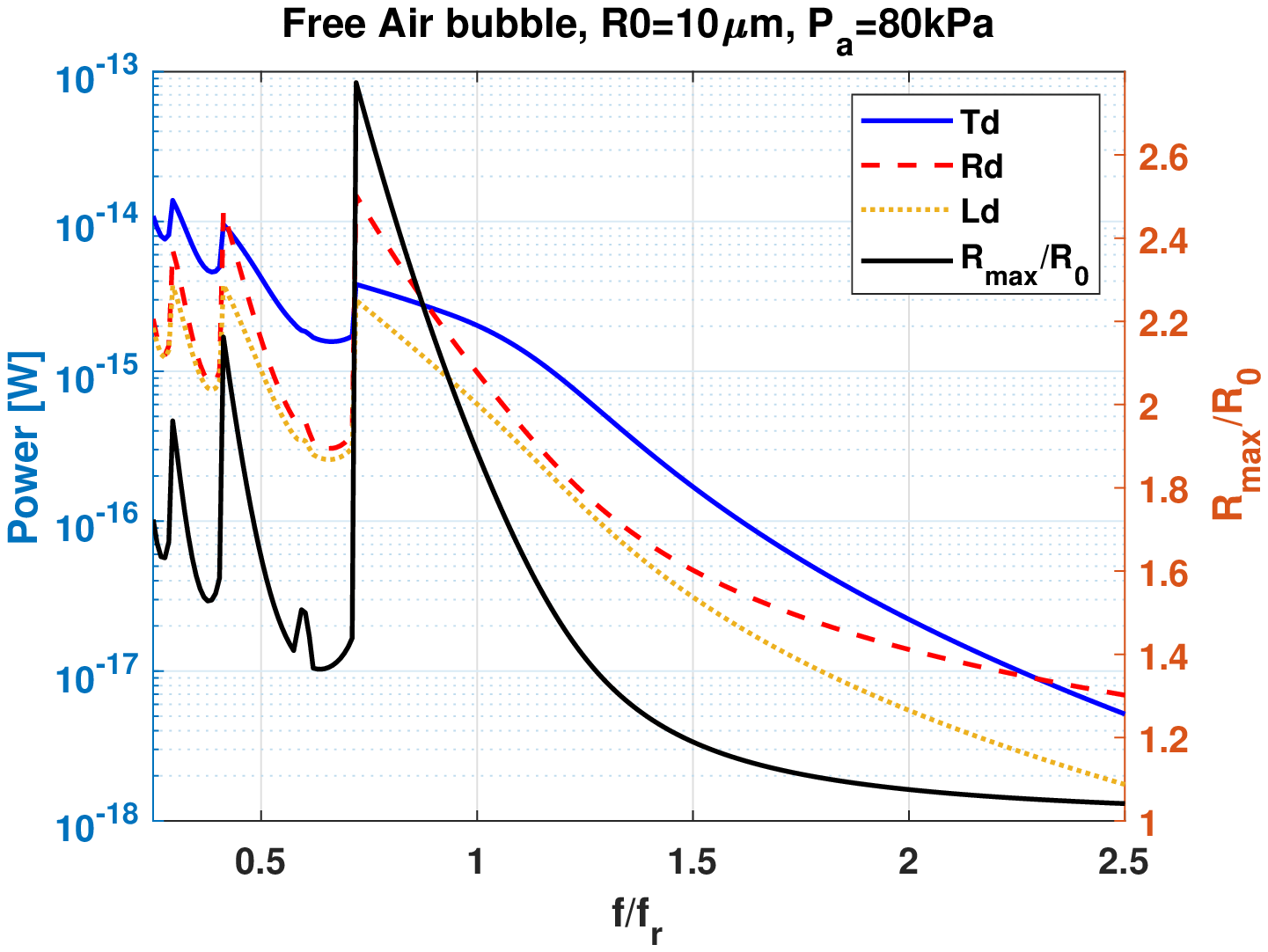}} \scalebox{0.43}{\includegraphics{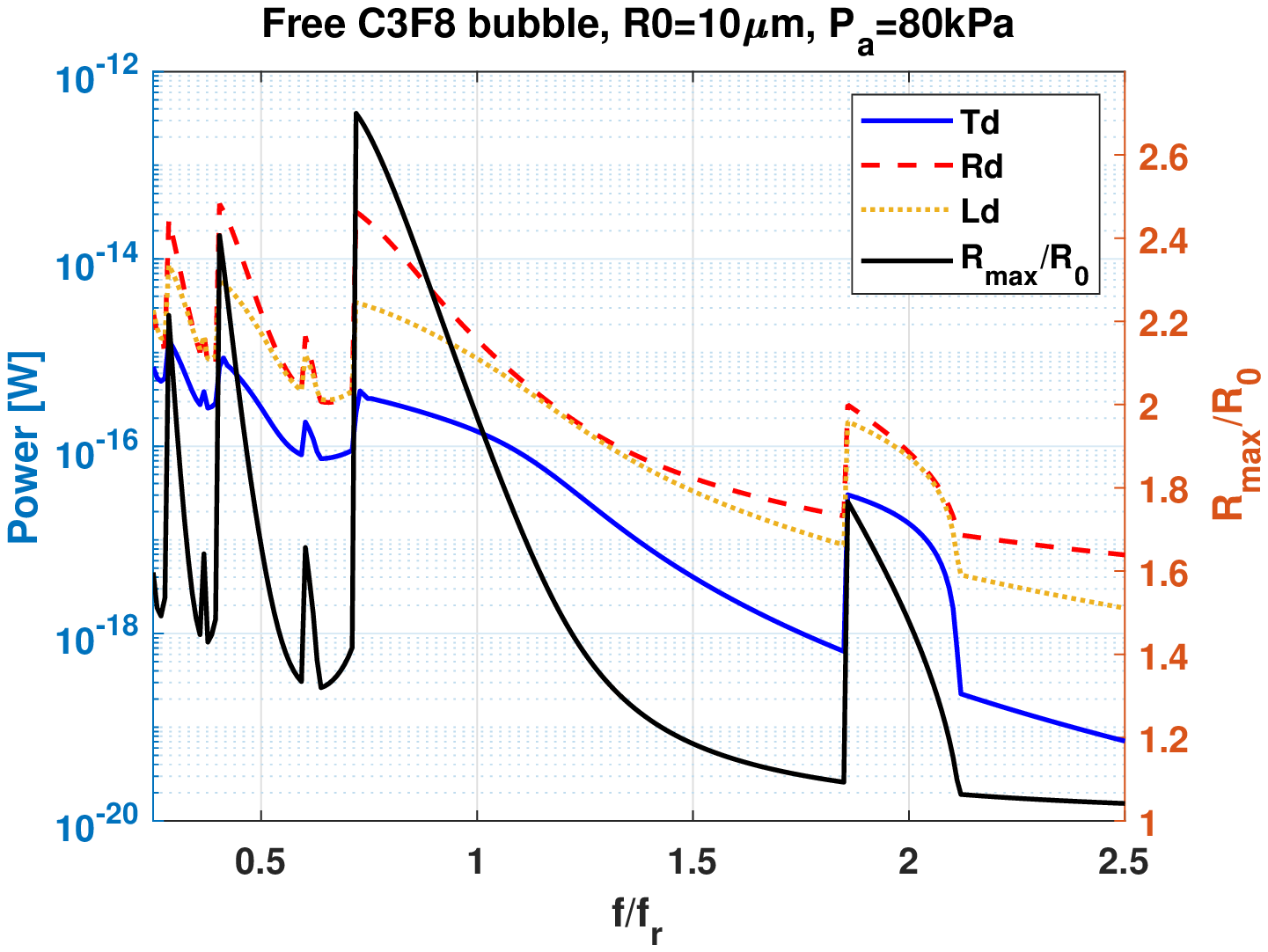}}\\
		\hspace{0.5cm} (e) \hspace{6cm} (f)\\
		\scalebox{0.43}{\includegraphics{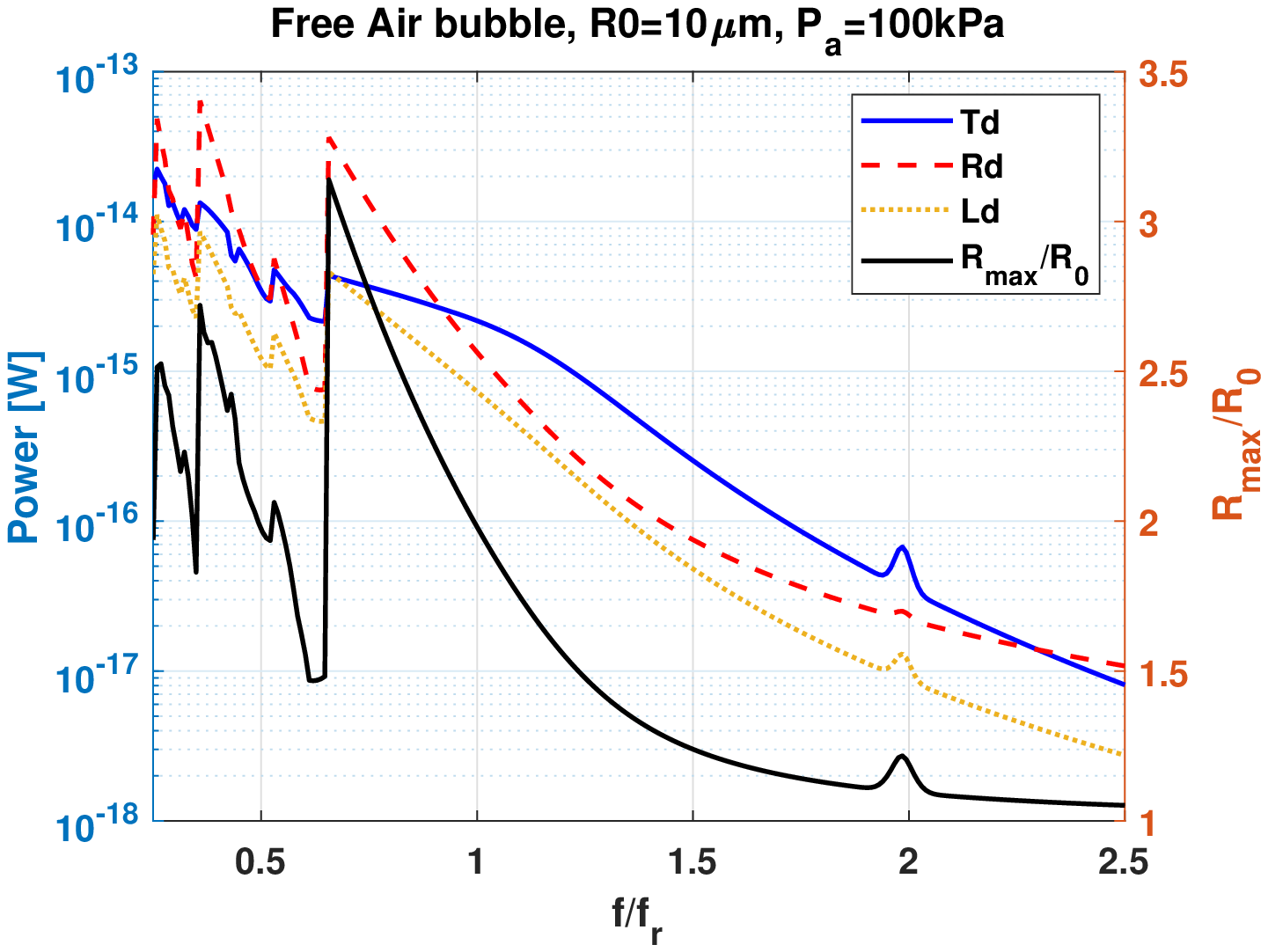}} \scalebox{0.43}{\includegraphics{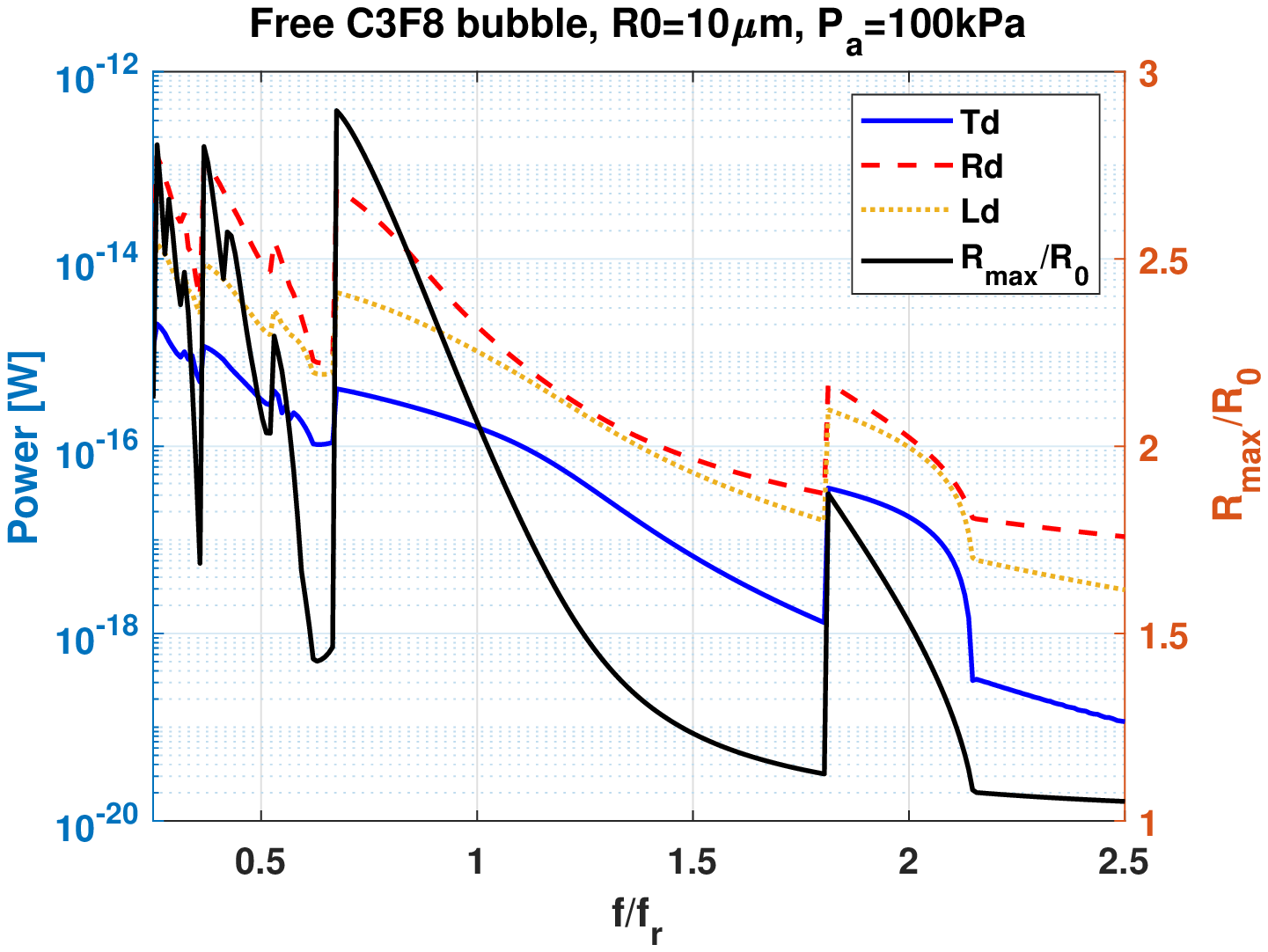}}\\
		\hspace{0.5cm} (g) \hspace{6cm} (h)\\
		\caption{Dissipated power due to Td, Ld and Rd as predicted by the full thermal model as a function of frequency for a free bubble with $R_0$= 10 $\mu$ m at various pressures (left column is a free Air bubble and right column is a C3F8 coated).}
	\end{center}
\end{figure*}

\subsection{Mechanisms of damping in uncoated bubbles at different pressure amplitudes and sizes}
\subsubsection{The uncoated bubble with $R_0=10 \mu m$}
In this section we investigate the dependence of Td, Ld, and Rd on pressure, frequency, bubble size and gas core composition. Fig. 3 shows Td, Rd, Ld and resonance frequency ($\frac{R_{max}}{R_0}$) of an uncoated bubble with $R_0=10\mu m$ as a function of frequency for different pressure amplitudes. The right column plots the data for an air bubble and the left column plots the data for the C3F8 gas core. All of the calculations are performed using the FTM. When the gas core is air (Fig. 3a), Td is the major energy dissipation factor for frequencies below $2f_r$. These results are consistent with analytical predictions \cite{50} where Td$>$Ld$>$Rd. For a C3F8 gas core (Fig. 3b) , Ld is the dominant damping factor at $0.5f_r<f<1.4f_r$ below which Td becomes dominant and above which Rd becomes dominant. The width of the curve is considerably narrower due to weaker damping effects; for the bubble with $R_0=10 \mu m$ thermal damping for the air gas core is very high and leads to widening of the curve. Due to smaller damping effects (Fig. 3b), the bubble is able to grow larger $\frac{R_{max}}{R_0}$ and achieve higher wall velocities, thus energy dissipation due to Ld becomes significant at resonance.\\
Figs. 3c-d show the case of $P_a$=40 kPa. For the air gas core bubbles (Fig. 3c) there is a  shift of the resonance frequency to lower values as well as the generation of 2nd and 3rd SuH resonance. Td is still the major damping factor for $f<2f_r$; however, Rd has grown  faster compared to Ld at resonance and is now stronger than Ld at pressure dependent resonance ($PDf_r$) and higher frequencies. This may increase the scattering to damping ratio (STDR); a parameter that we would like to maximize in imaging applications. For bubble with C3F8 gas core (Fig. 3d), we witness the generation of sub-harmonic (SH) resonance peak at $\approx 2f_r$. The SH peak occurs at lower pressures for C3F8 due to weaker damping  compared to air. Rd is the major damping factor at  $PDf_r$ and 2nd SuH resonance and Rd and Ld are comparable at $2f_r$ and 3rd SuH resonance. Due to a weaker Td (an order of magnitude smaller compared to Fig. 3c) in case of the C3F8 bubble (Fig. 3d) amplitude of the bubble oscillations are higher at the resonances.  Application of C3F8 as the gas core can potentially increase STDR in bigger bubbles by suppressing the effect of Td.\\
For air bubble, when $P_a$=80 kPa Rd dominates damping at frequencies below $PDf_r$ (Fig. 3e). This shows that compared to Ld and Td, Rd grows at a faster rate with $P_a$. Thus contrary to predictions of the linear models, we can find a pressure and frequency region in which Rd is higher than Td and Ld and thus the STDR is optimized. For C3F8 (Fig. 3f) and at $P_a$=80 kPa Rd is the strongest contributor to the dissipated power at the frequencies studied. At this pressure we also witness the generation of $\frac{3}{2}$ and $\frac{5}{2}$ UH resonance (e.g. the peak between $PDf_r$ and 2nd SuH) with Rd as the strongest damping factor.\\
For $P_a$=110 kPa, Fig. 3g shows that SH resonance peak appears at $f=\approx 2f_r$. Rd becomes the major damping factor at frequencies below $f_r$; however, due to the fact that $\frac{R_{max}}{R_0}$ have exceeded 2 the bubble can not sustain long lasting stable oscillations and will undergo destruction \cite{59,24}. We observe the same phenomenon in the C3F8 gas core bubbles (Fig. 3h) with Rd being the strongest damping mechanism. In case of the C3F8, Rd$>$Ld$>$Td and in case of Air gas core, Rd$>$Td$>$Ld. 

\subsubsection{The uncoated bubble with $R_0=2 \mu m$}
 Fig. 4 shows the dissipated power due to Td, Ld and Rd in the oscillations of a 2 $\mu m$ bubble. The left column represents an air bubble and the right column represents the C3F8 bubble.
 Fig. 4a shows that at $P_a=1 kPa$, Td and Ld are the major mechanisms of energy dissipation in the bubble oscillations. Because the bubble is smaller than the previous case (see Fig. 3a where Td is an order of magnitude greater than Ld), Td and Ld have equal contributions to damping. Fig. 4b represents the C3F8  bubble. Due to further weakening of the thermal effects, Ld is an order of magnitude higher than Td.\\
 Increasing $P_a$ to 40 kPa leads to the generation of 2nd and 3rd SuH resonance frequencies and the decrease in the fundamental resonance ($PDf_r$) (Fig 4c-d). Due to a weaker Td in the case of the C3F8 bubble (Fig. 4d), the amplitude of the bubble oscillations are higher at the resonances. For the air bubble Td and Ld are the major dissipation mechanisms for $f<2f_r$. For C3F8 $Ld>Rd>Td$ for frequencies of and above the $2nd SuH$; this suggests that the STDR can be higher for the C3F8 bubble.\\
 \begin{figure*}
 	\begin{center}
 		\scalebox{0.43}{\includegraphics{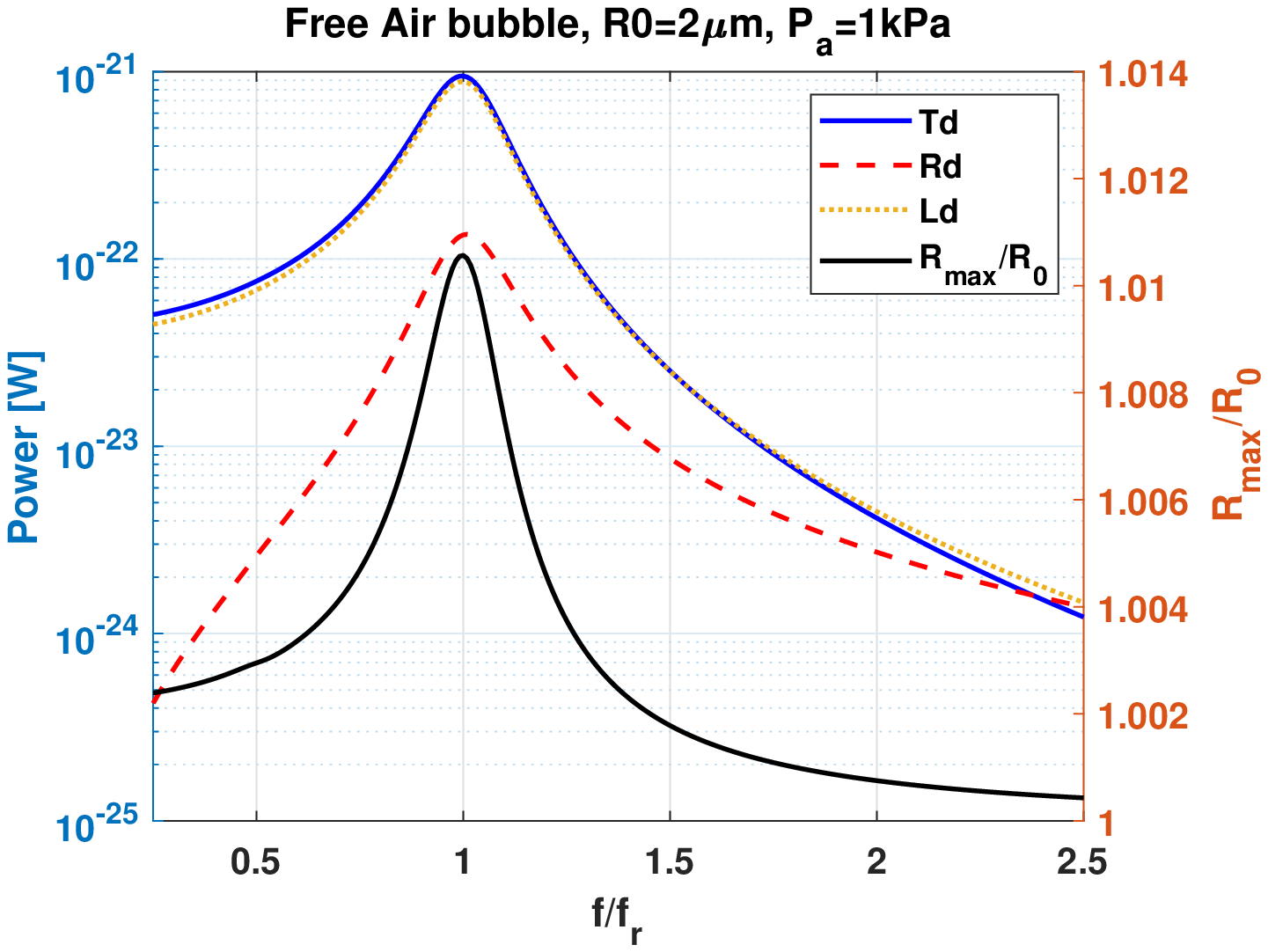}} \scalebox{0.43}{\includegraphics{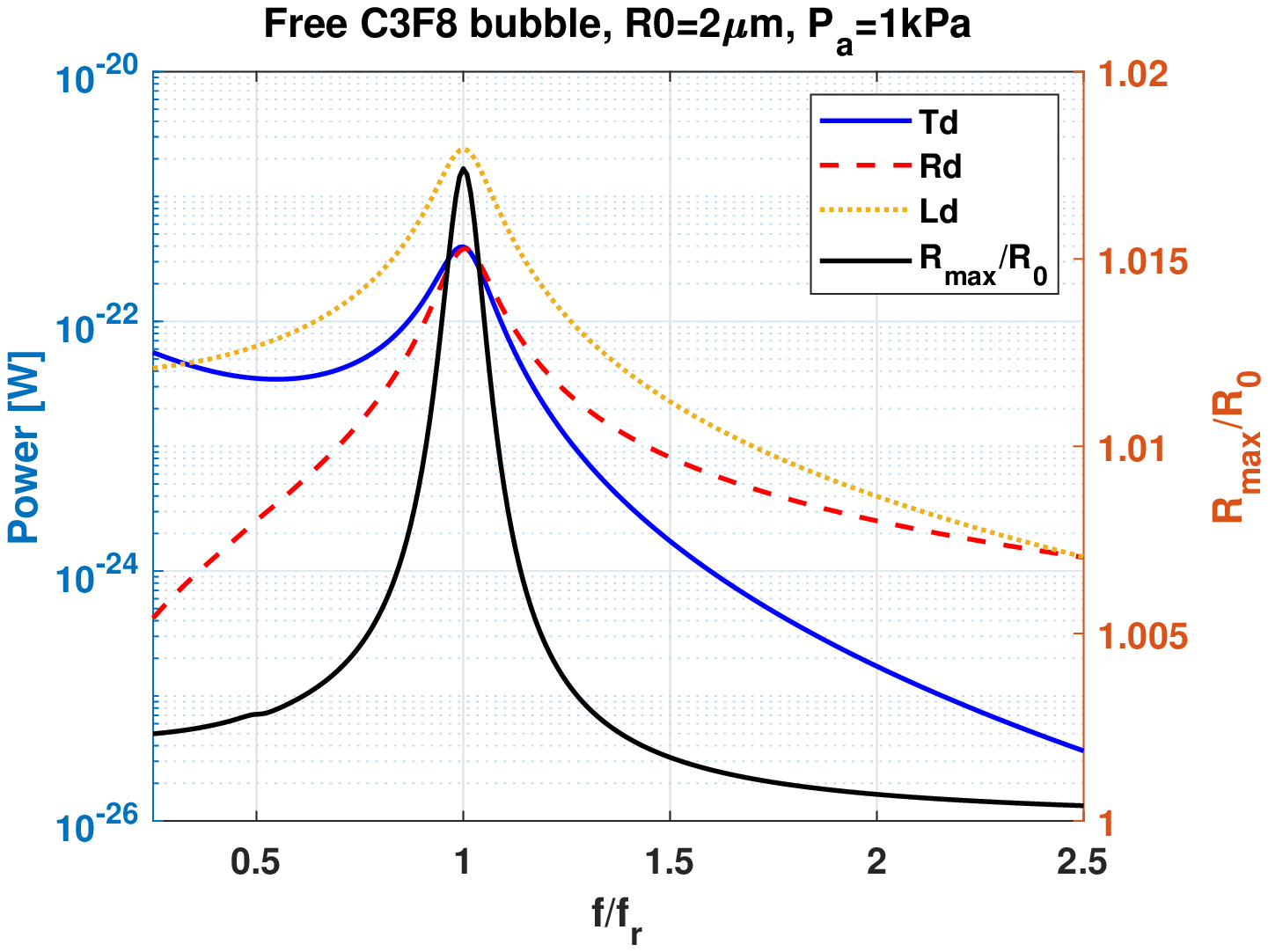}}\\
 		\hspace{0.5cm} (a) \hspace{6cm} (b)\\
 		\scalebox{0.43}{\includegraphics{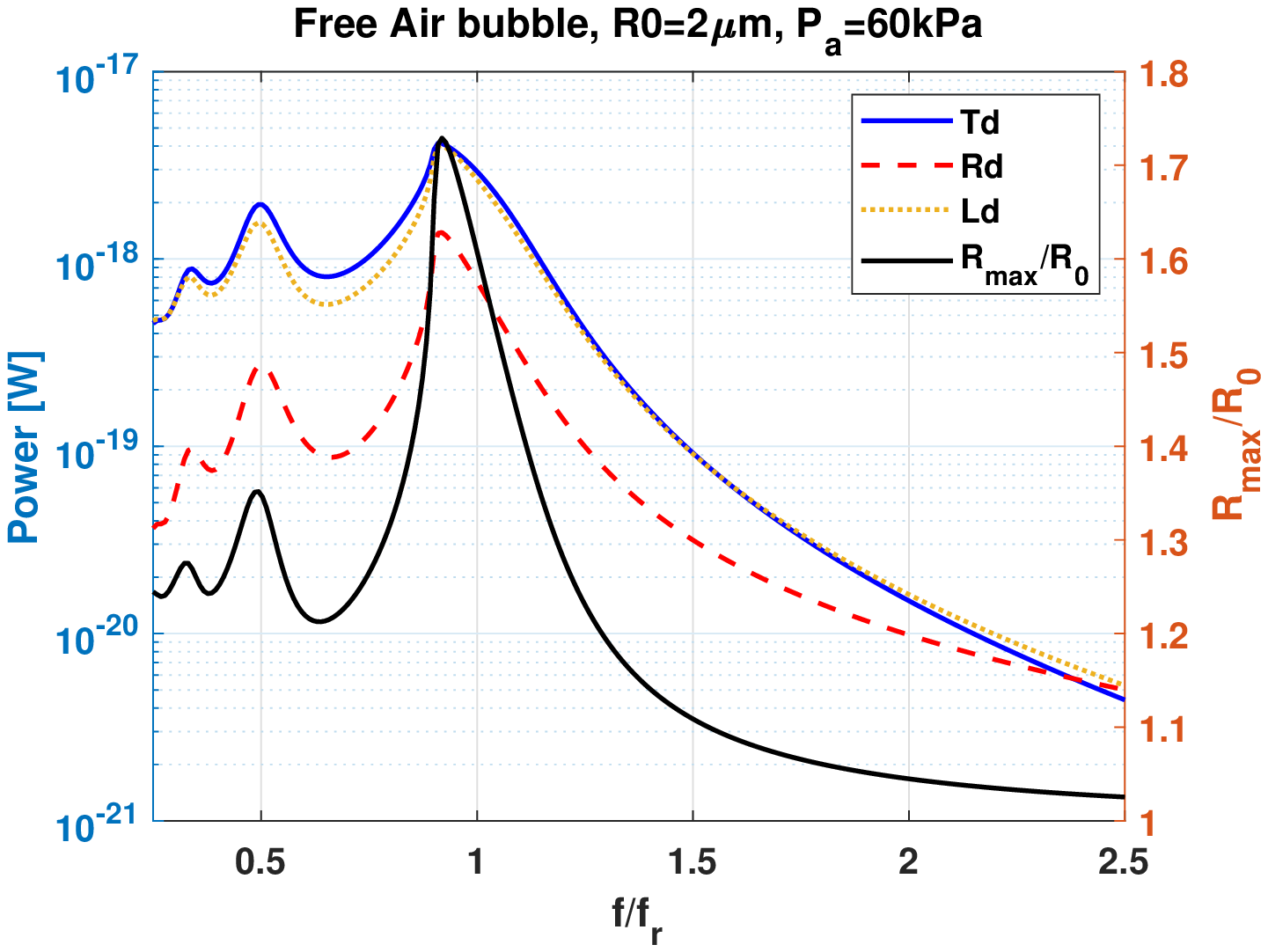}} \scalebox{0.43}{\includegraphics{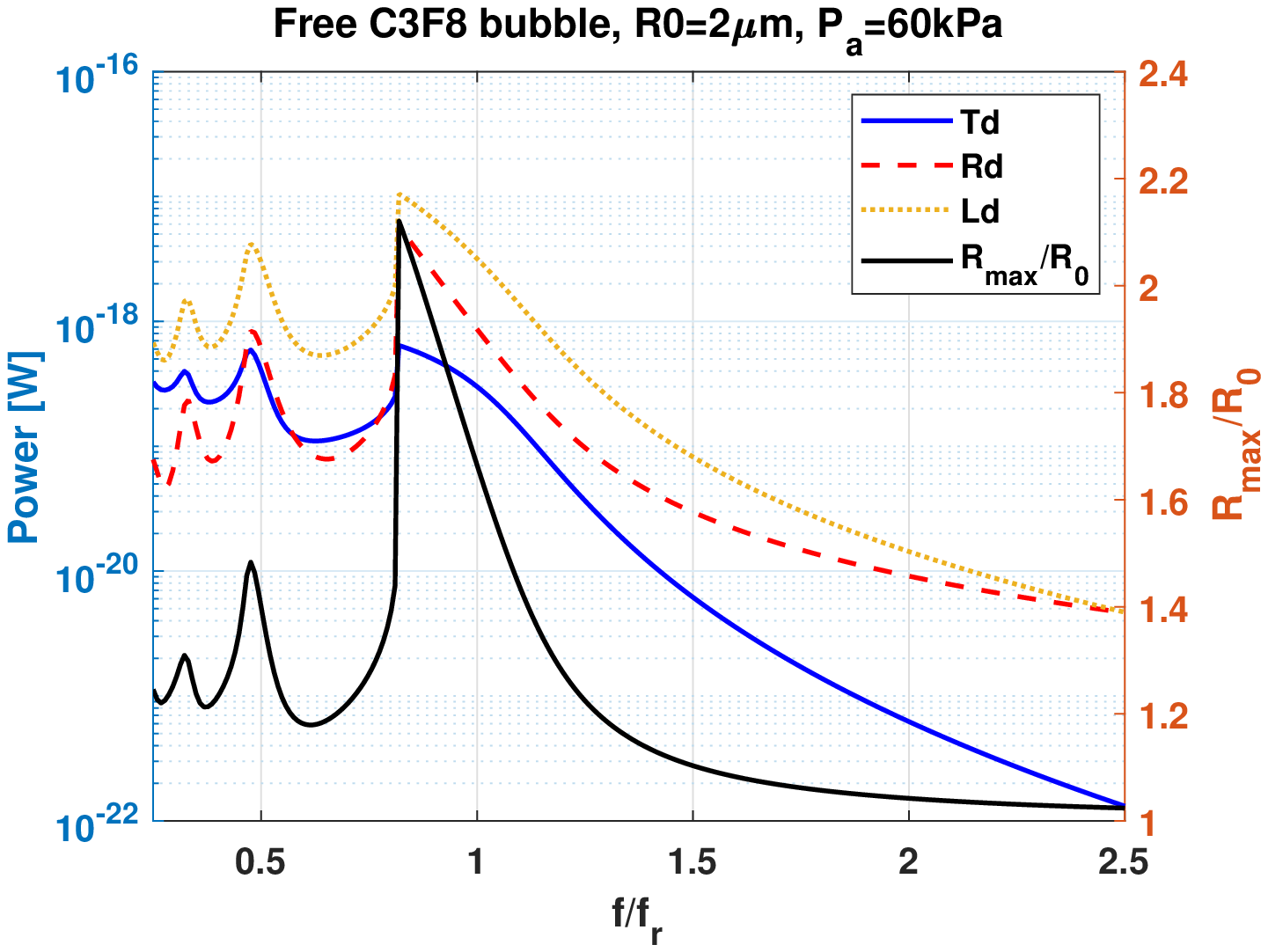}}\\
 		\hspace{0.5cm} (c) \hspace{6cm} (d)\\
 		\scalebox{0.43}{\includegraphics{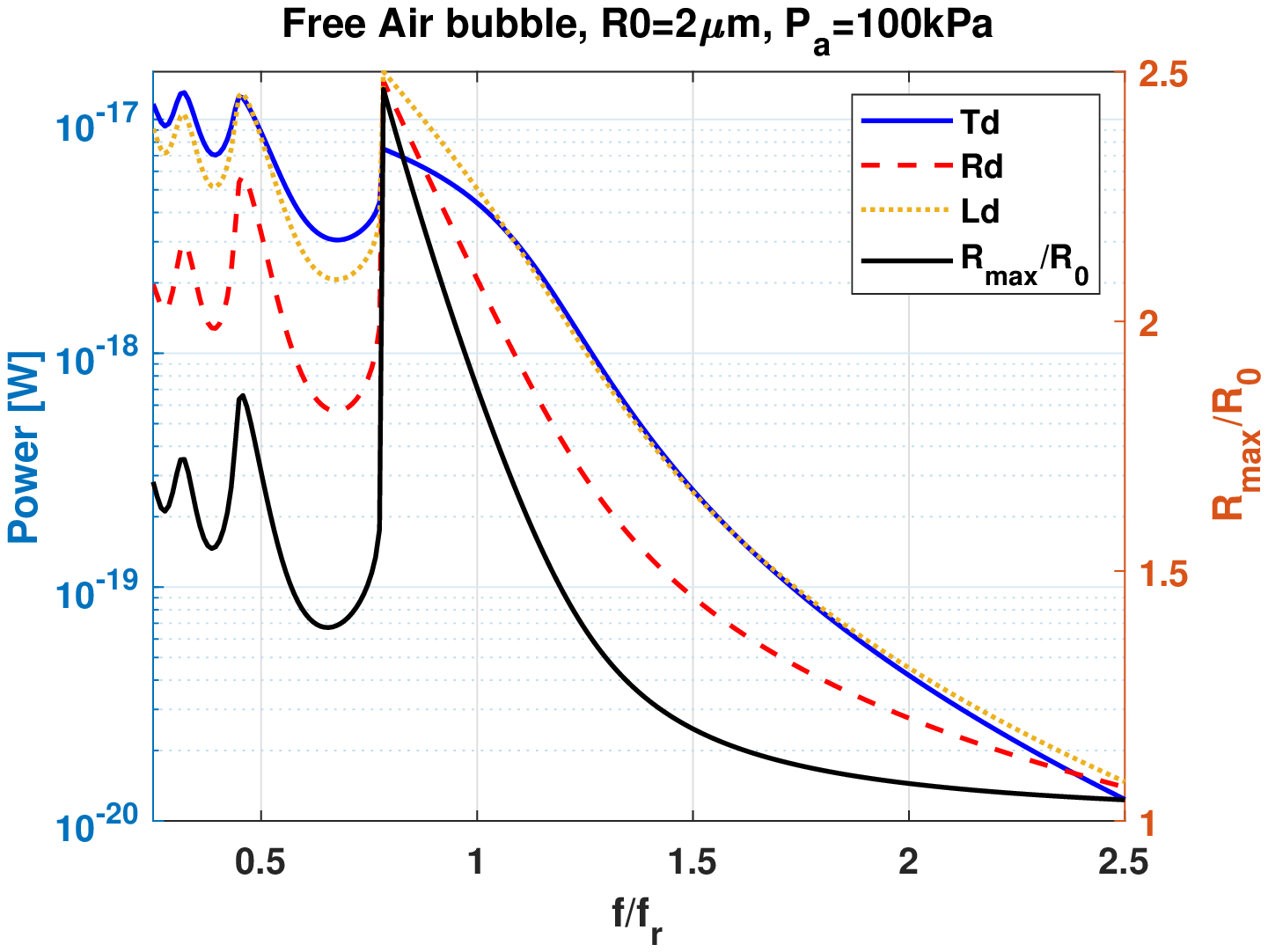}} \scalebox{0.43}{\includegraphics{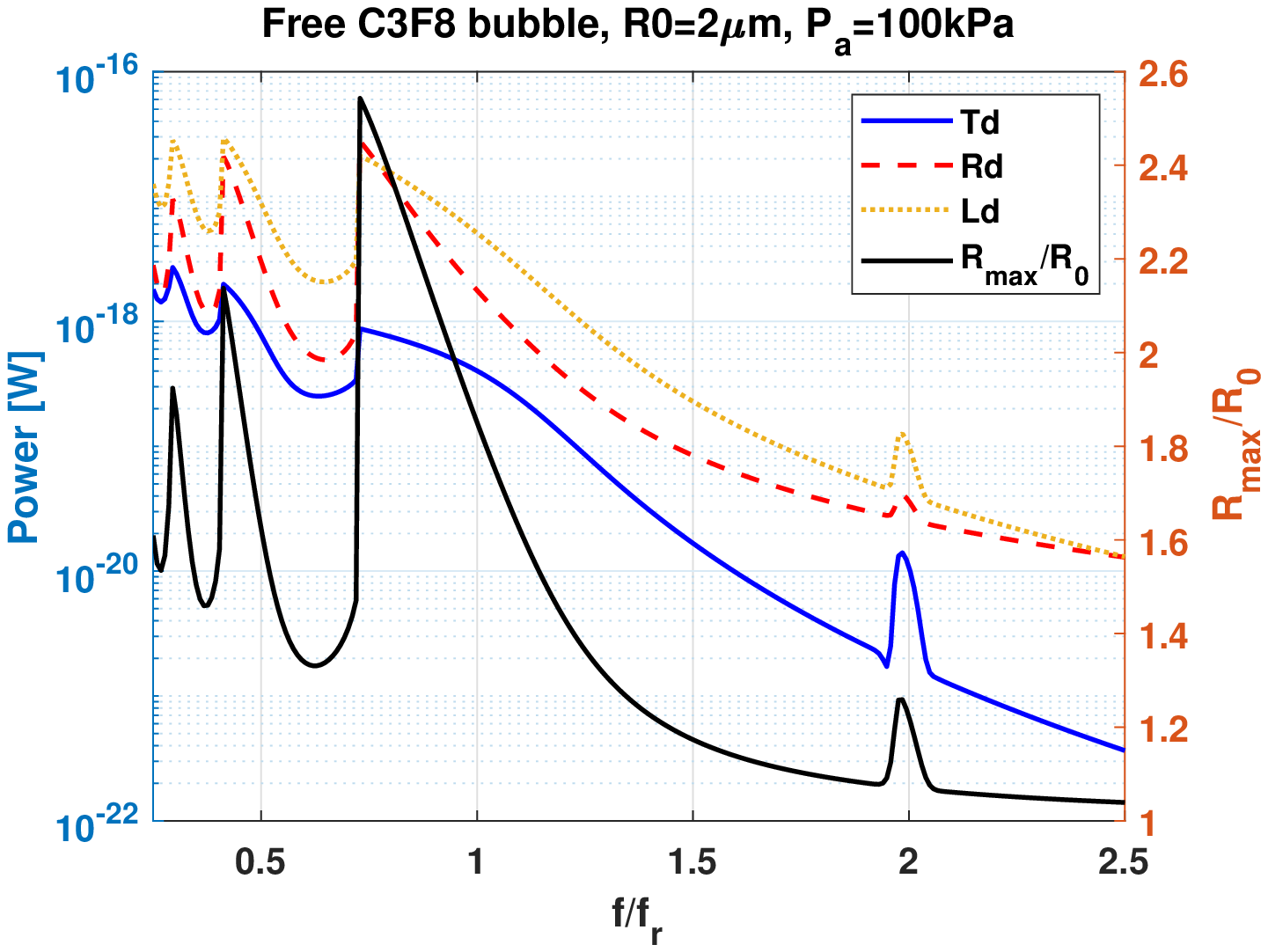}}\\
 		\hspace{0.5cm} (e) \hspace{6cm} (f)\\
 		\scalebox{0.43}{\includegraphics{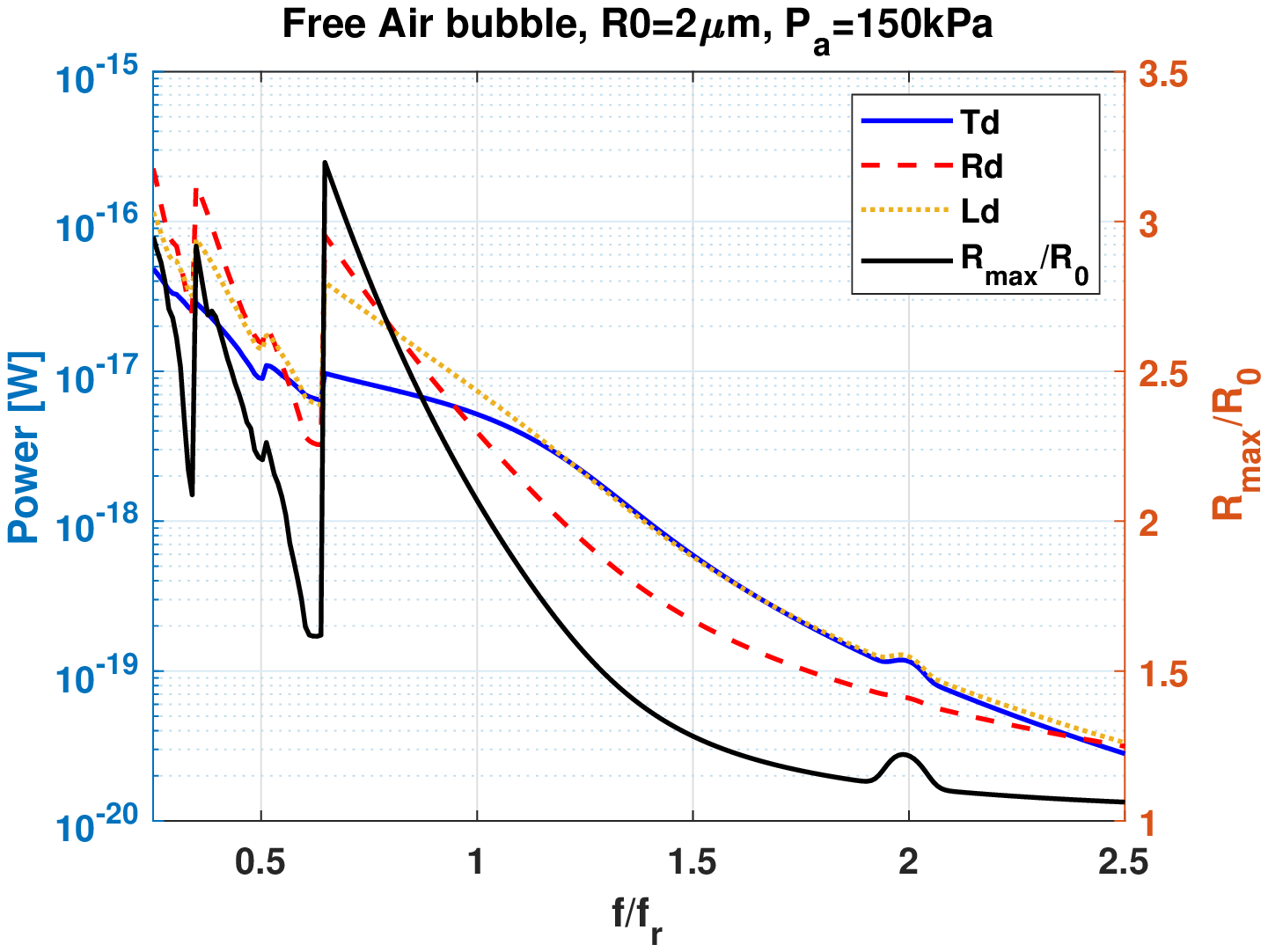}} \scalebox{0.43}{\includegraphics{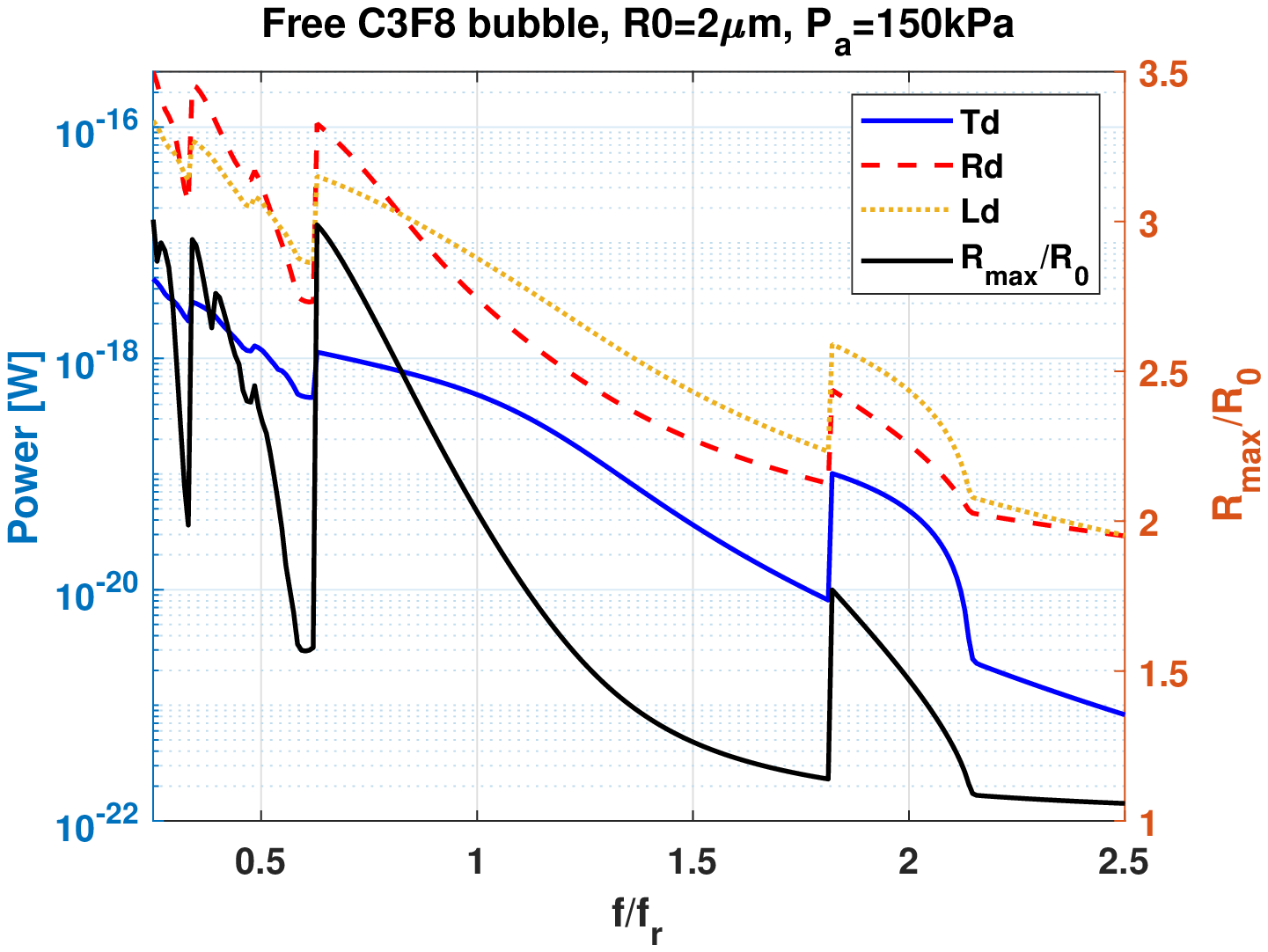}}\\
 		\hspace{0.5cm} (g) \hspace{6cm} (h)\\
 		\caption{Dissipated power due to Td, Ld and Rd as predicted by the full thermal model as a function of frequency for a free bubble with $R_0$= 2 $\mu$ m at various pressures (left column is a free air bubble and right column is a C3F8 coated).}
 	\end{center}
 \end{figure*}
 For $P_a$=100 kPa and for an air bubble (Fig. 4e) Td becomes less significant at f=$PDf_r$; with $Ld>Rd>Td$. This implies a higher STDR at this frequency and pressure. For f=2nd SuH and f=3rd SuH we witness a fast growth of Rd however still $Ld>Td>Rd$. These results indicate that as pressure increases Rd grows faster than Ld and Td. In case of C3F8 (Fig. 4f) bubble we see the generation of SH resonance at $f\approx 2f_r$ and $Ld>Sd\gg Td$. Because of a higher Rd, when the gas core is C3F8, a higher STDR may be expected. Further increasing the pressure to 150 kPa leads to the generation of SH resonance peak at $f\approx 2f_r$ (Fig. 4g) for the air bubble. The later appearance of the SH peak  is due to the higher Td in the air bubble. For an air bubble at $P_a$=100 kPa, $Rd>Ld>Td$ at $PDf_r$ , $2ndSuH$ and $3rdSuH$ and $\frac{3}{2} UH$ resonances. However, for frequencies above $f_r$ TD and Ld are the major damping factors. Results of Figs. 3 and 4 suggest a relationship between $\frac{R_{max}}{R_0}$ and the order of damping factors for the air bubble. With an increase in the incident pressure, Rd grows faster than other damping factors; at $\frac{R_{max}}{R_0} \approx 2$, Rd becomes stronger than the other damping factors. These results suggest that for case of air bubbles, in order to increase the STDR one needs to sufficiently increase the pressure; however we should also consider the lower threshold of bubble destruction which is $\frac{R_{max}}{R_0}>2$. At $P_a$=100 kPa and in case of C3F8 bubble (Fig. 4h), $Rd>Ld>Td$ for the studied frequency range. Also we see a stronger SH oscillations compared to the air bubble.\\
\subsubsection{The coated (encapsulated) bubble with $R_0=4 \mu m$}
Fig. 5 plots the dissipated power due to the damping from coating (Cd), Td, Rd and Ld for a coated bubble with $R_0$=4 $\mu$m ,$G_s$=45 MPa, $\theta$=9 nm and $\mu_{sh}=\frac{1.49(R_0(\mu m)-0.86)}{\theta (nm)}$. This was chosen as it is the upper limit of the coated bubbles that can be used in biomedical ultrasound and thus has the highest Td. The right column represents the air gas core and the left column represents the C3F8 gas core. When $P_a$=1 kPa and for the air as the filling gas (Fig. 5a) the damping due to Cd is the major damping factor $Cd>Td>ld>Rd$ at the main resonance and frequencies below. For frequencies above $1.5f_r$ Rd contributes more to damping with Rd$\approx$Cd at $f=2.5f_r$. At  $f \approx$ $f_r$, Cd is 3 times stronger than Td and Ld. For the bubble enclosing a C3F8 core (Fig. 5b) Cd is the major dissipation factor with $Cd>Ld>Rd>Td$; Cd is $\approx$ 3.8 larger than Ld and  40 times larger than Td. Thus use of C3F8 as the gas core significantly reduces the Td at resonance.\\
For the coated bubble enclosing an air core at $P_a = 80 kPa$ (Fig. 5c), we see the generation of 2nd and 3rd SuH resonances and a large shift of the main resonance to $PDf_r$ at $f=0.79f_r$. Cd is the major damping factor ($\approx$ 4.5 times larger than Ld) with $Ld\approx Rd \approx Td$. At frequencies below resonance $Cd>Td>Ld>Rd$ and at frequencies above $1.5f_r$ $Rd>Ld \approx Td$. For the bubble enclosing a C3F8 gas core (Fig. 5d) the shift in main resonance is more significant due to less thermal damping effects  ($PDf_r=0.76f_r$) and $Cd>(Ld\approx Rd)>Td$. Here Cd is 64 times stronger than Td at the $PDf_r$.\\
Fig. 5 displays the case of sonication with $P_a$= 160 kPa. For the Air bubble (Fig. 5e) $Cd>Rd>Ld>Td$ at frequencies near and below  $PDf_r$.  When the growth rate of the dissipation factors are compared in Fig. 5, we see that Rd grows the fastest with pressure increase and Td has the slowest growth rate. As pressure increases Rd grows faster and eventually becomes the second major dissipation factor while the initially strong Td lags behind the rest of the damping factors. Due to the high bubble oscillation amplitude $\frac{R_{max}}{R_0}>2$ the bubble may not sustain stable non-destructive oscillations for $f<f_r$ at 160 kPa. At this pressure and at $f_r<f<1.5f_r$ Ld=Td=Rd. For the C3F8 bubble, due to the smaller damping from thermal effects we can see the generation of SH resonance peak at $f \approx 2f$ (Fig. 5f). For the frequency range that is studied here, $Cd>Rd \approx Ld$. Cd is $\approx$ 100 times and 600 times stronger than Td, respectively at $PDf_r$ and $PDfsh$  (pressure dependent SH resonance at $\approx 1.86 f_r$).\\
At $P_a$ =240 kPa the SH resonance peak is seen for the coated air bubble (Fig. 5g). The bubble oscillation amplitude has exceeded the threshold of destruction $\frac{R_{max}}{R_0}>2$ for $f<f_r$. At f=PDfsh ($1.67f_r$ at $P_a$=240 kPa) $Cd>Rd>Td \approx Ld$. Cd is about 46 times stronger than Td. For the coated bubble enclosing C3F8 (Fig. 5h) thermal effects are much weaker at PDfsh ( $1.62 f_r$ at $P_a$=240 kPa) with $Cd>Rd>Ld>Td$ and Cd is about 72 times stronger than Td. For frequencies below $f_r$, the bubble can not sustain stability at this pressure.
 \begin{figure*}
	\begin{center}
		\scalebox{0.43}{\includegraphics{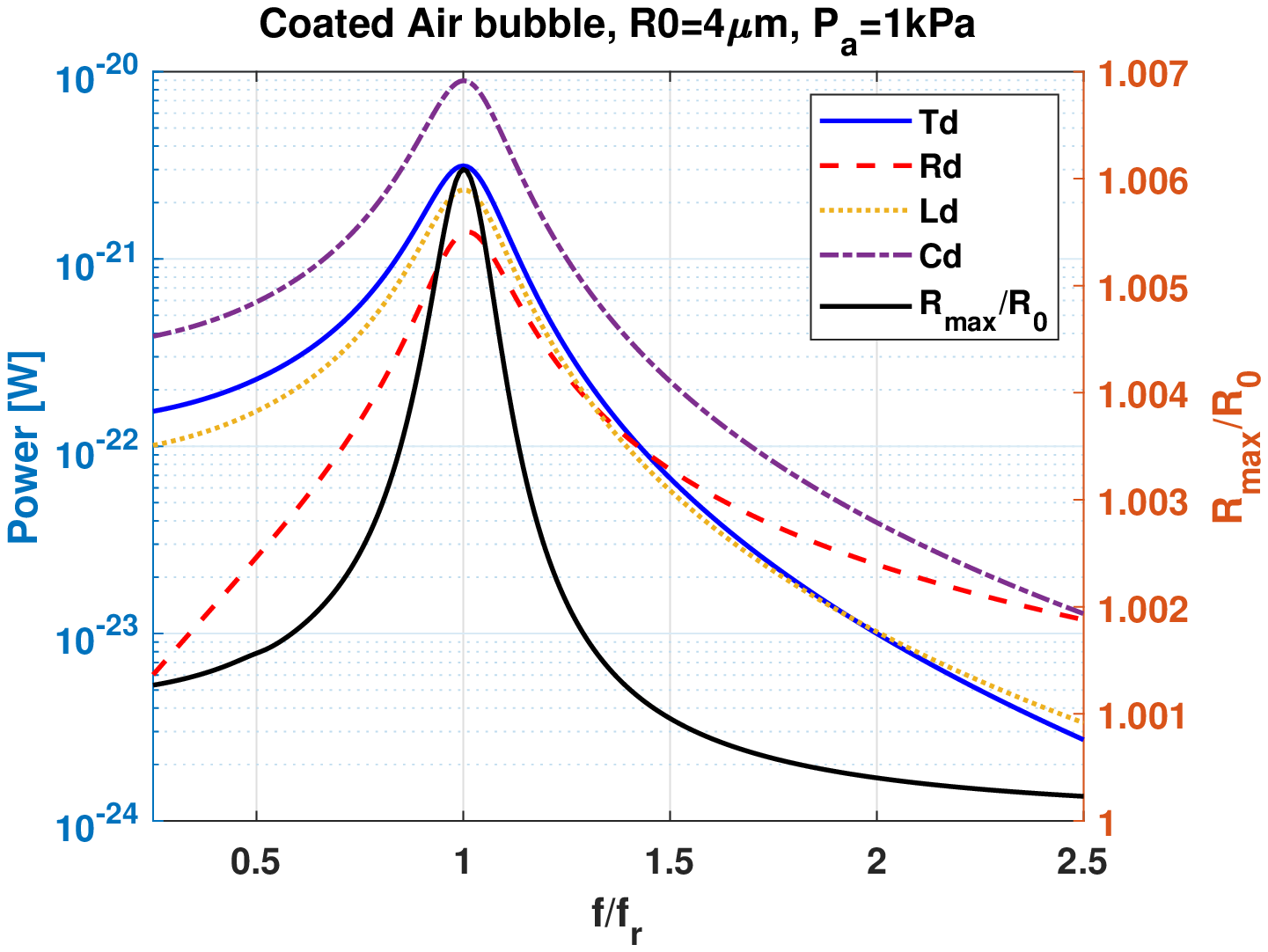}} \scalebox{0.43}{\includegraphics{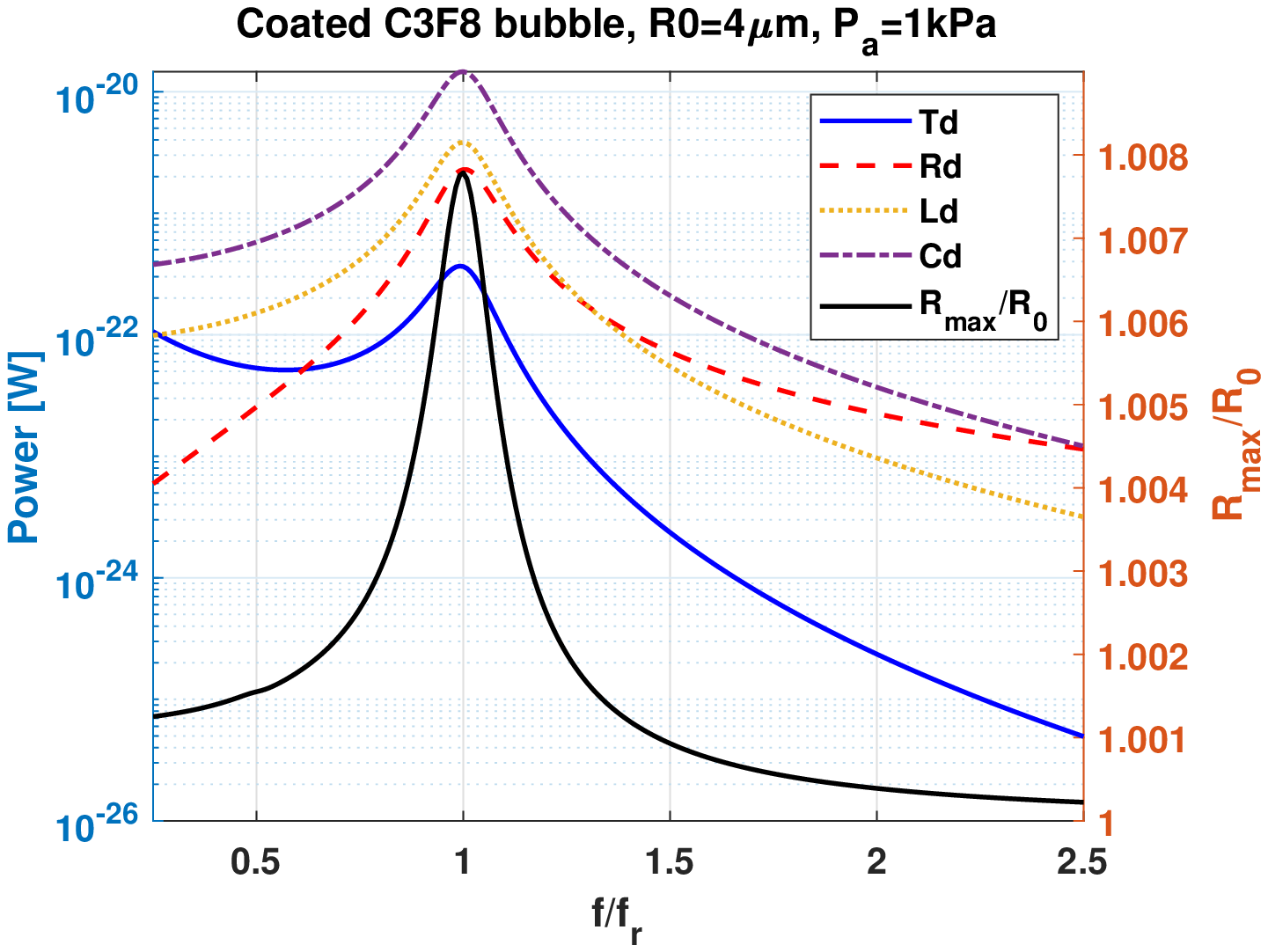}}\\
		\hspace{0.5cm} (a) \hspace{6cm} (b)\\
		\scalebox{0.43}{\includegraphics{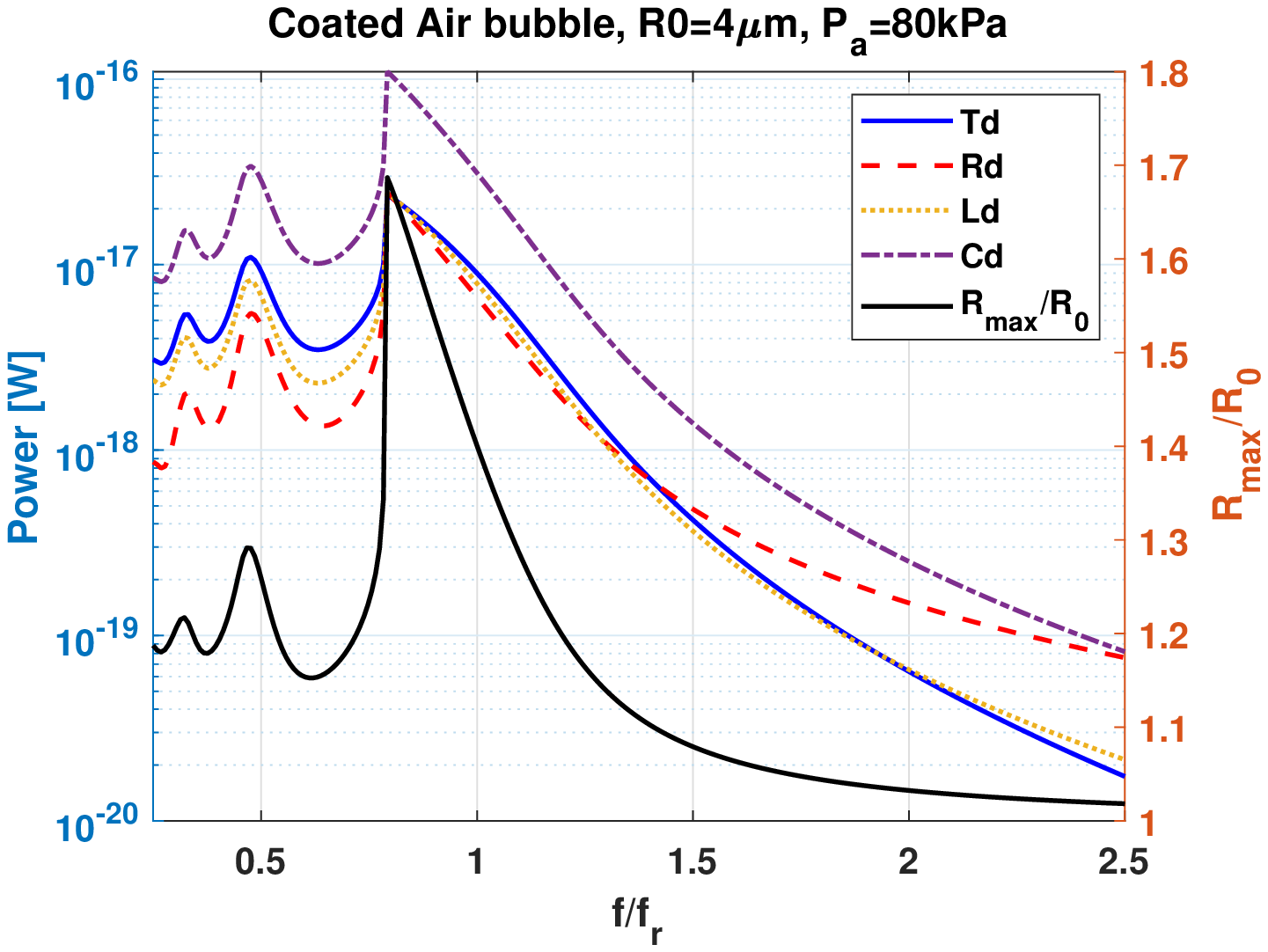}} \scalebox{0.43}{\includegraphics{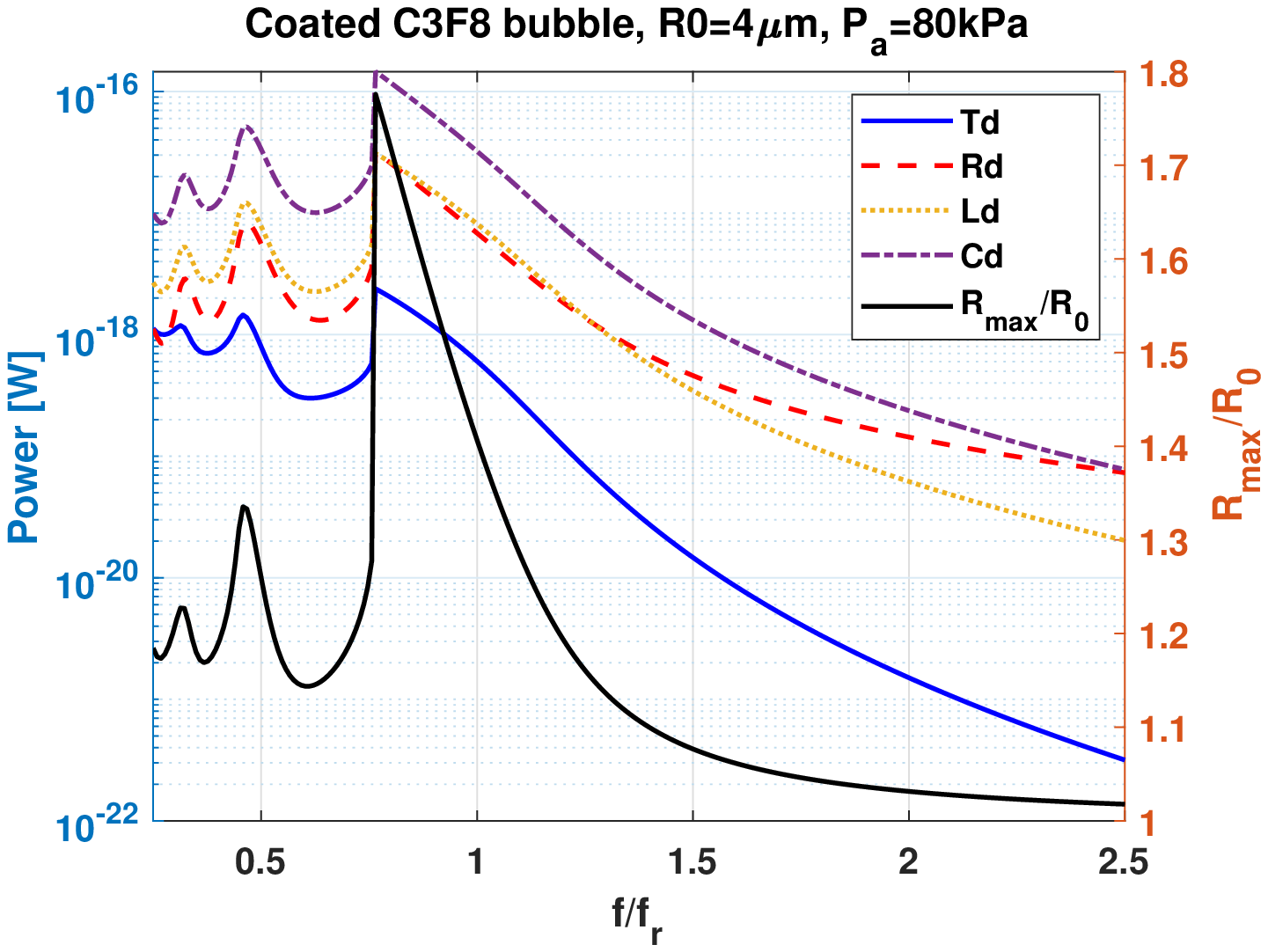}}\\
		\hspace{0.5cm} (c) \hspace{6cm} (b)\\
		\scalebox{0.43}{\includegraphics{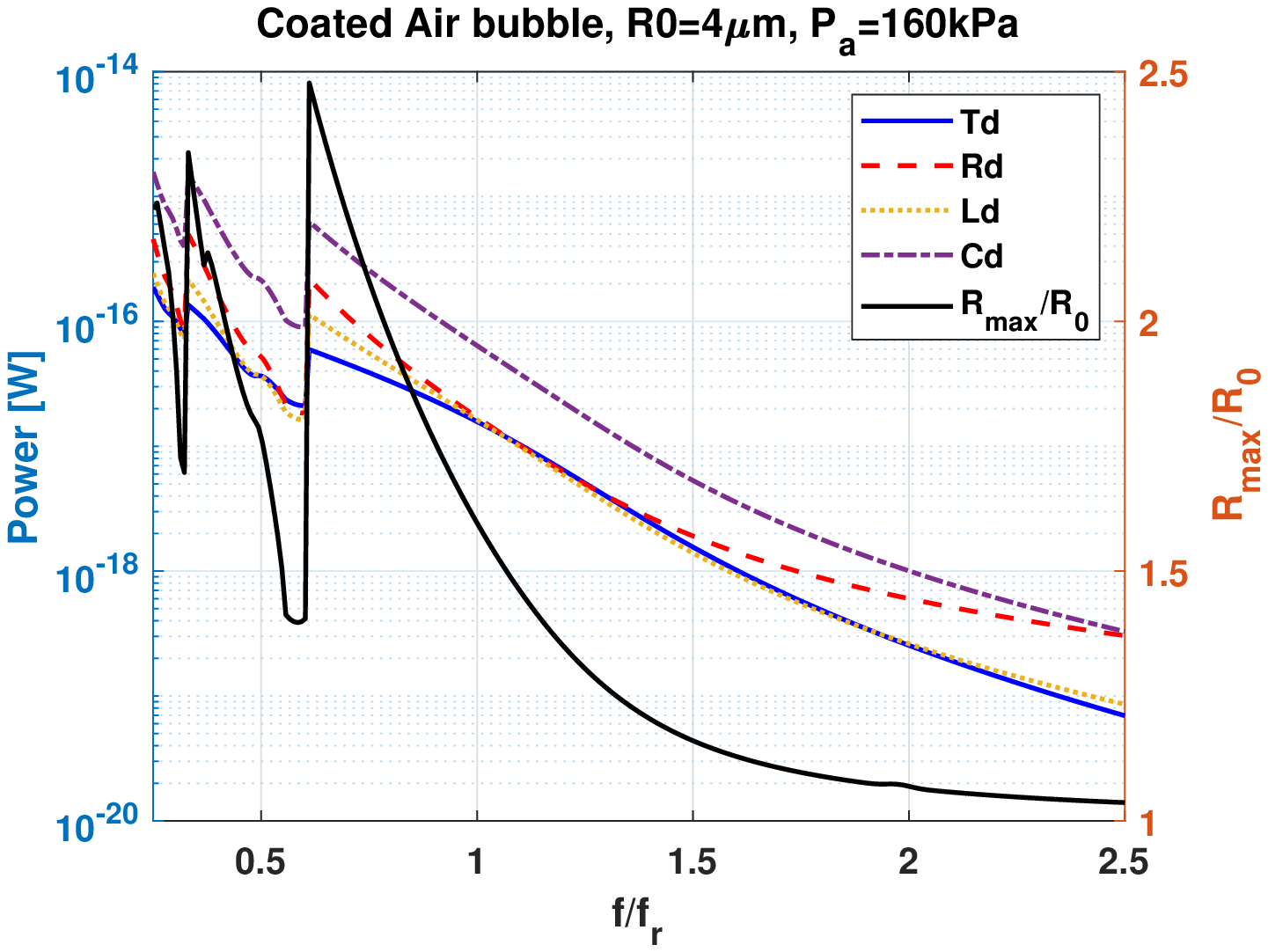}} \scalebox{0.43}{\includegraphics{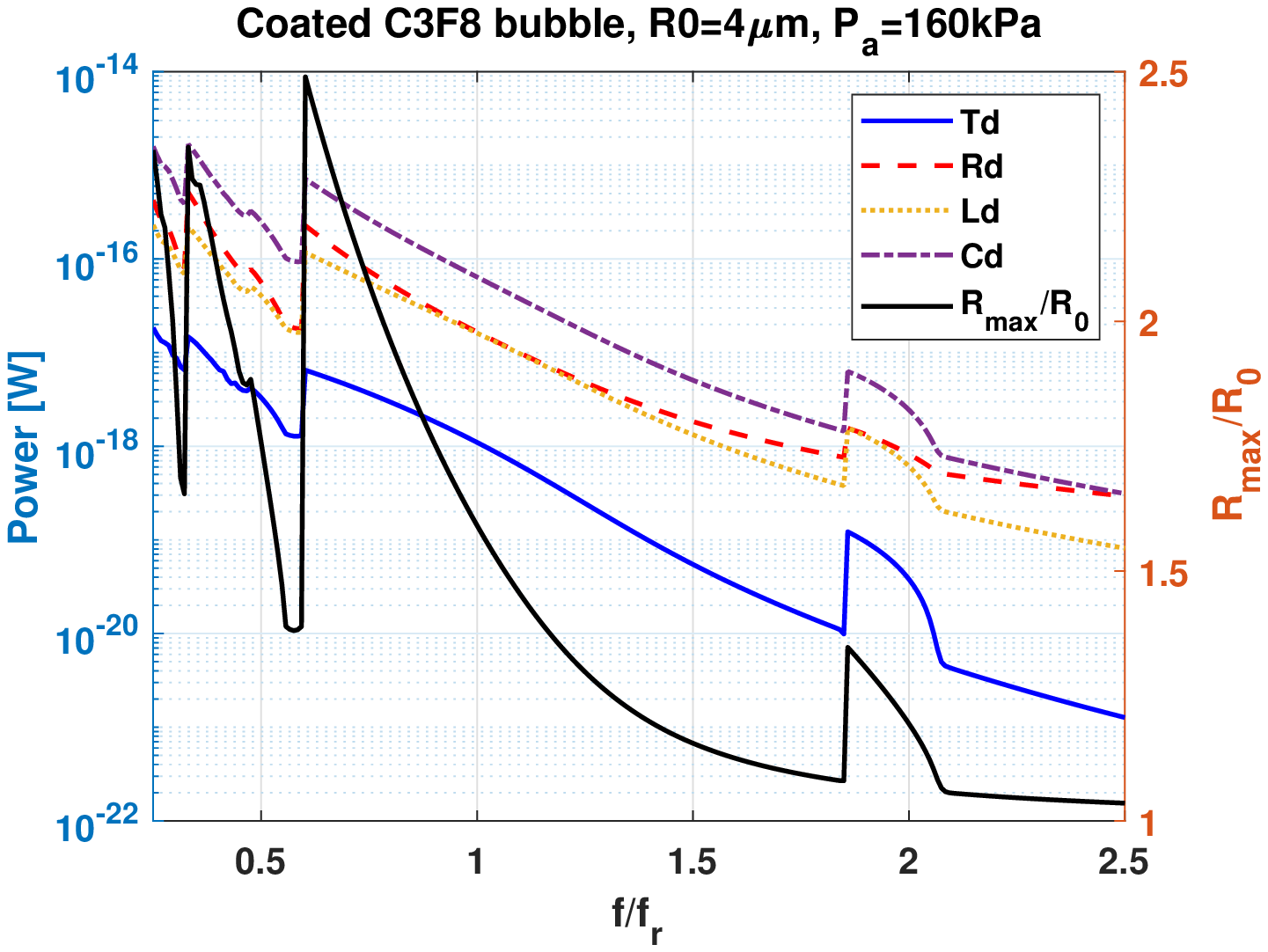}}\\
		\hspace{0.5cm} (e) \hspace{6cm} (f)\\
		\scalebox{0.43}{\includegraphics{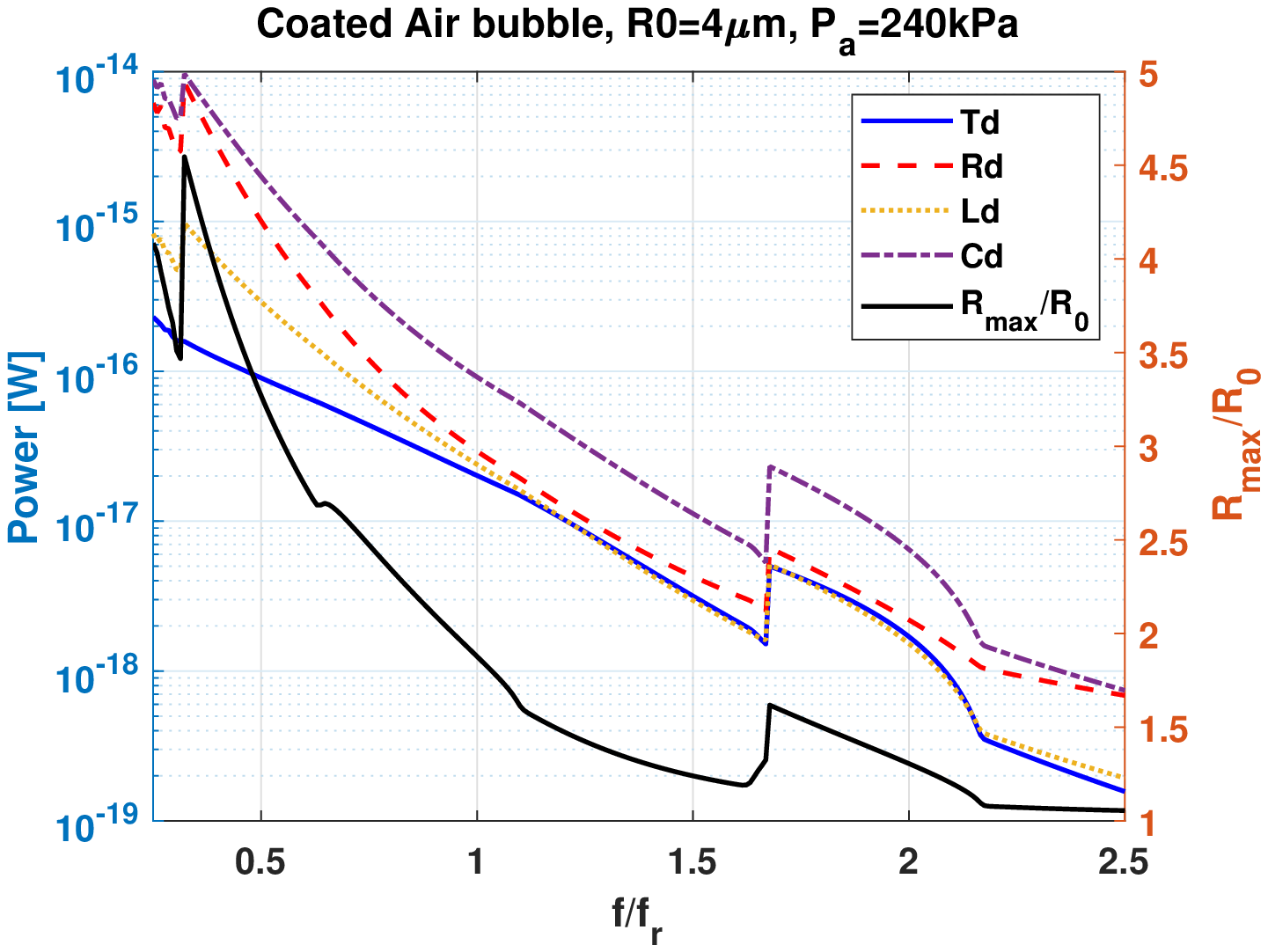}} \scalebox{0.43}{\includegraphics{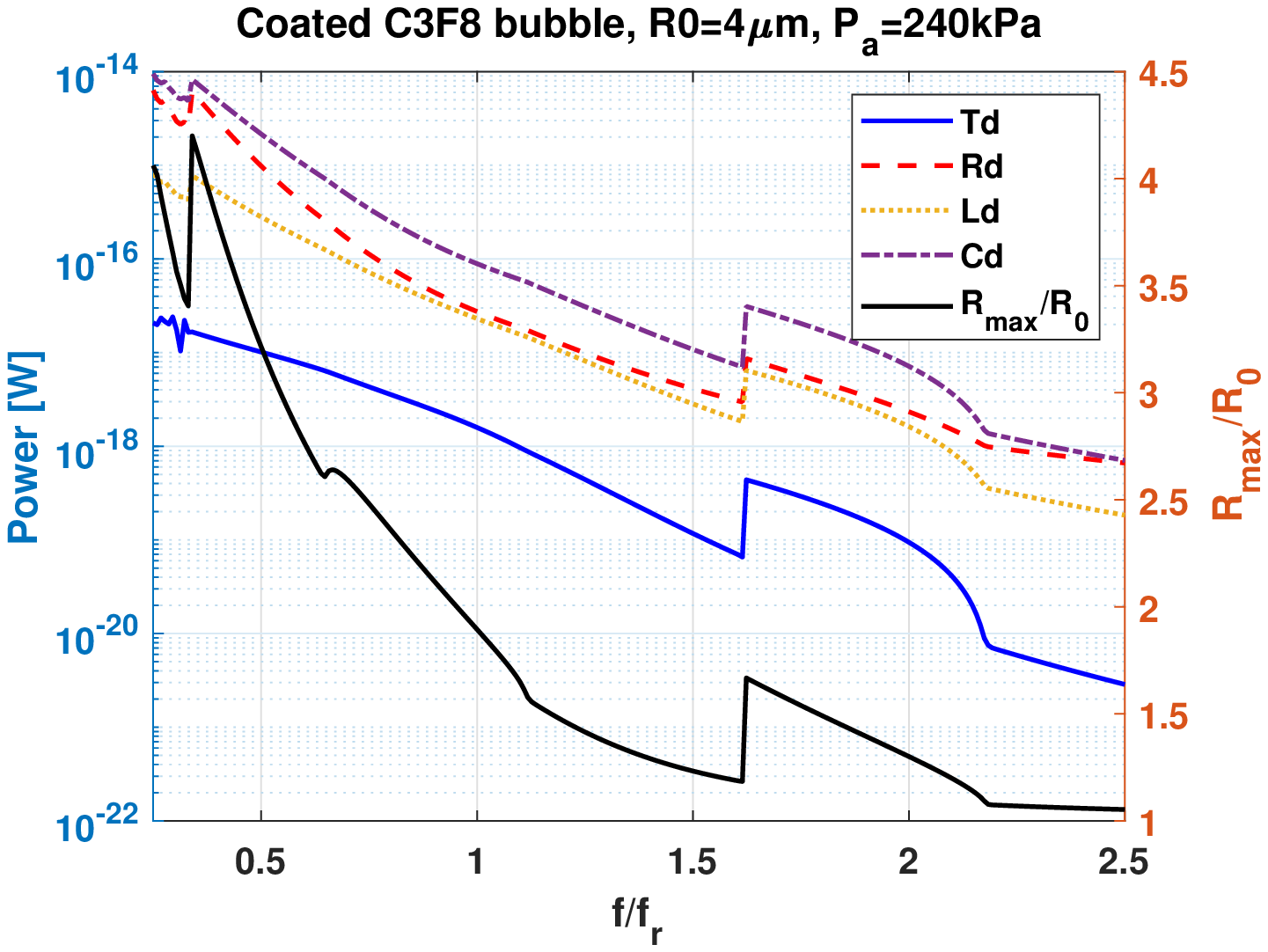}}\\
		\hspace{0.5cm} (g) \hspace{6cm} (h)\\
		\caption{Dissipated power due to Cd, Td, Ld and Rd as predicted by the full thermal model as a function of frequency for a coated bubble with $R_0$= 4 $\mu$ m at various pressures (left column is a free Air bubble and right column is a C3F8 coated).}
	\end{center}
\end{figure*}

\begin{figure*}
	\begin{center}
		\scalebox{0.43}{\includegraphics{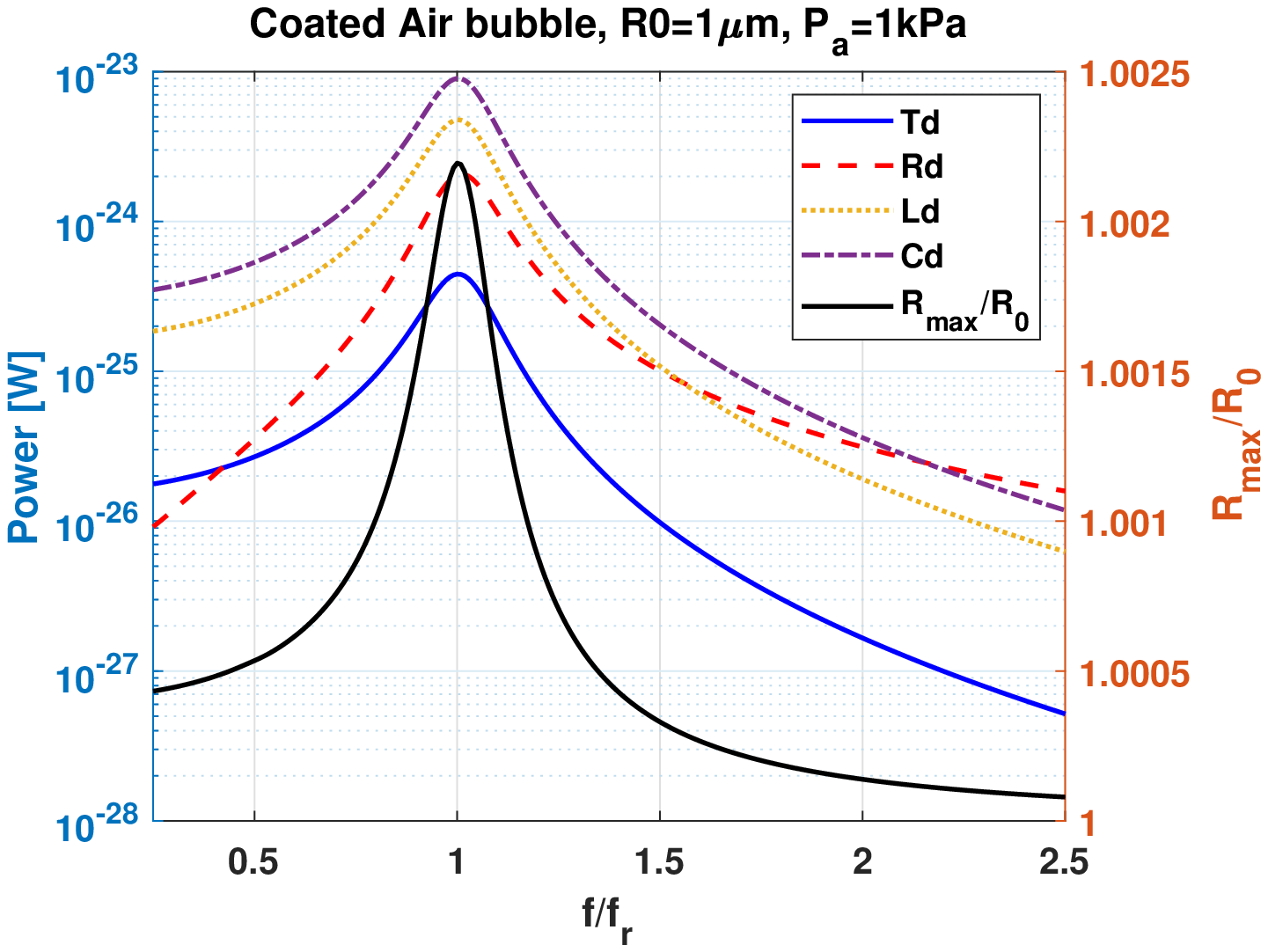}} \scalebox{0.43}{\includegraphics{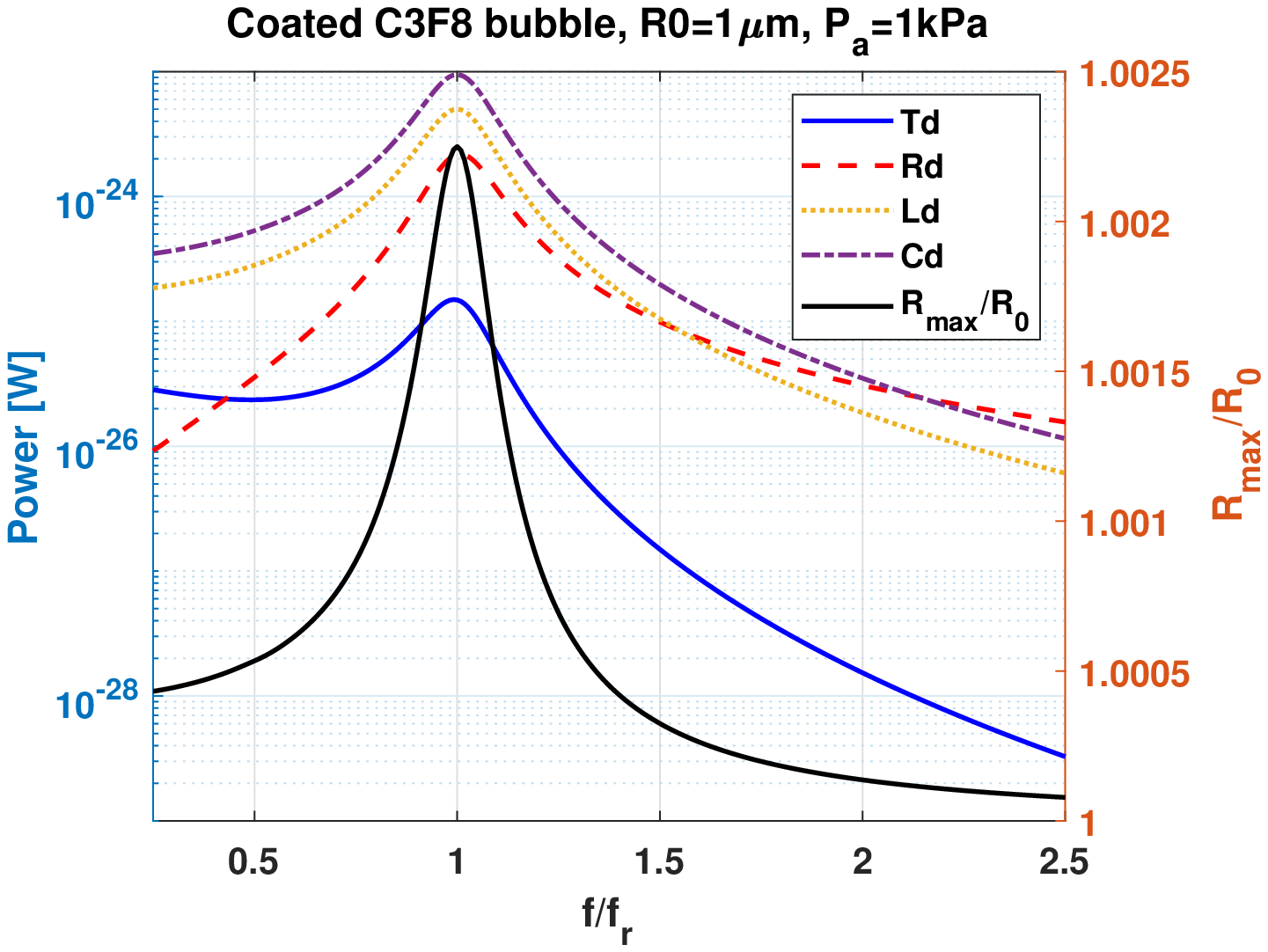}}\\
		\hspace{0.5cm} (a) \hspace{6cm} (b)\\
		\scalebox{0.43}{\includegraphics{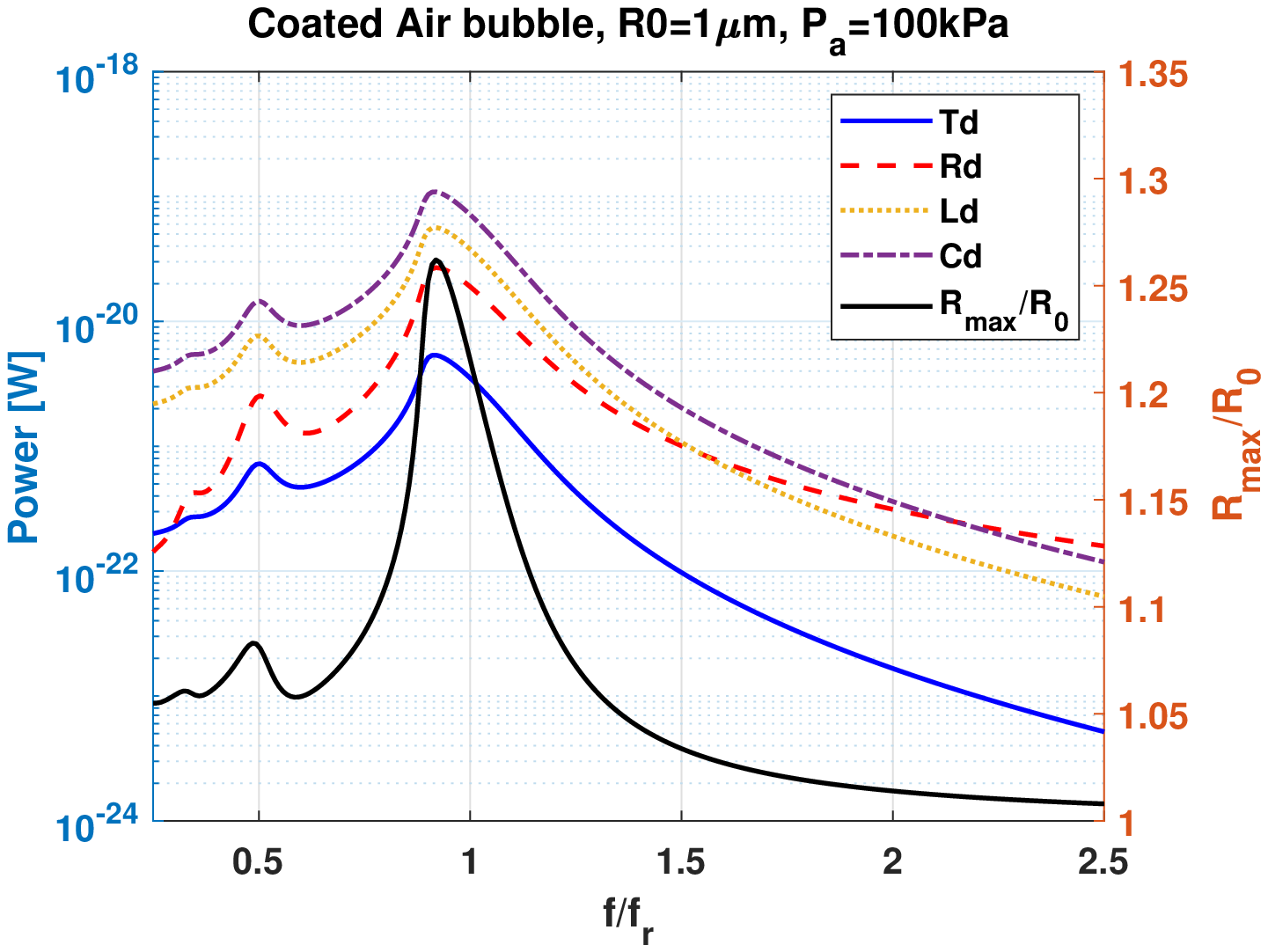}} \scalebox{0.43}{\includegraphics{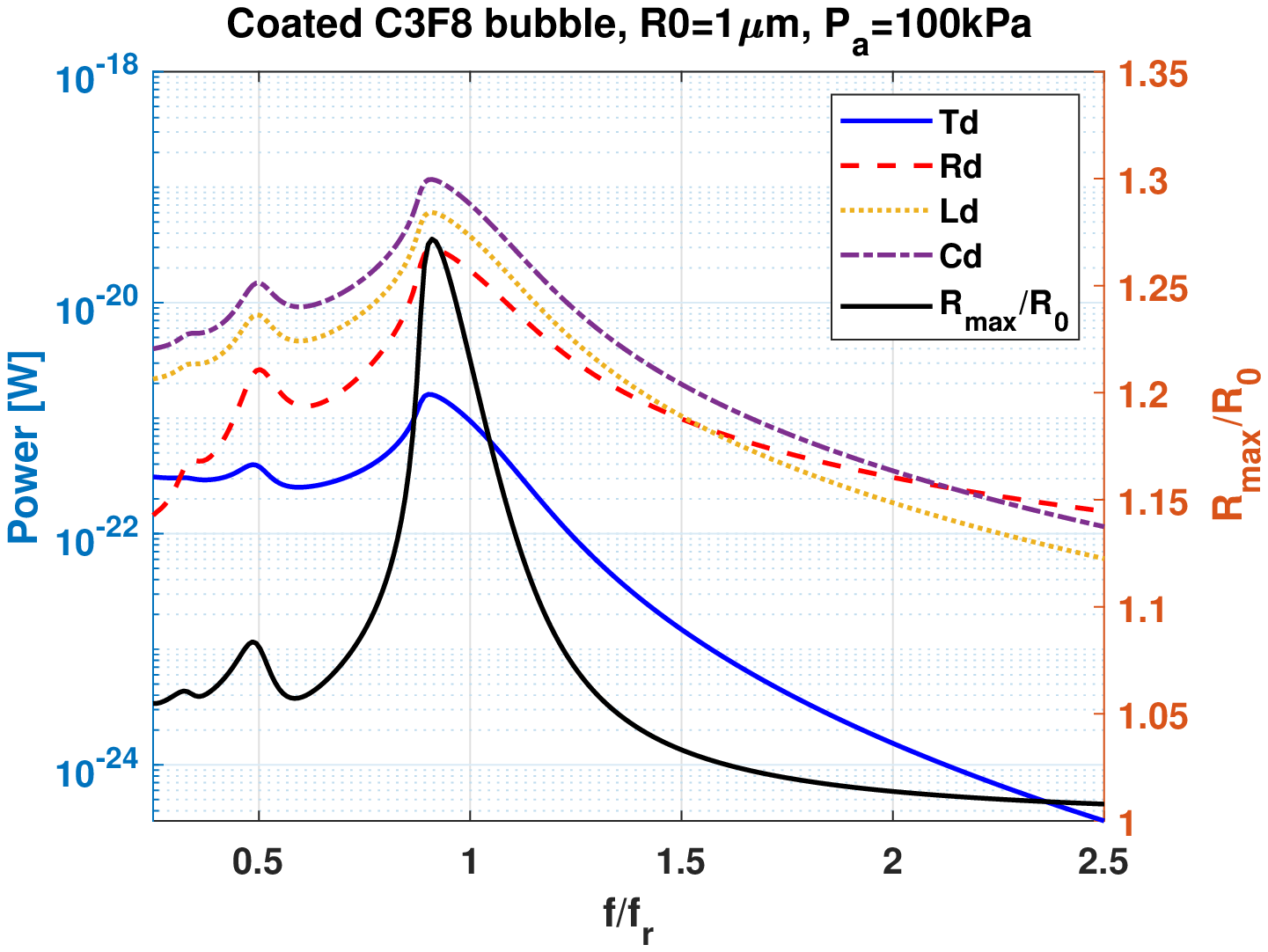}}\\
		\hspace{0.5cm} (c) \hspace{6cm} (d)\\
		\scalebox{0.43}{\includegraphics{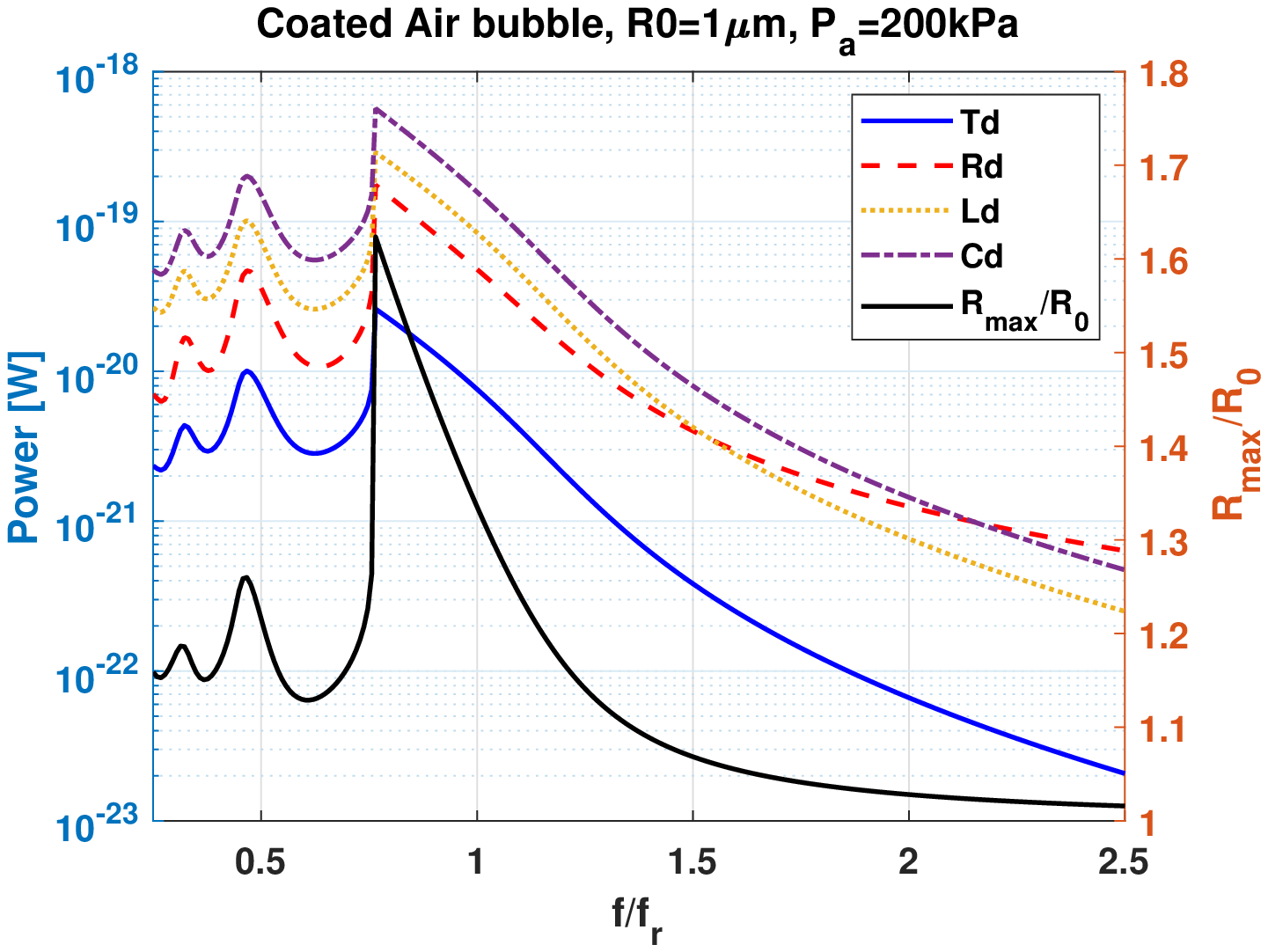}} \scalebox{0.43}{\includegraphics{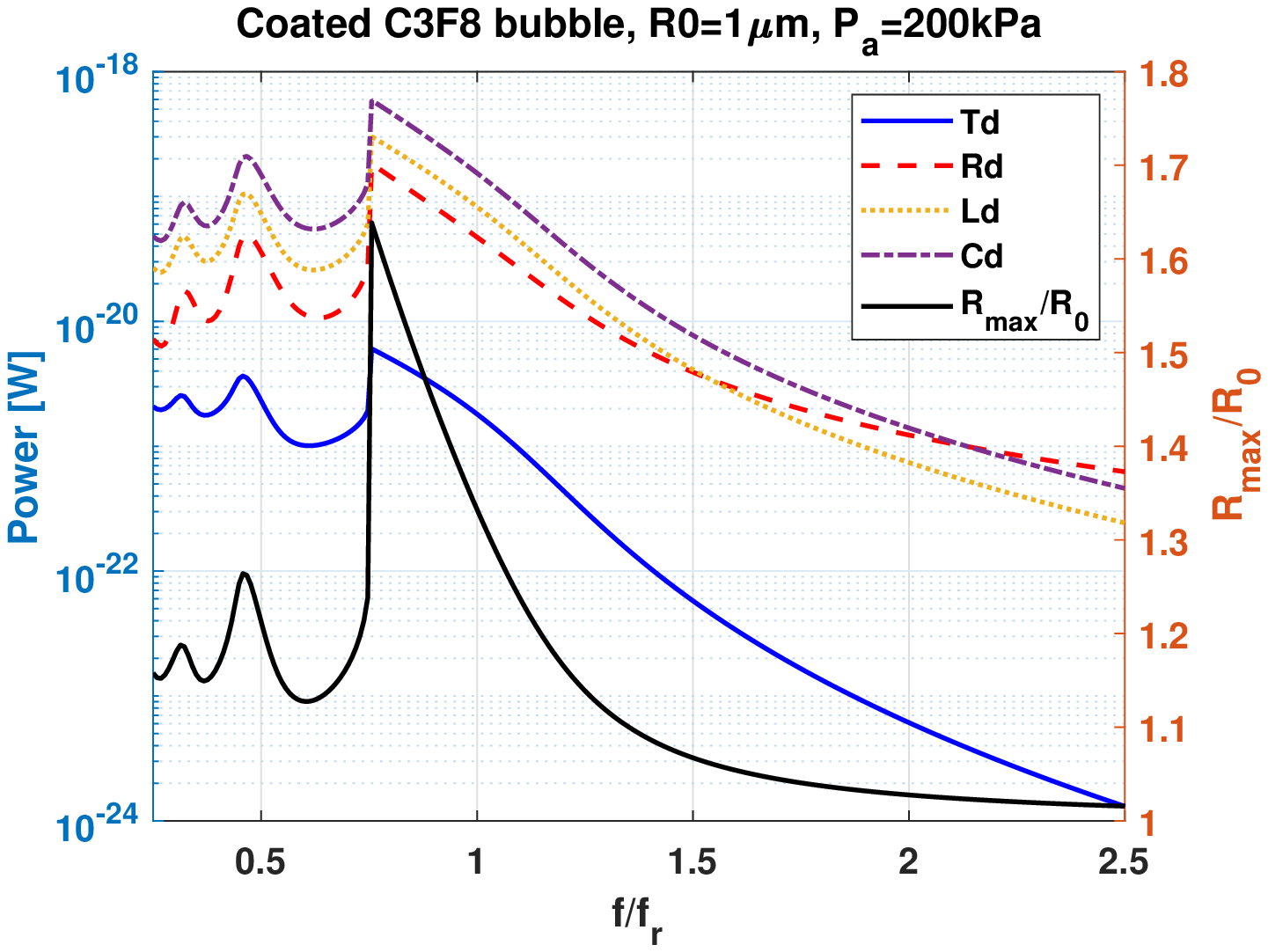}}\\
		\hspace{0.5cm} (e) \hspace{6cm} (f)\\
		\scalebox{0.43}{\includegraphics{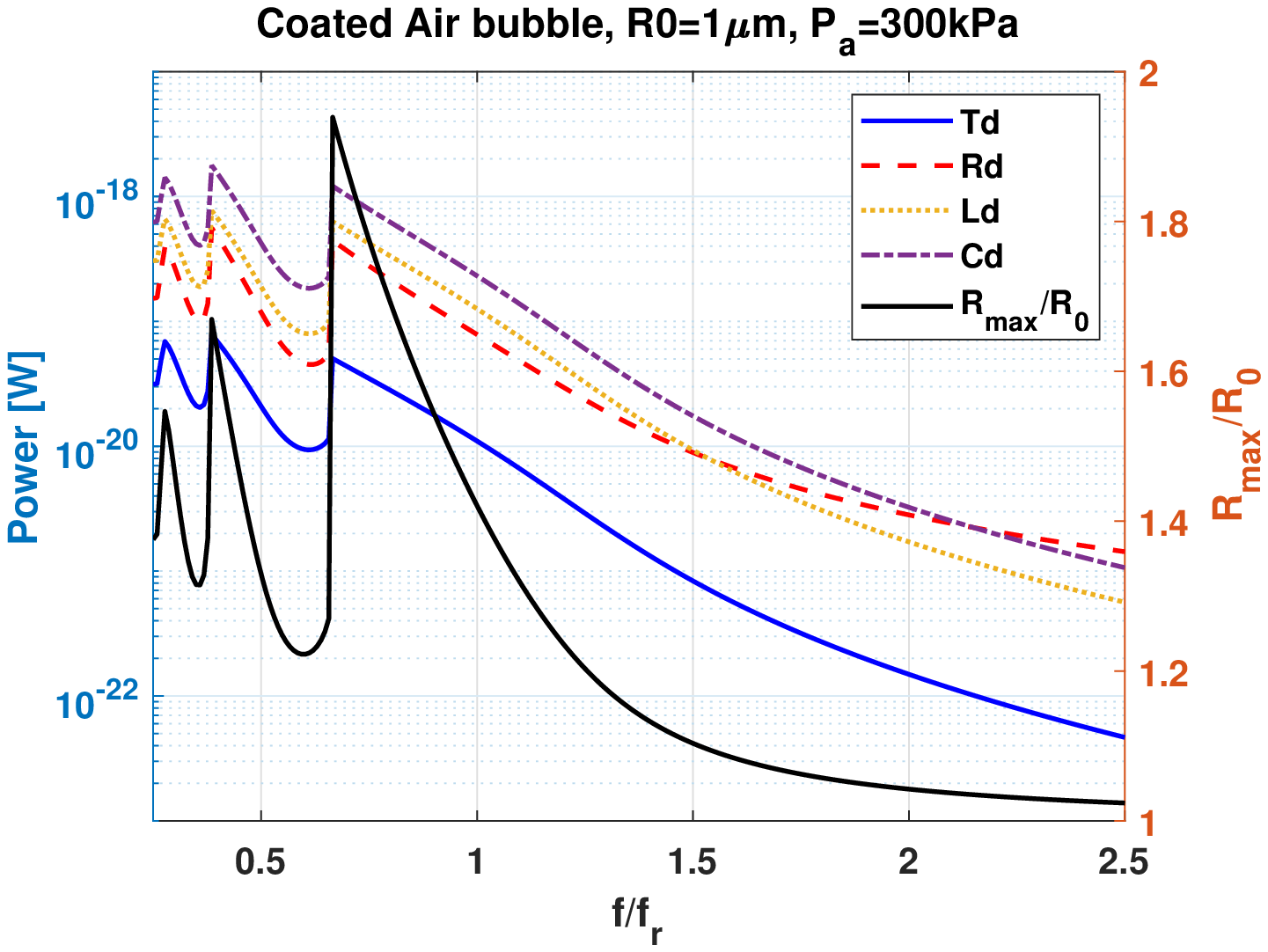}} \scalebox{0.43}{\includegraphics{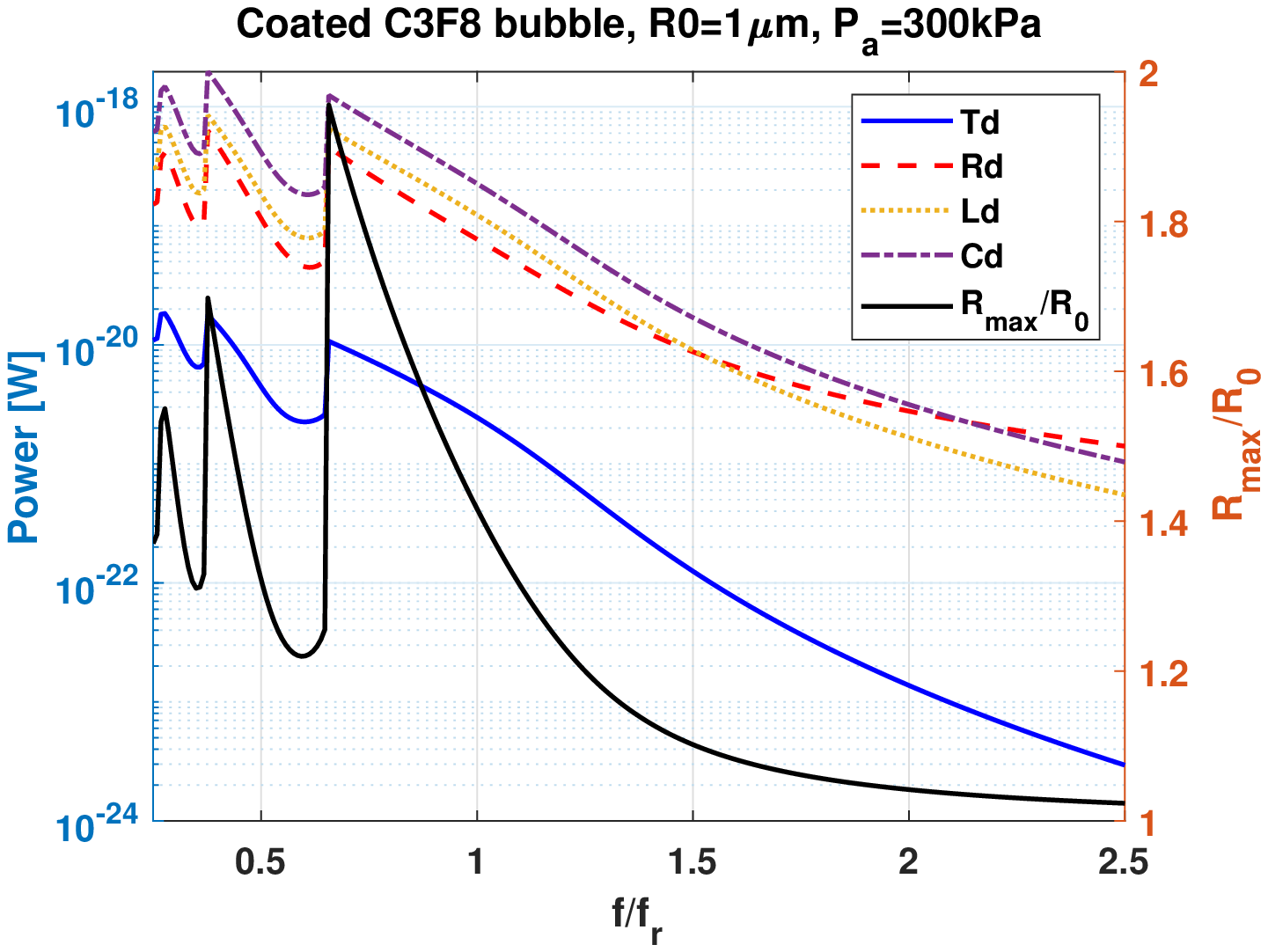}}\\
		\hspace{0.5cm} (g) \hspace{6cm} (h)\\
		\caption{Dissipated power due to Cd, Td, Ld and Rd as predicted by the full thermal model as a function of frequency for a coated bubble with $R_0$= 1 $\mu$ m at various pressures (left column is a free Air bubble and right column is a C3F8 coated).}
	\end{center}
\end{figure*}
\subsubsection{The coated (encapsulated) bubble with $R_0=1 \mu m$}
Fig. 6 shows the power dissipated due to Cd, Ld, Rd and Td for a coated bubble with $R_0=1 \mu m$, $G_s$=45 MPa, $\theta$=4 nm and $\mu_{sh}=\frac{1.49(R_0(\mu m)-0.86)}{\theta (nm)}$. The right column represents the air gas core bubble and the left column represents the C3F8 gas core bubble. Compared to Fig. 5, Td is further suppressed as the smaller bubble has smaller surface area for temperature exchange. At $P_a=1 kPa$ (Fig. 6a-b), we see a similar behavior for the Air and C3F8 gas core. $\frac{R_{max}}{R_0}$, Cd, Ld and Rd are similar for both cases;this is because the bubble with $R_0= 1\mu m$ Td is negligible and change of gas doesn't make a big difference in the oscillation amplitudes or the dissipated powers. $Cd>Ld>Rd>Td$ with Cd  $\approx$ 20 and 62 times larger than Td for the air and C3F8 bubble respectively.\\
Increasing the pressure to 100 kPa (Fig. 6c-d) results in the generation of 2nd SuH frequency with $Cd>Ld>Rd>Td$ for $0.2f_r<f<1.5f_r$. At higher frequencies ($f>2f_r$) Rd becomes the strongest damping factor.\\
When $P_a$=200 kPa a shift in the fundamental frequency occurs (f=$f_r$ at 1kPa) to $PDf_r$ (0.756$f_r$ and 0.76$f_r$ for air (Fig. 6e) and C3F8 (Fig. 6f) bubbles, respectively). Furthermore, a 3rd SuH resonance also appears below the 2nd SuH resonance frequency.  For $f<1.5f_r$ $Cd>Ld>Rd>Td$ with Rd becoming the major damping factor at $f>2f_r$. At this pressure Cd is 22.5 and 100 times larger than Td for Air and C3F8 respectively.\\ 
Increasing the pressure to  300 $kP_a$ (Figs 6g-h) leads to a higher $\frac{R_{max}}{R_0}$ at 3rd SuH, 2nd SuH and $PDf_r$ and a decrease in the value of the resonances of the system. For example at 300 kPa $PDf_r$ is $\approx$ $0.65f_r$ and $0.66f_r$ respectively for air and C3F8. For $f<1.5f_r$, Cd is the major damping factor with $Cd>Ld>Rd>Td$.  As the pressure increases, Rd grows faster and approaches Ld and will eventually have a greater contribution to the total damping compared to Ld. Td on the other hand has the slowest growth with $P_a$ increase at the main resonance. At $P_a$ =300 kPa unlike at the lower pressures, the highest dissipation occurs at 2nd SuH resonance and not at the main resonance.\\ 
\subsubsection{Scattering to damping ratio (STDR)}
Bubble scattered pressure ($P_{sc}$) can be measured in real time alongside with attenuation. $P_{sc}$ is dependent on $R(t)$, $\dot{R(t)}$ and $\ddot{R(t)}$; thus there is a direct relation between the enhanced bubble activity and increased $P_{sc}$.  Power loss due to radiation ($P_{sc}$) can be calculated using Eq. 25 or 27 and is related to the scattered pressure measured in real time in applications. Total damping can be calculated using equations 25 and 27 and attenuation which is directly related to damping \cite{34,35,37} can be calculated using the total damping \cite{34,35,37}.  Attenuation can also be measured in real time during applications. The scattering to damping ratio ($\frac{Rd}{W_{total}}$ where $W_{total}$ is the total dissipated power due to bubble oscillations) can be defined as a dimensionless parameter that can be used to assess the ratio of the energy re-radiated by the bubble as a fraction of the total attenuation. For example in applications like bubble enhanced heating in high intensity focused ultrasound (HIFU), the presence of the pre-focal bubbles can limit the energy that can reach the bubbles in the target region; moreover, additionally enhanced activity of bubbles distributed pre-focally can lead to undesirable heating in healthy tissue. A successful treatment will limit the bubble activity and attenuation in the pre-focal region while maximizing activity at the target for enhanced heating. In this instance, the STDR can be a good parameter to asses and optimize the sonication parameters.\\
We have presented the nonlinear energy loss formulas in the previous sections and showed that dissipation due to Cd, Ld, Rd and Td are pressure dependent. We showed that as pressure increases, Rd can grow faster than other damp- ing factors and can become the dominant factor to total damping. This increased STDR would benefit applications in which greater scattering is desired. In this section we investigate the pressure dependence of STDR and present the regions for which STDR can be enhanced.\\
 \begin{figure*}
 	\begin{center}
 		\scalebox{0.43}{\includegraphics{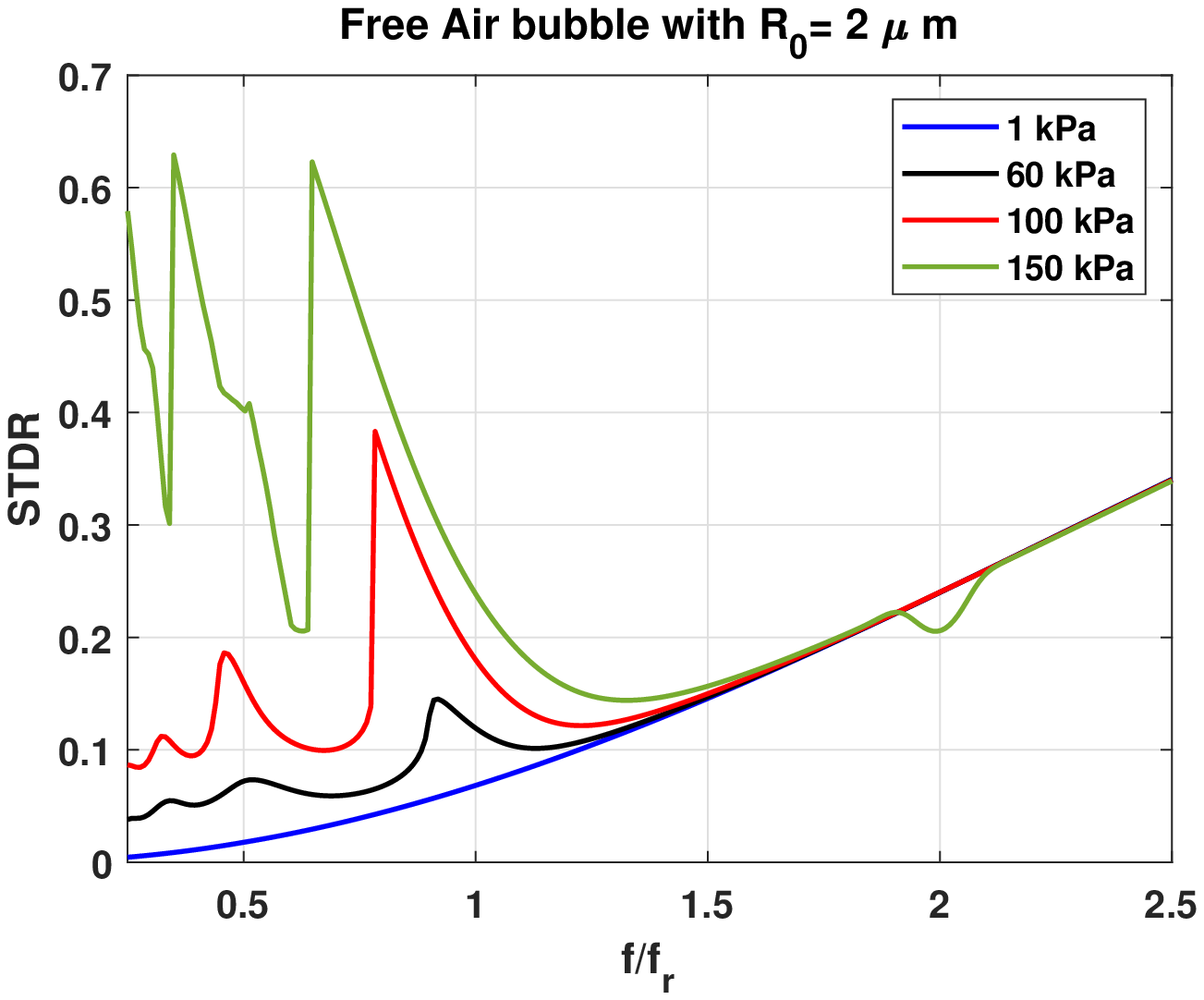}} \scalebox{0.43}{\includegraphics{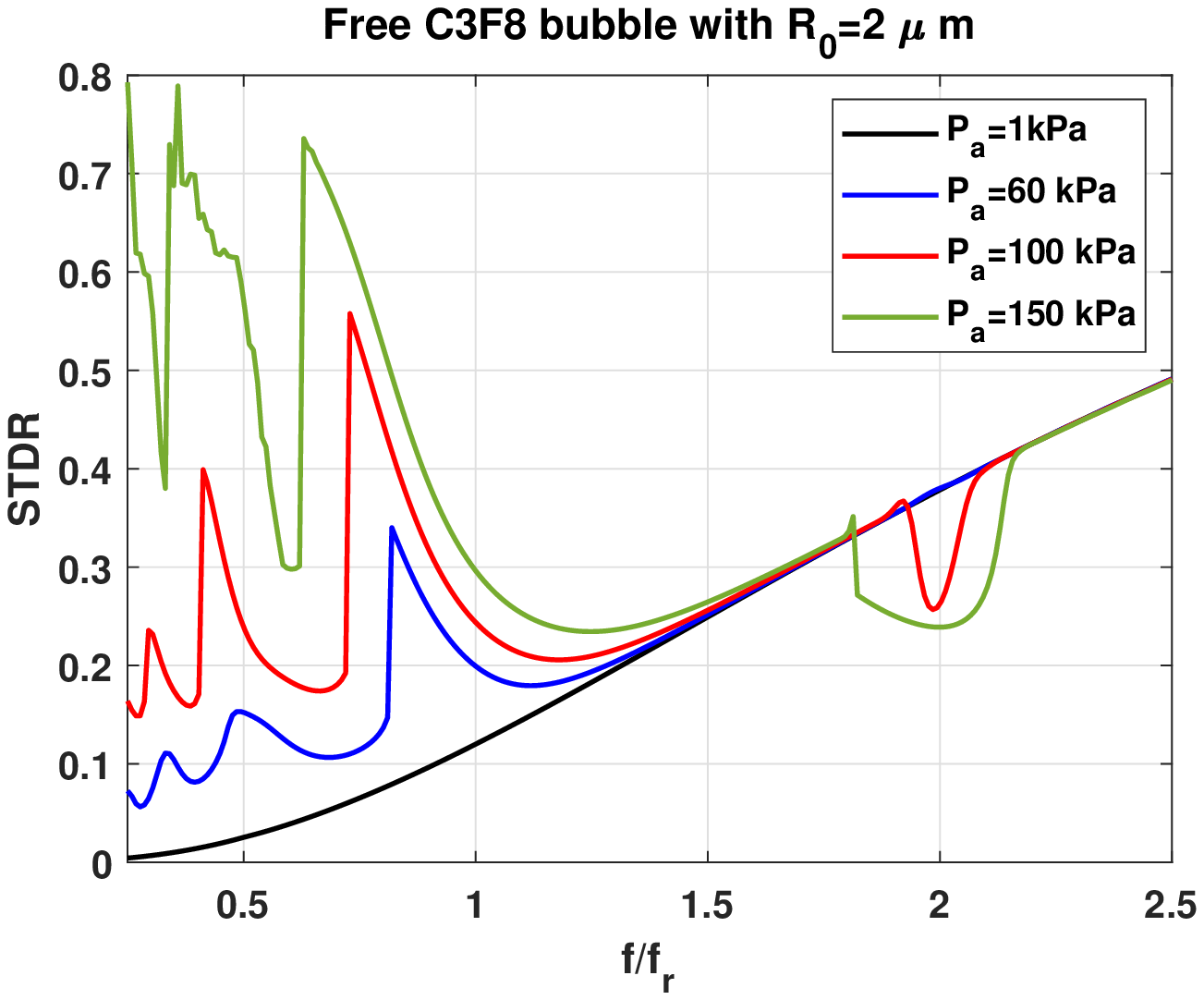}}\\
 		\hspace{0.5cm} (a) \hspace{6cm} (b)\\
 		\scalebox{0.43}{\includegraphics{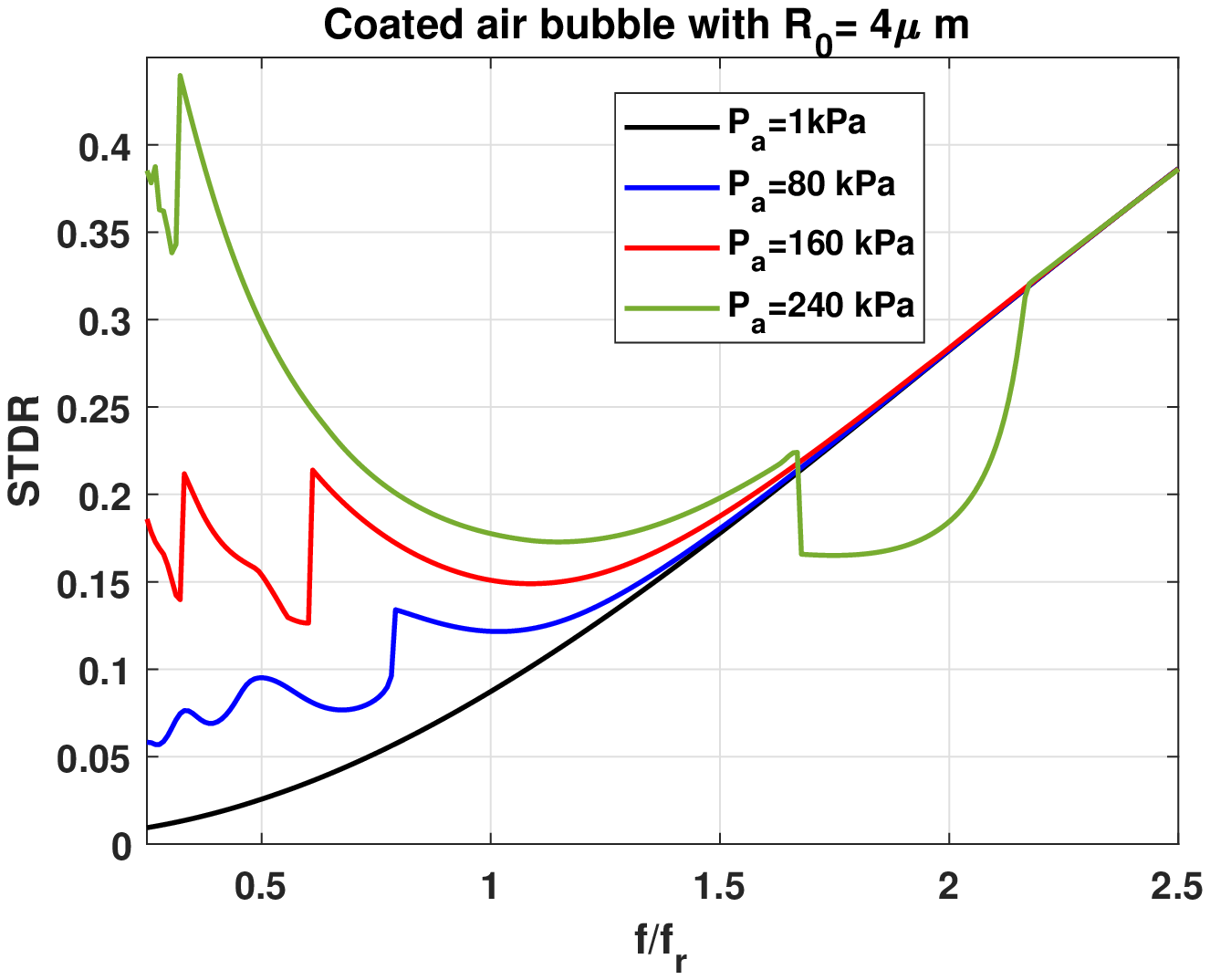}} \scalebox{0.43}{\includegraphics{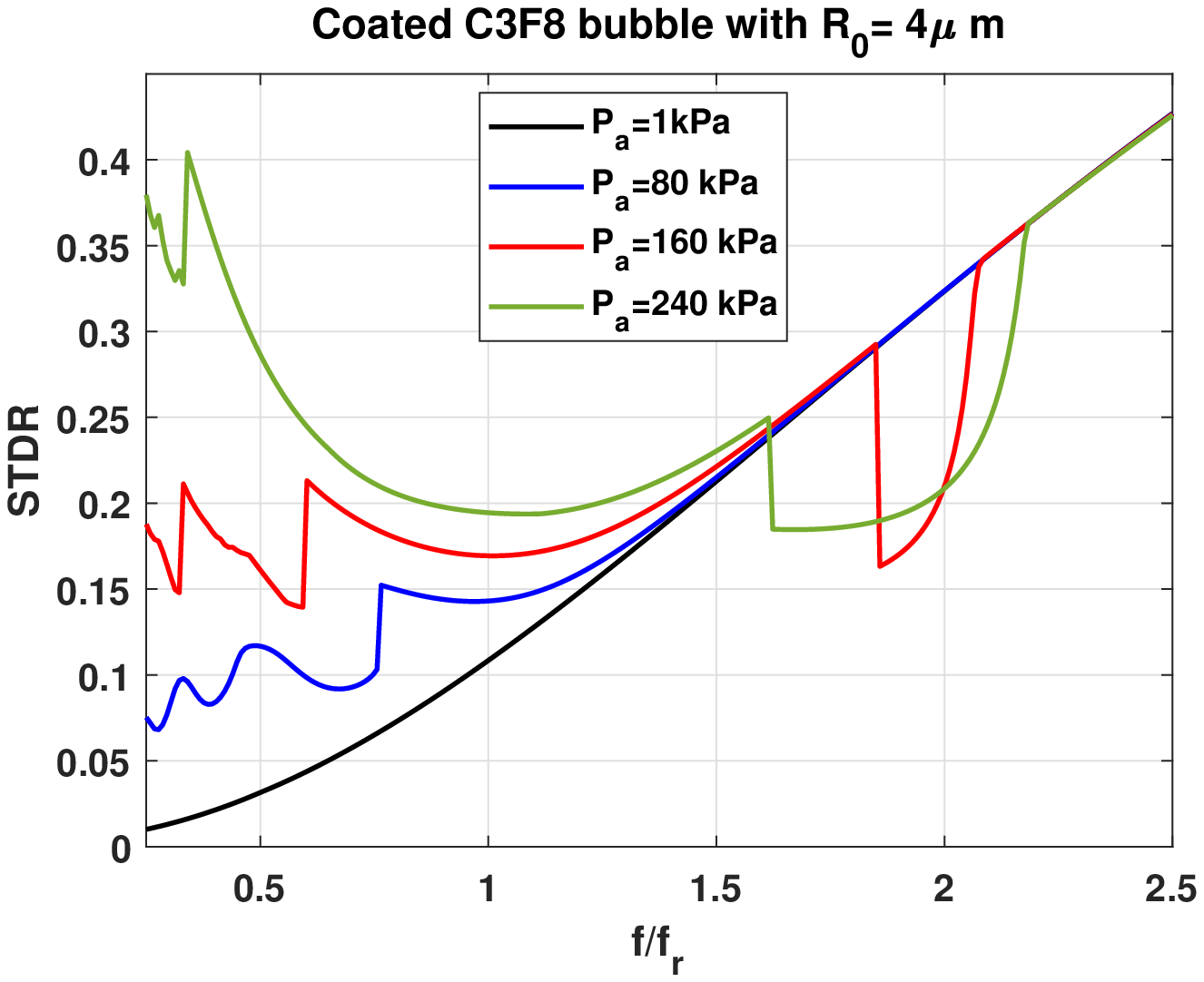}}\\
 		\hspace{0.5cm} (c) \hspace{6cm} (d)\\
 		\caption{Scattering to damping ratio (STDR) as a function of frequency at various excitation pressures. Top row is for an uncoated bubble with $R_0= 2 \mu m$ and bottom row is for a coated bubble with $R_0= 4 \mu m$. (left column is a free Air bubble and right column is a C3F8 coated ).}
 	\end{center}
 \end{figure*}
Fig. 7a shows the STDR of an air uncoated bubble with $R_0= 2 \mu m$ as a function of frequency at various pressures. At 1 kPa, the STDR diagram doesn't show any distinct peak; STDR grows as a function of frequency. At higher frequencies the bubble oscillations are very weak; thus Ld and Td are negligible compared to active scattering by Rd. This is because even for weak bubble oscillations, the bubble re-radiates and scatters sound waves. However, the very weak oscillation leads to near zero wall velocities thus Ld becomes very small. Similarly, due to the near zero temperature elevation inside bubble, Td becomes very small as well. Thus, the STDR is higher at higher frequencies and lower excitation pressures. An increase in pressure leads to increase in the STDR. The STDR is very high at $PDf_r$ \cite{24} (e.g. 0.4 at 100 kPa). As pressure increases, the STDR at SuH resonance frequencies and $PDf_r$ increases. This is due the faster growth of Rd as incident pressure increases when compared to other damping factors. At 150 kPa and concomitant with the generation of $\frac{1}{2}$ SH resonance \cite{60} the peak STDR decreases. This is because when the bubble oscillation amplitude increases, Ld and Td become significant, decreasing the STDR. We have shown the same effect for the case of sonication with pressure dependent SH resonance \cite{61} when $\frac{1}{3}$ \cite{62,63} order SH resonance occurs \cite{36}. For the investigated bubble when  $f>1.5f_r$, the STDR does not change as pressure increases, unless SH oscillations appear (in which case STDR decrease by pressure increase).\\
Fig 7b shows the STDR of a C3F8 uncoated bubble with $R_0= 2 \mu m$ as a function of frequency at various pressures. The evolution of STDR with pressure is similar to that of the air bubble; however, SH oscillations occur at lower pressures than the Air gas core due to smaller effects of Td. Furthermore, the STDRs at $PDf_r$ and SuH resonances are stronger than those for the air bubble due to weaker contributions from Td. Thus, for uncoated bubbles one may get larger STDRs if a gas like C3F8 is used.\\
Fig 7c shows the STDR as a function of frequency of a coated air bubble with $R_0= 4 \mu m$. The evolution of the STDR with pressure increase is very similar to the uncoated bubble; however, the STDR is considerably reduced compared to the uncoated counterpart. This is because of the increased damping due to coating friction. In Figs.5 and 6 we showed that Cd is the major damping factor for the coated bubble. This is similar to the bubble with a C3F8 gas core (Fig. 7d). The values of STDR are very close to the ones of the air gas core for $f<1.5 f_r$. This is because Cd is the major contributor to total damping and changes in Td do not affect the $W_{total}$ significantly. At higher frequencies however, the C3F8 bubble exhibit SH oscillations at a lower pressure.         

\section{Discussion}
Bubbles are the building blocks of several applications from material science and sonochemsitry \cite{6,7,8,9,10} to oceanography \cite{4,5} and medicine \cite{11,12,13}. Dynamics of bubbles are nonlinear and complex \cite{15,17,18,19,20} and to achieve their full potential in applications we need to have detailed understanding about the bubble response to exposure parameters of the acoustic field. The presence of bubble however, changes the acoustic properties of the medium \cite{31,32,33,34,35,36,37}. Changes in the acoustic properties are nonlinear and is a function of bubbles size, excitation frequency and pressure \cite{34,35,36,37}. Because of the changes in the acoustic properties of the medium (e.g. increased pressure dependent attenuation \cite{34,37,43,45}), knowledge of just the nonlinear bubble behavior is not enough to control and optimize applications. One must have sufficient knowledge on the relationship between the nonlinear bubble oscillations and changes in the acoustic properties of the medium. For example, increased attenuation from bubbles in the beam path can limit the energy that the bubbles at the target are exposed to \cite{34,45} and decrease the efficacy of sonochemical applications \cite{34,35,37}, drug delivery \cite{65} and enhanced heating in HIFU \cite{44}. Increased attenuation of bubbles can also create shielding of the post-focal tissue and bubbles in contrast enhanced diagnostic ultrasound and deteriorate the ultrasound images \cite{65,66}. In bubble characterization applications like oceanography studies \cite{4,5} or characterization of ultrasound contrast agents \cite{38,39,40,41,42,43} pressure dependent changes in the attenuation are of significant importance and should be understood in detail.\\    
In this study we present a simple model for coated bubbles that accounts for the compressibility effects up to the first order of Mach number. The model is called CHKM model as it is a hybrid of Church-Hoff model \cite{38} and Keller-Miksis (KM) model \cite{47}. The goal of incorporating the compressibility effects was to investigate the dissipation of acoustic energy  due to re-radiation (Rd) effects \cite{35,36,37}. Rd is a important pressure dependent dissipation mechanism \cite{35,36,37}; however, it is simplified \cite{48} or fully neglected \cite{38} in models that are used to study the coated bubbles behavior. To model the oscillations of coated bubbles and uncoated bubbles CHKM and KM models were receptively solved; in each case three forms were considered for the gas pressure inside the bubble. First, CHKM was coupled to nonlinear ordinary differential equations (ODEs) that model the thermal behavior of gas bubbles \cite{49,51,52,53}.  Second, thermal effects were incorporated using linear approximations that include the thermal effects \cite{33,53}. The linear thermal model has been widely used in studies related to coated bubble characterization and characterization of bubbles in oceanography \cite{4,5,39,40}. Third, thermal effects were fully neglected.\\
Using our previous approach \cite{36}; we derived the nonlinear terms for dissipation of energy from thermal effects (Td), viscosity of coating (Cd), liquid viscosity (Ld) and damping due to re-radiation (backscatered pressure $P_{sc}$) by bubbles.\\
In order to investigate the effect of thermal damping on bubble oscillations and choose the proper model for bubble oscillations, the total dissipated acoustic energy was modeled using the three models described above for both free and coated bubbles of different sizes. For each bubble size energy curves were displayed as a function of frequency at various pressures and for two gas cores of Air and C3F8.\\ Results showed that in case of uncoated bubbles thermal effects are significant and cannot be neglected. Furthermore, linear approximations are only valid for small excitation pressures  (e.g. 1 kPa). For higher pressures (e.g. $>$ 10 kPa) predictions of the linear thermal model and the full thermal model deviate and full thermal model should be used to model bubble oscillations. For the moderate pressures used in this  study predictions of the full thermal model in ODE form have been shown to be in good agreement with the thermal model employing the full partial differential equations (PDES)  \cite{51}. Thus studies related to bubble characterization in oceanography \cite{4,5} and free bubble oscillations \cite{33,34,35,36,37} would benefit from considering the full ODEs describing the thermal effects.\\ 
In case of coated bubbles with Air gas core full thermal model is a more precise model to study the bubble behavior at higher pressures. Because of very high damping due to Cd and Ld, damping effects due to Td (for Air gas core) were negligible at frequencies $<1.5f_r$ even at higher pressures. This was the case even for the largest possible coated bubble that can be used in medical ultrasound ($R_0=4 \mu m$) although for larger bubbles Cd and Ld are smaller and Td is higher. For frequencies $>1.5f_r$ and for pressure and frequency ranges where SH oscillations occur (e.g. \cite{60,61}), Td becomes important and should be incorporated.
In case of C3F8 gas core, and for the pressure and frequency ranges that are used here, thermal effects can fully be neglected. The pressure ranges studied here (1kPa-100 kPa) are often used to characterize the shell parameters of coated bubbles in medical ultrasound. The previous studies \cite{39,67,68}; considered an artificial viscosity term and added it to the liquid viscosity to consider the thermal effects in the bubble oscillations. The viscosity is usually twice that of the water viscosity \cite{39,67,68}. Here , we show that such approximation leads to considerable overestimation of the Td especially in case of C3F8 or C4F10 bubbles. Thus fitted values for the viscosity of coating \cite{39,67,68} and pressure threshold of SH oscillations may be inaccurate.\\
In the second part of the study we investigated the dependence of Cd, Ld, Rd and Td as a function of frequency and at different pressures. We showed that the dissipated energy depends on size, frequency and pressure. In case of uncoated bubbles Td is very important for the larger Air bubbles (e.g. $10 \mu m$); however at higher pressures Rd can exceed Td. For C3F8 uncoated large bubbles Td is only the major damping factor for $f<f_r$ and lower pressures (e.g. $P_a<10 kPa$). Rd is the major damping factor at $f>1.5 f_r$ and for all the pressures studied here. For the uncoated C3F8 large bubble (e.g. $10 \mu m$) Rd becomes the major damping factor for $f<f_r$ (e.g. $PDf_r$ and super harmonic resonance ($SuHf_r$)) as pressure increases ($P_a>40 kPa$). For smaller uncoated Air and C3F8 bubbles ($R_0<2 \mu m$) Td is not anymore the major damping factor. Rd, Ld and Td grow as pressure increases; Rd exhibits the fastest growth rate and Td the slowest. At higher pressures (e.g. $P_a>100 kPa$) Rd and Ld become an order of magnitude higher than Td (for the studied frequency range here $0.25f_r<f<2.5f_r$).\\
For coated bubbles Cd is the major damping factor for the pressure and frequency range that was studied here except for $f>2f_r$ for the coated bubble with $R_0=1 \mu m$ where Rd is the most significant dissipation mechanism. Similar to the uncoated bubble, Rd grows faster than other damping factors with pressure increases while Td has the slowest growth rate. Because of the very strong effect of Cd on the damping, Td has less significant effect on the dissipation in case of the coated bubbles. Dissipation effects of Td become weaker for smaller bubbles and when the gas core is C3F8.\\
We showed that scattering to damping ratio (STDR) is pressure dependent; STDR grows with pressure at pressure dependent resonance frequencies ($PDf_r$ \cite{24}) and at super harmonic resonances of the system ($SuHf_r$ \cite{23}). STDR is generally higher at $f>f_r$. For the studied bubbles and frequency and pressure ranges here,  elevation in excitation pressure  does not change the value of STDR for $f>1.5 f_r$ only until the occurrence of SH resonance which leads to a decrease in STDR. For the stable non-destructive bubble oscillations $\frac{R_{max}}{R_0}$, STDR is higher than $PDf_r$ and $SuHf_r$ even when SH oscillations occur.\\
Coated bubbles are used as ultrasound contrast agents (UCAs) in ultrasound imaging \cite{11,12,13,40, 45, 65, 66, 67,68}  where higher scattering and small attenuation are desired. The scattering defines the ability of UCA to enhance the echogenicity of the target. However, the presence of bubbles increases the damping of the acoustic energy in the beam path and reduces the energy available for the bubbles to be activated at the focal point. Thus higher scattering ability of UCAs and lower total damping is desired. In this regard STDR expresses the ability of the UCA to enhance the visualization of the tissue containing the UCAs and the underlying tissues \cite{45,65}. In therapeutic applications like drug delivery shear stress resulted from micro-streaming due to bubble oscillations are used to increase the drug delivery to cells in the target \cite{64}. In drug delivery higher wall velocities and higher $P_{sc}$ are desired. Compared to imaging applications, in therapeutic applications higher concentration of bubbles are used to achieve desirable effects \cite{64}; thus damping effects are more significant \cite{13,44}. Higher STDR and $P_{sc}$ are good indicators of a faster bubble wall velocity and lower total damping. In High Intensity Focused Ultrasound the re-radiated energy by bubbles is used to enhance the heating in the target area \cite{44}; re-radiated pressure often have strong higher harmonics content which have higher absorption rate in tissue \cite{44}. This results in localized efficient enhanced heating \cite{44}. In these applications as well, higher $P_{sc}$ and lower energy damping due to pre-focal bubble oscillations are desired. Same conclusions can be made to active bubbles in sonochemistry and other applications. Decreasing the pre-focal power dissipation becomes more important in treatment of locations where delivery of energy is more difficult (e.g. presence of pre-focal bone in blood brain barrier opening \cite{69,70}).\\
In this study we provided the fundamental equations and approaches that can be used to analyze the energy dissipation for coated and uncoated bubbles. It should however be noted that STDR by itself is not a sufficient indicator for efficient bubble application. We have shown here and in \cite{36} that STDR is higher at frequencies above $f_r$, however, at higher frequencies bubbles are only active when pressure exceeds a threshold (e.g. above the pressure threshold of P3 or P4 oscillations \cite{62,63,71,72,73}). Thus, depending on the application \cite{36}, STDR must be used in tandem with $P_{sc}$ or $Rd$ to determine the regions of higher bubble activity and lower total damping. In this regard, we have introduced two parameters of PmSTDR and RdSTDR in \cite{36} that can be used to design the exposure parameters for a desired outcome.\\
\section{Summary and conclusion}
Despite the importance of radiation effects, majority of the models for coated bubble oscillations neglect or simplify radiation effects. Moreover, thermal effects are often approximated using linear models or are fully neglected. Additionally, pressure dependent dissipation effects during bubble oscillations are not fully understood. In this work we introduced a simple and comprehensive model for coated bubble oscillations. The model incorporates the compressibility of the medium to the first order of Mach number. The equations for the dissipation effects due to thermal damping, liquid viscous damping, coating viscous damping and radiation damping are derived with no linear simplifications. The dissipation mechanisms were then studied as a function of frequency and at different excitation pressures for the first time. We showed that radiation effects are important and can not be neglected. Radiation damping becomes more important with pressure increase. Even at frequencies below resonance, dissipation due to radiation can become the major dissipation mechanism with pressure increase.  This is in stark contradiction to the predictions of linear models. For uncoated bubbles, thermal effects are very important and can not be neglected. With pressure increase, predictions of the linear thermal model loses accuracy and inclusion of the full thermal effects are recommended. In case of coated bubbles that encapsulate gas cores similar to C3F8, thermal effects are not important and can be neglected even at higher pressures. We also showed that scattering to damping ratio is pressure dependent and there exists frequency and pressure ranges in which STDR is higher. The basic equations provided here and the presented fundamental information can be used to optimize the exposure parameters in applications and for more accurate characterization of bubbly media and coated bubbles.
\section{Acknowledgments}
The work is supported by the Natural Sciences and Engineering Research Council of Canada (Discovery Grant RGPIN-2017-06496), NSERC and the Canadian Institutes of Health Research ( Collaborative Health Research Projects ) and the Terry Fox New Frontiers Program Project Grant in Ultrasound and MRI for Cancer Therapy (project $\#$1034). A. J. Sojahrood is supported by a CIHR Vanier Scholarship.

\end{document}